# eROSITA Science Book:
# Mapping the Structure of the Energetic Universe

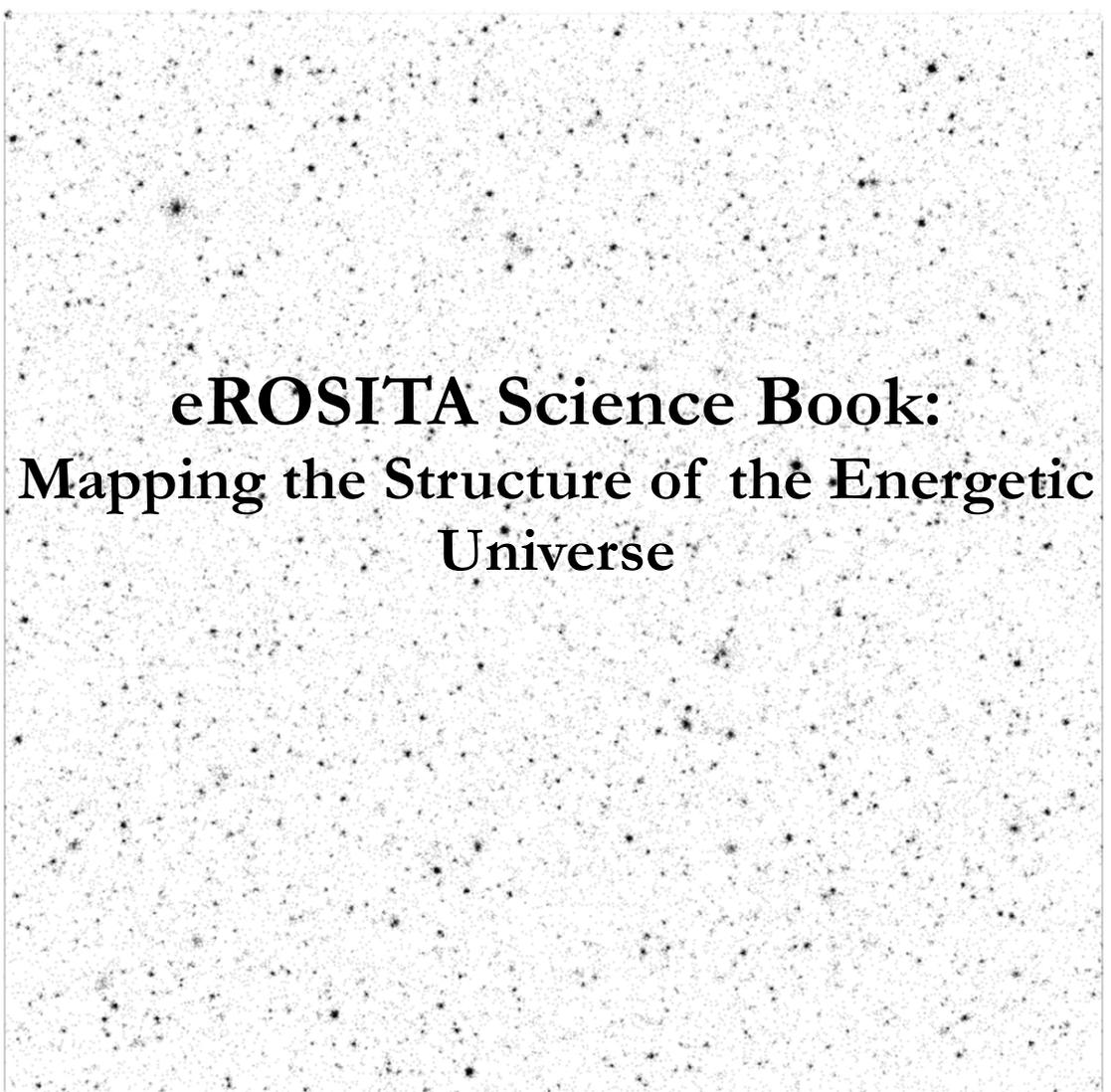

A. Merloni[1], P. Predehl[1], W. Becker[1], H. Böhringer[1], T. Boller[1], H. Brunner[1], M. Brusa[1], K. Dennerl[1], M. Freyberg[1], P. Friedrich[1], A. Georgakakis[1,2], F. Haberl[1], G. Hasinger[3], N. Meidinger[1,4], J. Mohr[5], K. Nandra[1], A. Rau[1], T.H. Reiprich[6], J. Robrade[7], M. Salvato[1], A. Santangelo[8], M. Sasaki[8], A. Schwope[9], J. Wilms[10]
and the German *eROSITA* Consortium

Edited by S. Allen[11,12], G. Hasinger[3], K. Nandra[1]

[1] Max-Planck-Institut für extraterrestrische Physik, Gießenbachstr. Postfach 1312, 85741 Garching, Germany
[2] National Observatory of Athens, I. Metaxa & V. Paulou, Athens 15236, Greece
[3] Institute for Astronomy, University of Hawaii, 2680 Woodlawn Drive, Honolulu, HI 96822-1839, USA
[4] MPI Halbleiterlabor, Otto-Hahn-Ring 6, 81739 München, Germany
[5] Department of Physics, Ludwig-Maximilians Universität, Scheinerstr. 1, 81679 München, Germany
[6] Argelander-Institut für Astronomie der Universität Bonn, Auf dem Hügel 71, 53121 Bonn, Germany
[7] Hamburger Sternwarte, Universität Hamburg, Gojenbergsweg 112, 21029 Hamburg, Germany
[8] Institut für Astronomie und Astrophysik, Universität Tübingen, Sand 1, 72076 Tübingen, Germany
[9] Leibniz-Institut für Astrophysik Potsdam, An der Sternwarte 16, 14482 Potsdam, Germany
[10] Dr. Karl Remeis-Sternwarte & ECAP, Universität Erlangen-Nürnberg, Sternwartstr. 7, 96049 Bamberg, Germany
[11] Kavli Institute for Particle Astrophysics and Cosmology, Department of Physics, Stanford University, 452 Lomita Mall, Stanford, CA 94305-4085, USA
[12] SLAC National Accelerator Laboratory, 2575 Sand Hill Road, Menlo Park, CA 94025, USA

# Executive Summary

*eROSITA* (extended ROentgen Survey with an Imaging Telescope Array) is the primary instrument on the Russian Spektrum-Roentgen-Gamma (*SRG*) mission. *eROSITA* is currently being built, assembled and tested under the leadership of the Max-Planck Institute for extraterrestrial Physics (MPE). In the first four years of scientific operation after its launch, foreseen for the year 2014, *eROSITA* will perform a deep survey of the entire X-ray sky. In the soft X-ray band (0.5-2 keV), this will be about 20 times more sensitive than the *ROSAT* all sky survey, while in the hard band (2-10 keV) it will provide the first ever true imaging survey of the sky at those energies. Such a sensitive all-sky survey will revolutionize our view of the high-energy sky, and calls for major efforts in synergic, multi-wavelength wide area surveys in order to fully exploit the scientific potential of the X-ray data. The all-sky survey program will be followed by an estimated 3.5 years of pointed observations, with open access through regular announcement of opportunities for the entire astrophysical community. With on-axis spatial resolution similar to the *XMM-Newton* one, a comparable effective area at low energies, and a wider field of view, *eROSITA* will provide a powerful and highly competitive X-ray observatory for the next decade.

The design-driving science of *eROSITA* is the detection of very large samples (~$10^5$ objects) of galaxy clusters out to redshifts $z \gtrsim 1$, in order to study the large scale structure in the Universe, test and characterize cosmological models including Dark Energy (DE). *eROSITA* is also expected to yield a sample of around 3 millions Active Galactic Nuclei, including both obscured and un-obscured objects, providing a unique view of the evolution of supermassive black holes within the emerging cosmic structure. The survey will also provide new insights into a wide range of astrophysical phenomena, including accreting binaries, active stars and diffuse emission within the Galaxy, as well as solar system bodies that emit X-rays via the charge exchange process. Finally, such a deep imaging survey at high spectral resolution, with its scanning strategy sensitive to a range of variability timescales from tens of seconds to years, will undoubtedly open up a vast discovery space for the study of rare, unpredicted, or unpredictable high-energy astrophysical phenomena.

In this living document we present a comprehensive description of the main scientific goals of the mission, with strong emphasis on the early survey phases. The following sections are the outcome of years of work of scientists in the German *eROSITA* Consortium, responsible for the definition, management, planning, implementation and operation of the *eROSITA* telescope. After a brief outline of the historical development of the telescope, we will describe the main technical characteristics and expected performances of the *eROSITA* telescope array (Section 2), the planned mission strategy (Section 3) and the simulation work done in order to make accurate predictions and forecasts (Section 4). The main science goals of the mission are then described in Section 5, while section 6 will be devoted to a discussion of the role of *eROSITA* in the context of current and future multi-wavelength wide-area surveys.

*Cover image: Simulated equatorial 3.6×3.6 deg² eROSITA field, with galaxy clusters, AGN and particle background.*
*Courtesy of C. Schmid and N. Clerc*

# Table of Content

# 1. Historical development

The *eROSITA* telescope concept is based on a long series of previous scientific and technological developments, dating back to the very successful German/US/UK *ROSAT* mission (1990-1999; Trümper 1982), developed and managed under the leadership of the Max-Planck Institute for extraterrestrial Physics (MPE). *ROSAT* carried out the first complete survey of the sky with an imaging X-ray telescope (in the energy range between 0.1 and 2.4 keV) and performed tens of thousands of pointed observations. The deepest of such pointed observations could resolve into individual sources the majority of the Cosmic X-ray Background (CXRB), the discovery of which motivated the Nobel Prize to R. Giacconi in 2002. These sources are mostly Active Galactic Nuclei (AGN), signposts of the growth of massive black holes, which populate the nuclei of almost all massive galaxies.

At the time of the *ROSAT* mission, it had however become clear that the Quasars and AGN observable in the soft-X-ray energy range, as well as in the optical part of the electromagnetic spectrum, were just a relatively small fraction of the entire population of accreting black holes. To explain the very hard shape of the CXRB spectrum, peaking at an energy of about 30 keV, the majority of AGN must have their light absorbed in passing through gas and dust in the central parts of their host galaxies, allowing thus only photons with energies greater than approximately 2 keV to escape. The following generation of large X-ray telescopes, the NASA-led *Chandra* Observatory and the ESA mission *XMM-Newton*, required the development of mirror systems with longer focal length (7.5-10 meters) in order to bring harder X-ray radiation to focus. These observatories, both launched in 1999, could however only perform pointed observations; due to their limited field of view, it was unfeasible to perform large area surveys with those telescopes. This led to the proposal of a hard X-ray imaging telescope with all-sky surveying capabilities. The Astrophysikalisches Institut Potsdam (AIP), MPE and the University of Tübingen (IAAT) proposed therefore the mission *ABRIXAS* (A BRoad-band Imaging X-ray All-sky Survey).

From the very beginning, the *ABRIXAS* mission concept was developed through the coherent adaptation of mirror and detector technologies developed for *XMM-Newton* for a national, small-scale satellite mission. Consequently, the development timescale for the project was relatively short (~ 3 years) and the overall costs moderate. The satellite project and construction was led by the OHB company in Bremen, the optics consisted of seven compact Wolter-1 telescopes (tilted by 7.25 degrees with respect to each other), each equipped with 27 thin Nickel mirror shells produced via Galvanoplastic technology. These mirrors were produced by the company Carl Zeiss, following in the steps of the previous *XMM-Newton* experience. The seven mirror modules shared in the focal plane an identical copy of the pn-CCD camera developed for *XMM-Newton*, and, with their small focal length of 1.6 meters, were ideal for a small satellite. Unfortunately, because of a design failure in the power supply, the satellite lost its main battery soon after launch, in April 1999, and could never be used for scientific purposes. The hard- and software developments enabled by the *ABRIXAS* mission have however been extremely useful in the framework of a number of subsequent projects.

Despite the unfortunate outcome of the *ABRIXAS* mission, the appeal of the original scientific goal of an imaging hard X-ray all-sky survey remained high, with no other planned astrophysical mission with similar objectives on the horizon. With the help of the Semiconductor Laboratory (HLL) for the production of high-sensitivity detectors of the Max-Planck Institutes for Physics and extraterrestrial Physics, a new project to further develop the highly successful *XMM-Newton* pn-CCD technology was launched. In March 2002, all participating institutes proposed to ESA to accommodate the *ROSITA* (ROentgen Survey with an Imaging Telescope Array) telescope on an external platform of the International Space Station (ISS). The seven mirror modules would have been built identical to the *ABRIXAS* ones (with varying tilt between 4 and 6 degrees), but each of them would have been equipped with its own, newly developed, frame-store pn-CCD in the focal plane. In September 2002, such a proposal was supported by ESA with the highest scientific priority, and recommended for phase-A study. Launch, however, would have been possible only from 2011 onwards, due to the occupation of the external platforms of the ISS. It emerged, subsequently, that the ISS would not have been a viable option for *ROSITA*, first of all because NASA decided to terminate its Shuttle flights to the ISS in 2010, and furthermore, because a contamination-experiment on the ISS revealed that its environment was not safe from contamination for the sensitive X-ray mirrors and detectors of *ROSITA*.

At the turn of the millennium, the wealth of new cosmological observations led to a cascade of new findings and discoveries. Among those, the observations of supernovae type Ia by two independent groups (awarded with the Nobel Prize for Physics in 2011 for their discoveries) revealed that the expansion of the Universe is accelerating, a fact that may suggest the existence of a "cosmological constant" $\Lambda$, whose effect is best described by the vacuum energy density $\Omega_\Lambda$. The subsequent measures, taken by *Boomerang* and *WMAP,* of the tiny temperature fluctuations of the microwave background radiation out of which galaxies, clusters of galaxies and the overall large scale structure of the Universe originated, pointed towards a "flat Universe", whereby the total matter plus energy content of the cosmos attains almost exactly the critical value (i.e., the sum of matter and energy density in the Universe obeys the relation $\Omega_\Lambda + \Omega_M =1$). In fact, studies of the baryon fraction in X-ray selected (mainly from *ROSAT*) clusters of galaxies throughout the '90s had already shown compelling evidence that $\Omega_M <1$ (Briel et al. 1992; White et al. 1993). The further surprising discovery of very massive (about $10^{15}$ $M_\odot$) X-ray emitting clusters of galaxies at redshift z>0.8, whose mere existence demonstrated that the matter density in the Universe must be well below the critical value, was also a critical milestone in this cosmological revolution. These independent measures all contributed to convince a large fraction of astronomers of the existence of a still not understood "Dark Energy" component, that, in the currently favored "concordance cosmology" dominates the energy density budget of the Universe ($\Omega_\Lambda \sim 0.7$). Its effect only becomes apparent at scales comparable with the size of the Universe itself, and is otherwise so small that only very wide area astronomical surveys can be used to obtain detailed information on its nature.

The fact that very large sample of clusters of galaxies can be particularly useful for precision-cosmology stimulated many different groups to conceive dedicated large area clusters surveys. In April 2003 members of the *ROSITA* team participated in the proposal of *DUO*, a NASA SMEX-mission, based on a modification of the *ROSITA* telescope design. Together with other four missions, *DUO* was selected among 36 competing proposals for a phase-A study, carried out in 2004. NASA, however, did not select the *DUO*-project for further development, with the only mission among the five finally executed within the SMEX program being the hard X-ray focusing telescope *NuSTAR*, launched in 2012.

*DUO* would have surveyed an area of the sky of about 6.000 deg$^2$, overlapping that explored by the optical Sloan Digital Sky Survey (SDSS). In that way, about 10.000 clusters of galaxies could have been detected, that would have been able to provide constraints on the Dark Energy density accurate at the 10% level. In February 2005 the "Astronomy and Astrophysics Advisory Committee" (AAAC), founded by the National Science Foundation (NSF), NASA and the Department of Energy (DOE), and the "High Energy Physics Advisory Panel" (HEPAP), founded by NSF and DOE, established a "Dark Energy Task Force" (DETF) with the task of advising NSF, NASA and DOE on the optimal strategies for the future of Dark Energy research. In particular, the DETF evaluated and compared the various ground- and space-based instrumentation and observational methods. Within this framework, it became clear that a *DUO* survey of about 10,000 galaxy clusters would have been insufficient for a detailed study of Dark Energy. In particular the measurement of "Baryonic Acoustic Oscillations" (BAO) in the cluster power spectrum, that could lead to a model-independent determination of cosmological parameters, would require a much larger number of objects (see e.g. Angulo et al. 2005). A dedicated "White Paper" (Haiman et al. 2005) reached the conclusion that, within available technologies at the time, it would have been possible to gather a sample of about 100,000 X-ray selected clusters of galaxies, that would have provided very stringent constraints on the fundamental parameters of the cosmological model of the Universe.

The construction of a sample of galaxy clusters of that magnitude is the primary objective of the *eROSITA* telescope. To achieve this goal, the *ABRIXAS* mirror design had to be modified to include 27 additional shells, doubling the diameter of each telescope module and increasing the effective area at low energies by a factor of five. With such a fundamental re-scope, *eROSITA* will likely be the first Stage IV Dark Energy probe to be realized and will outperform the *DUO*-like Stage IV probe originally considered by the DETF.

In June 2006 the funding proposal for *eROSITA* was submitted to the German Space Agency (DLR). Five German institutes (MPE, IAAT and AIP, the Hamburg Observatory and Dr. Remeis-Sternwarte in Bamberg - the Astronomical Institute of the Erlangen-Nürnberg University) agreed to work together to develop, build and organize the scientific exploitation of the instrument and formed the German *eROSITA* Consortium. They were later joined by three more Institutes: the Argelander-Institut für Astronomie at the University of Bonn, the Max-Planck Institut für Astrophysik (MPA), and the Universitäts-Sternwarte München (LMU). In March 2007 DLR approved the project and funded the *eROSITA* telescope, while a Memorandum of Understanding was signed between DLR and the Russian Space Agency (Roskosmos) to ensure that *eROSITA* would be launched in the framework of the Spektrum-Roentgen-Gamma mission. Soon afterwards, the construction of the telescope began. In September 2008 Roskosmos came to a final decision on the payload orbit and launcher. *eROSITA* will be launched by a Zenit-Fregat launcher together with the Russian hard-X-ray telescope *ART-XC* on a L2 orbit to ensure maximal efficiency for a survey instrument. Eventually, additional funds from the Max-Planck Society and DLR were secured for *eROSITA* in July 2009, to compensate the cost-driving L2 orbit. A detailed agreement between Roskosmos and DLR was signed the following month. Since fall 2009 *eROSITA* entered phase C/D. With all major contracts placed, instrument assembly, testing and calibration work proceeds towards the expected launch in 2014.

# 2. Technical characteristics

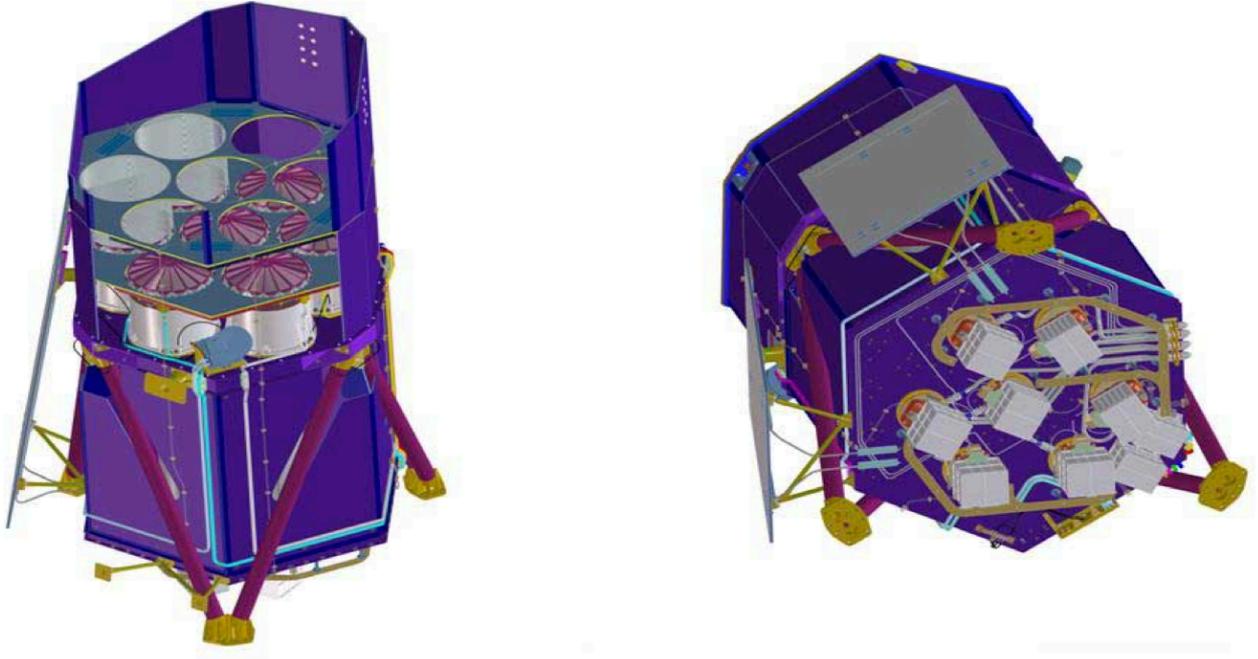

*Figure 2.1.1: Schematic view of the eROSITA telescope. On the left, where the front cover is omitted, the seven telescopes with their X-ray baffles are visible. On the right, the view from the bottom reveals the seven CCD housings.*

## 2.1 *eROSITA* **X-ray optics and imaging capabilities**

The *eROSITA* mirror system consists of 7 identical mirror modules with 54 mirror shells each, and a baffle in front of each module. Compared to a large single mirror system, the advantages of a multiple mirror system are mainly a shorter focal length (and thus reduced instrumental background), and a reduced pileup when observing bright sources. Obviously, such a configuration allows a more compact telescope, and multiple identical cameras (see Section 2.2 below), which automatically provide a 7-fold redundancy. Mirror production is based on a replication technique developed for *XMM-Newton* and then applied to *ABRIXAS*, which had scaled the *XMM-Newton* telescopes focal length down by a factor of about 4. As the *eROSITA* focal length is kept at the same value (1.6 meters), the *ABRIXAS* optical design and manufacturing process have been adopted for *eROSITA*, too, and the inner 27 mirror shells being in fact identical. Unlike on *ABRIXAS*, however, the seven optical axes are co-aligned (Figure 2.1.1), ensuring an identical field of view for the seven telescopes of the array.

*Table 2.1.1: eROSITA Telescope*

| Mirror Modules | 7 |
|---|---|
| Mirror shells per module | 54 |
| Energy range | ~0.2-10 keV |
| Outer diameter | 358 mm |
| Inner diameter | 76 mm |
| Length of shells | 300 mm |
| Angular resolution (HEW@ 1.5 keV) | < 15" on axis |
| Grazing angles | 20' < α < 96' |
| Wall thickness | 0.25 - 0.54 mm |
| Micro-roughness | < 0.5 nm RMS |
| Mirror coating | Au (>50 nm) |
| Focal length | 1.6 m |
| Weight (one module, incl. baffle) | <50 kg |
| Material of shell | Nickel |

The mirror shells are manufactured by replication from super-polished mandrels. The reflective layers of gold copy the surface of the mandrel to get the required X-ray optical quality. The carrier material of the reflecting surface is electroformed nickel, and the wall thickness of the mirror shells varies between 0.25 and 0.54 mm. All shells are adjusted and bonded to a supporting spider wheel. The main characteristics of the mirror system are summarized in Table 2.1.1.

The capabilities of the X-ray mirror system can be roughly characterized by three numbers: the effective area, the vignetting function, and the Point-Spread Function (PSF). As it is typical for Wolter-type geometry mirrors, the PSF (or the Half Energy Width; HEW) is rapidly degrading (increasing) at higher off-axis angles (see left panels of Fig. 2.1.3). Since the entire field of view (FoV) is used for the surveys to maximize the efficiency with which *eROSITA* can cover the whole sky down to a given limiting flux, the effective average spatial resolution of the telescopes will be given by an average over the FoV.

During the all-sky survey (see section 3 below), as the telescope scans the sky, each source will move on a track in the detector plane. A source in a typical (low ecliptic latitude) field will be observed up to six times for each of the 8 all-sky surveys. Figure 2.1.2 shows, as an illustrative example, a simulated very bright (0.5-2 keV flux of $10^{-9}$ erg cm$^{-2}$ s$^{-1}$) source as it moves across the detector plane in different scans. The left panel shows the tracks of six successive scans in one visit of the region of the sky containing he source during one 6-months all-sky survey, while the right panel shows all the tracks of that particular source over the course of 4-years (8 all-sky surveys). The obvious degradation of the PSF at large off-axis angles causes the tracks to widen substantially at the edges of the detector. The right panel of Figure 2.1.3 shows the expected HEW at ~1 keV, as a function of off-axis angle, both for a pointed observation (red curve) and averaged over the FoV in survey (scanning) mode, calculated assuming the detectors will be placed in focus on axis. The averaging procedure, including the effects of shadowing and vignetting, causes the effective PSF of *eROSITA* in survey (scanning) mode to be as large as ~28" in the soft band and ~40" in the hard band.

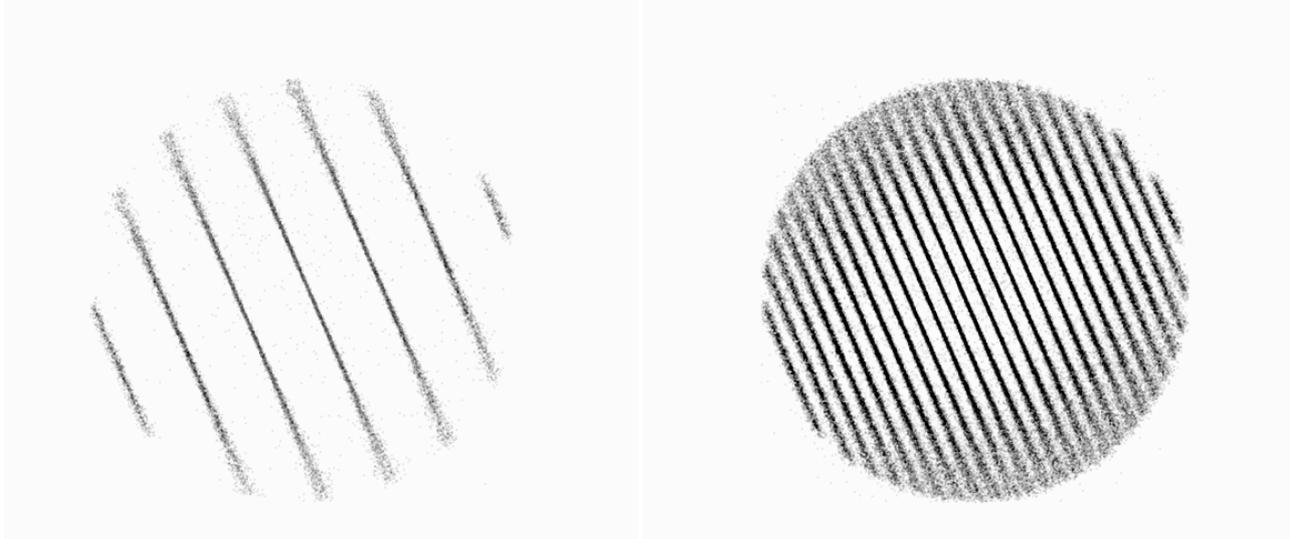

*Figure 2.1.2: Simulated images of a series of eROSITA scans over a very bright, point-like source at low ecliptic latitude (0.5-2 keV flux of $10^{-9}$ erg cm$^{-2}$ s$^{-1}$). The left panel shows the tracks of six successive scans in one visit of the region of the sky containing he source during one 6-months all-sky survey, while the right panel shows all the tracks of that particular source over the course of 4-years (8 all-sky surveys). Courtesy of N. Clerc.*

A system to prevent stray light from reaching the detectors had to be developed, including telescope baffles and a blocking filter on the CCD (details of which are still to be decided). The worst-case analysis indicated the visible photon background is $1 \times 10^{-2}$ photons/s/pixel. The baffle tubes are thermally isolated from the temperature controlled mirror modules by GFRP isolations. An additional X-ray baffle (54 concentric cylindrical shells) will reduce the amount of single reflections from the rear end of the hyperboloid, if a bright source is just outside the field of view. The mirror system must be maintained at 20±2 °C during operation to avoid deformation and image degradation.

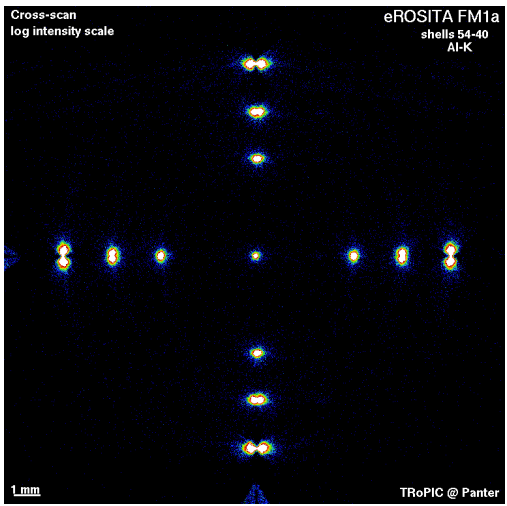 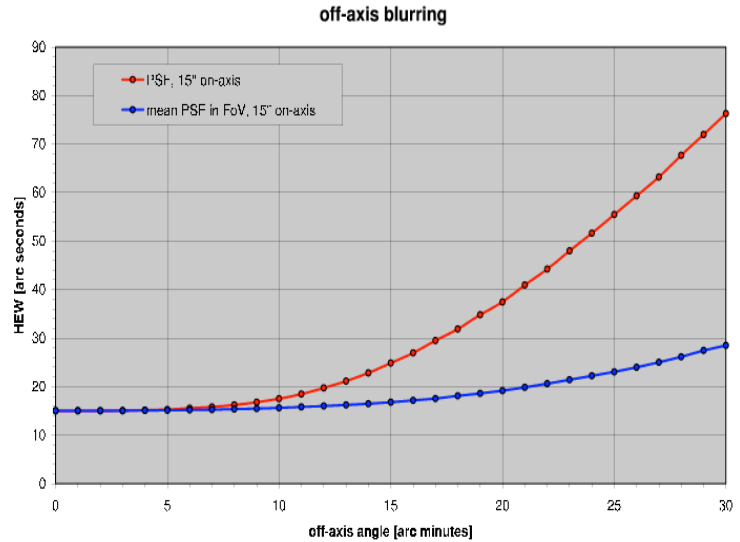

*Figure 2.1.3: Left: PANTER test of eROSITA Flight Mirror Module FM1, with 15 inner mirror shells integrated. Shown is the PSF response of the mirror module at various off-axis angles to a point-like Al-K (1.5 keV) source. On-axis HEW is ~13.1 arcsec. Right: Computed HEW from a ray-tracing calculation. Red Curve: the PSF HEW as function of the off-axis angle in pointing mode. Blue Curve: the soft energy band PSF HEW averaged within an encircled off-axis angle during a scan, as a function of this off-axis angle (including vignetting): the average over the whole FoV is about 28".*

## 2.2 *eROSITA* cameras

Since more than 20 years, the HLL semiconductor laboratory manufactures radiation sensitive detectors for space applications. Outstanding achievements are the pn-CCD camera onboard *XMM-Newton* and the sensors operating on the Mars-rovers Spirit and Opportunity. The pn-CCD camera onboard *XMM-Newton* (Strüder et al. 2000) has been working without reduction of its performance for more than ten years. In recent years, the concept has been further developed, providing the following major improvement for the *eROSITA* cameras (Meidinger et al. 2011):

- The CCD has been extended by a frame store area which allows the fast shift from the image area to reduce so-called "out-of-time events", photons which are recorded during charge-transfer.
- The pixel size has been reduced to 75μm to better adapt to the resolution of the *eROSITA* telescope.
- The use of 6-inches silicon wafers with 450μm thickness gives higher quantum efficiency at higher energies.
- The low energy response and energy resolution have been improved by a modification of the wafer-processing. In principle, the *eROSITA*-CCD could work at operating temperature as high as -60°C (for comparison, *XMM-Newton*: -90°C). However, we conservatively aim for a temperature of -90° because we do not yet have an experience with radiation damage at L2.

Each of the seven mirror modules has its own camera in its focus, equipped with a CCD-module and a processing electronics. The *eROSITA*-CCD has 384×384 pixels or an image area of 28.8mm×28.8mm, respectively, for a field of view of 1.03° diameter. The 384 channels are read out in parallel by three modified CAMEX-ASICs (amplification and readout chips, Figure 2.2.1). The nominal integration time for *eROSITA* will be 50 ms. The integrated image can be shifted into the frame store area within 130μsec before it is read out within about 10 ms. The CCD, together with the three CAMEX and the (passive) front-end electronics are integrated on a ceramic printed circuit board (= CCD-module) connected to the "outer world" by a flexlead. A flight-batch of 32 CCDs has been already fabricated. The CCDs are tested and parameters have been studied in detail in the HLL semiconductor lab. The CCD-module is already designed, fabricated, tested, and qualified (thermal vacuum, vibration). A prototype of the *eROSITA* camera (TRoPIC) is working in the PANTER X-ray test facility since March 2007.

The design of the camera housing follows particular thermal constraints to reduce parasitic thermal loads (Figure 2.2.2). While the active load (primarily CAMEX) is 0.74W, the total heat load is of the order of 1.5W per camera. A temperature control maintains the stability of the CCD-temperature within ±0.5K. The CCD is shielded against particle radiation by a massive copper housing surrounding the entire CCD-Module. Fluorescence X-ray radiation generated by cosmic particles is minimized by a graded shield consisting of aluminium and boron carbide or beryllium, respectively. For calibration purposes, each camera contains a radioactive $^{55}$Fe source and an Al/Ti target providing three major spectral lines at 1.5keV (Al K$\alpha$), 4.5keV (Ti K$\alpha$), and 5.9keV (Mn K$\alpha$), plus the Ti K$\beta$ and Mn K$\beta$ lines.

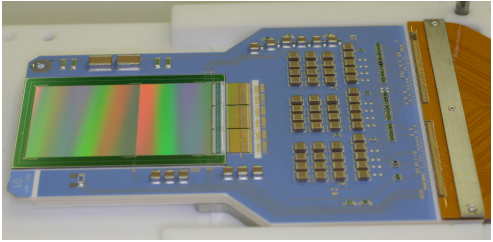

*Figure 2.2.1: eROSITA CCD-Module. The CCD with its image area (3×3 cm²) and the slightly smaller frame store area (right) is connected via 384 bond wires with three CAMEX read out chips. They are mounted, together with the (passive) front end electronics, on a ceramic printed circuit board (blue). The flexlead on the right connects the CCD-Module with the experiment electronics.*

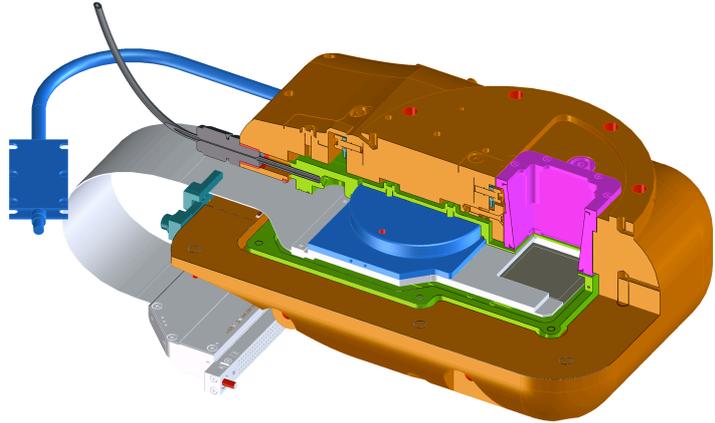

*Figure 2.2.2: Schematic representation of the eROSITA Camera with the (warm) Cu-proton shield (brown), the housing (cold, green), and the CCD-Module (grey) which is thermally connected to heat pipes by a Ti-block (blue). An additional graded shield (magenta) suppresses fluorescent X-rays generated in the instrument and leading to an enhanced background.*

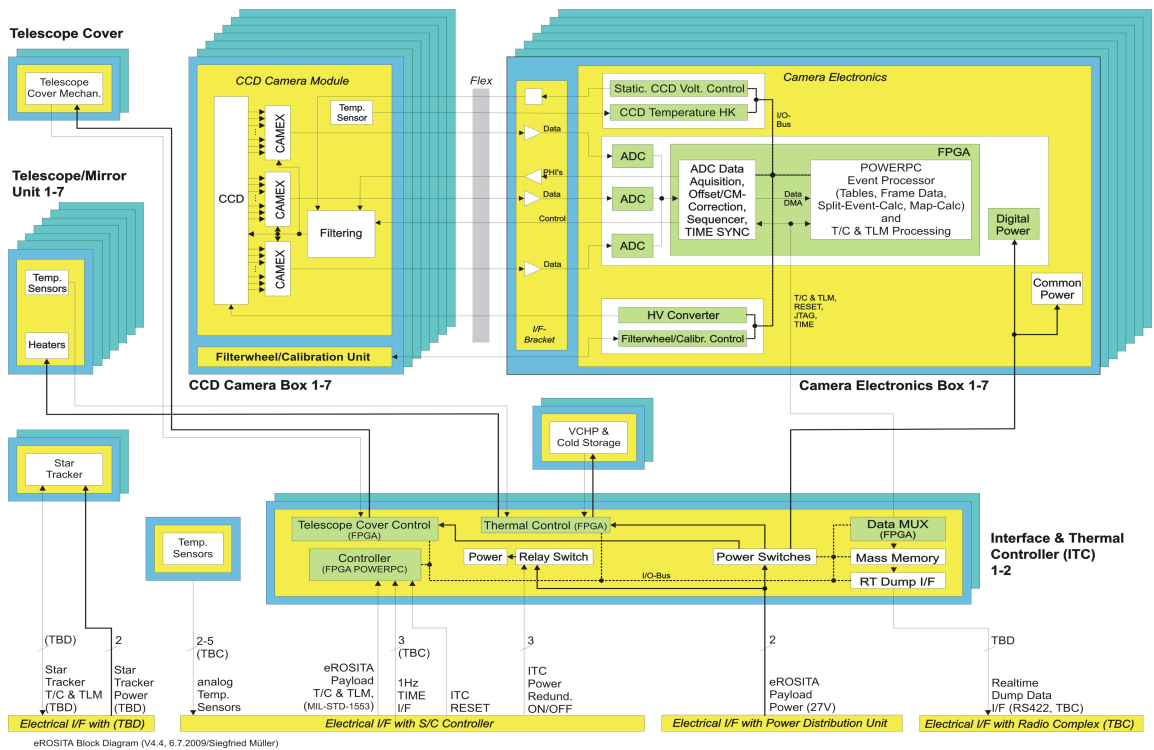

*Figure 2.2.3: Scheme of the Camera Electronics. Each camera has its own (identical) electronics. An 8th electronics box is for (thermal) control of the instrument.*

The experiment electronics (see figure 2.2.3), separated in seven electronics boxes has the following tasks:

- A "sequencer" consisting of a FPGA logic provides the correct timing signals for CCD, CAMEX and the two ADC.
- Two 14-bit ADC for each CCD-module digitize the CAMEX output signal.
- The main event processing is performed by the FPGA logic which incorporates also two "Power-PCs". This comprises the subtraction of an offset-map, the correction of common modes, the application of pixel-wise thresholds using a previously stored noise map, and the delivery of event data frames to the controller. The FPGA acts also as control unit, which collects telemetry information (both, events and housekeeping), executes commands, and forms the interface to the spacecraft (S/C).

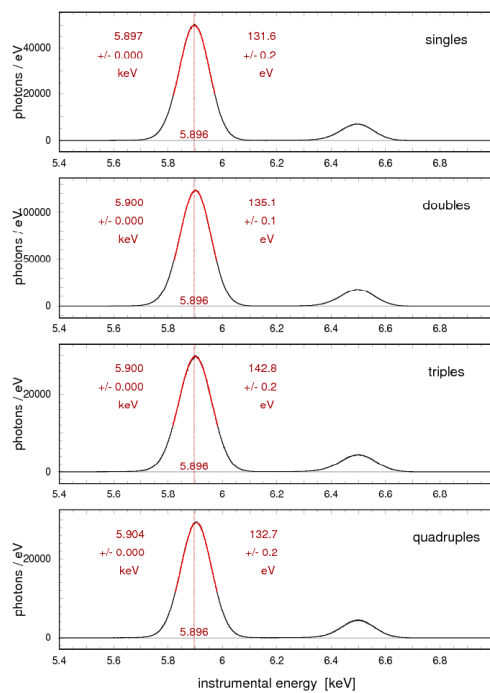

We use radiation-hard components for the entire electronics. The average data rate is determined by the brightness of the extragalactic sky background. We assume to detect between 30 and 40 photons per second (all modules). Since most of the detected photons produce charges in more than one pixel (split-events), and each event is coded using 30 bits, we will have ~3.6kbit/s. With some overhead and together with HK data, the data rate will be below 10kbits/s. A bright source (e.g. Cyg X-2) will increase this number to 100kbits/s, for a short time.

Commanding of the *eROSITA* cameras is simple because only a few operational modes are implemented:
- Standby: camera completely switched off, survival heaters are controlled by S/C;
- Checkout, Test: Electronics switched on, CCD still off;
- Normal: camera is completely switched on and working; this includes also calibration.

*Fig. 2.2.4 Spectra for all singles and recombined valid patterns, obtained from HK111205.045 at Mn-Kα during the tests of the eROSITA CCD Engineering model. For single pixel events, a spectral resolution of ~132 eV at 5.9 keV is achieved.*

At the time of writing, the engineering model of the *eROSITA* detector has been extensively tested. A FWHM for the Mn-Kα line at 5.9 keV of ~132 eV has been measured (see Figure 2.2.4), which fulfills the project specification. The mean detector read noise has a value of 2.6 electrons ENC rms, a factor of two better compared with the *XMM-Newton* pn-CCD. First tests also indicate a very low CTI of ~1.1e-5. Finally, a sub-pixel reconstruction algorithm has been developed and applied to *eROSITA* mirror module measurements at PANTER. Under specific conditions, this might improve the spatial resolution from the native 9.6" pixel size to ~2" (Dennerl et al. 2012).

## 2.3 Telescope structure and thermal control

The optical bench forms the mechanical interface to the S/C bus and connects the mirror system and the baffles on one side with the focal plane instrumentation on the other side. It has to meet the tough requirements concerning cleanliness, which mandate a front cover, closed during ground operations and liftoff. The stiffness of the structure requires particular emphasis because the weight of the telescope is dominated by the mirror system which is far above the S/C structure. The dimensions of the telescope structure will be of the order of 1.9 m diameter x 3.1 m height (launch configuration). The total weight of *eROSITA* is specified to 800 kg maximum

The CCD-Modules will be cooled down to about -90°C passively by means of two radiators and a sophisticated system of cryogenic heatpipes. The temperature stability of ± 0.5° is maintained by so-called 'variable conductance' heatpipes. Additional heaters on the mirrors will keep their temperature at 20°C±2°. The complete thermal systems will be controlled by an Interface and Thermal Control (ITC) Electronics which forms another electronics box of *eROSITA*. This electronics also provides the data link to and from all cameras, as well as to the spacecraft. This ITC is cold redundant; *eROSITA* therefore comprises 9 electronics boxes in total.

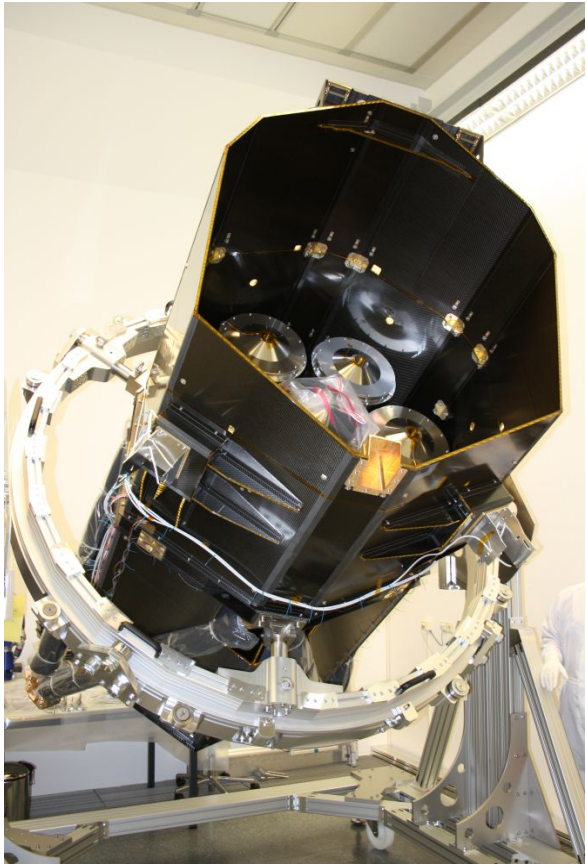

*Figure 2.4.1: The eROSITA telescope structure during the integration phase at MPE, in spring 2012. The telescope is mounted on a ring support for ease of manipulation, and shows, without its front cover, the mirror modules.*

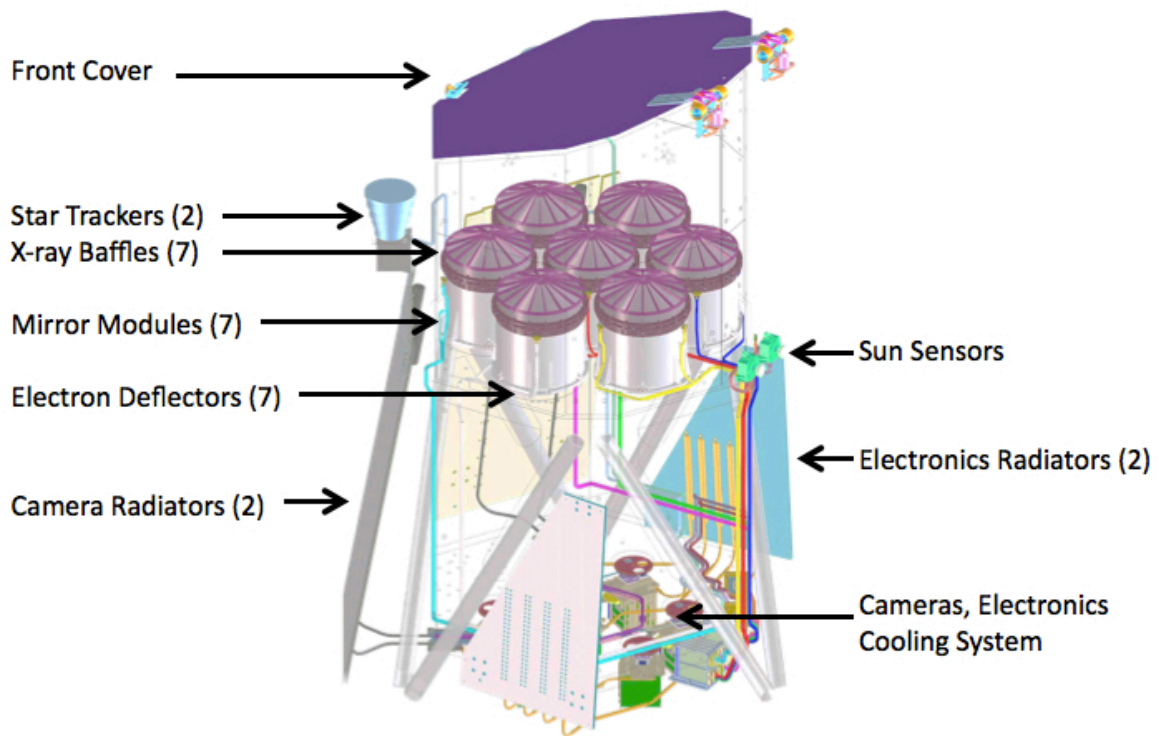

*Figure 2.4.2: Schematic diagram of the eROSITA telescope structure.*

# 3. Observing strategy

The *SRG* satellite will be launched by a Zenit-Fregat launcher from Baikonur in Kazakhstan and inserted into a trajectory to the second Lagrangian point L2. The satellite will be placed in an orbit with a semi-major axis of about 1 million kilometers and an orbital period of about 6 months. The L2 point is rapidly establishing itself as a preeminent location for advanced space probes, with a number of missions that do and will make use of this orbital 'sweet-spot' in the coming years. L2 is home to ESA missions such as *Herschel* and *Planck*, and there will fly *GAIA*, the *James Webb Space Telescope* and *Euclid*.

L2 is one of the so-called Lagrangian points, discovered by mathematician Joseph Louis Lagrange. Lagrangian points are locations in space where gravitational forces and the orbital motion of a body balance each other. Therefore, they can be used by spacecraft to 'hover'. L2 is located 1.5 million kilometers directly 'behind' the Earth as viewed from the Sun. It is about four times further away from the Earth than the Moon ever gets and orbits the Sun at the same rate as the Earth. A spacecraft in L2 would not have to make constant orbits around the Earth, passing in and out of the Earth's shadow and causing it to heat up and cool down, distorting the image quality of any mirror system. Free from this restriction and far away from the heat radiated by Earth, L2 provides a much more stable viewpoint; moreover the absence of Earth occultations allows long continuous observations. A disadvantage of L2 is the substantial higher load by cosmic radiation compared with an equatorial low earth orbit (see Perinati et al. 2012, and section 4.2). This results in a higher background and radiation damages of the CCDs. The latest pn-CCD radiation tests on ground, however, have shown that these are not limiting their performance.

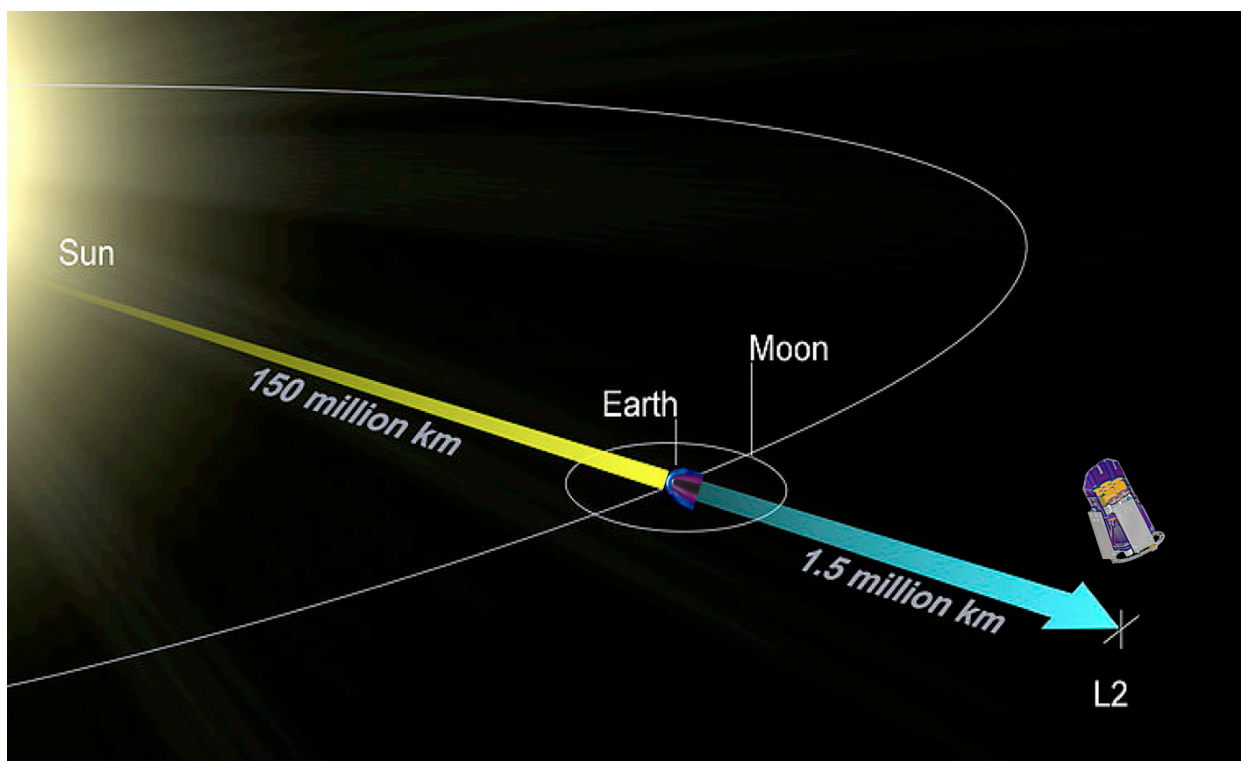

*Figure 3.1.1: Schematic view of the location of the L2 orbit of SRG. At the position of the earth, a scaled picture of the geocoronal emission is shown. Unlike any other X-ray satellite launched to date, eROSITA will become the first telescope to observe the X-ray sky from L2, unaffected by geocoronal X-ray emission (composite image courtesy of K. Dennerl).*

After a ~3 months cruising phase to L2, during which commissioning tests and calibration will be carried out, *eROSITA* will start its 4-years long all-sky survey, composed of 8 successive passages over the entire celestial sphere (in the following, *eROSITA* All-Sky Survey, or eRASS:1-8, with numbers denoting the increasing exposure from the first to the eighth passage). The current baseline mission scenario is based on a simple scan geometry, having the satellite rotation axis facing towards the Sun. This leads to an overlap of all great circles at the ecliptic poles, allowing a full coverage of the entire sky in about 180 days, with relatively uniform coverage and two deeply exposed regions at the ecliptic poles. Small variations of such a scanning law are envisaged, possibly moving regularly the scanning axis a few degrees away from the sun in order to avoid hitting confusion limits in the deeply exposed polar areas (see section 3.1 below).

The final combined survey (eRASS:8) will cover the entire sky to orders of magnitude deeper fluxes than any existing or foreseen X-ray instrument, both for point-like and extended sources (see Figure 4.1 for a comparison of *eROSITA* with existing surveys in the area-sensitivity plane).

All-sky survey data will be proprietary for the *eROSITA* team for a fixed period of time (less than 2 years), and will be split into two equal, non-overlapping sky parts for the German and Russian *eROSITA* Consortia, respectively. The division line cuts the Galactic plane in two halves, through the Galactic Center and the north and south galactic poles (see Figure 6.1.1 below). At the end of the survey period, the *SRG* satellite will be operational for at least 3.5 more years, during which regular pointed observations will be possible.

## 3.1 All-sky survey scanning law

During the four year all-sky survey, the *eROSITA* telescopes will scan the sky in great circles with a scanning speed corresponding to one full circle being completed every four hours. The scanning axis is either pointed directly towards the sun or alternatively a few degrees away from it (see next subsection for details). As the satellite moves around the sun, the plane of the scans is thus advanced by about one degree per day, resulting in a full coverage of the sky every half year. The scanning law is also slightly modulated since the satellite is located in an L2 halo orbit with a semi-major axis of about 1,000,000 km and an orbital period of about 6 months.

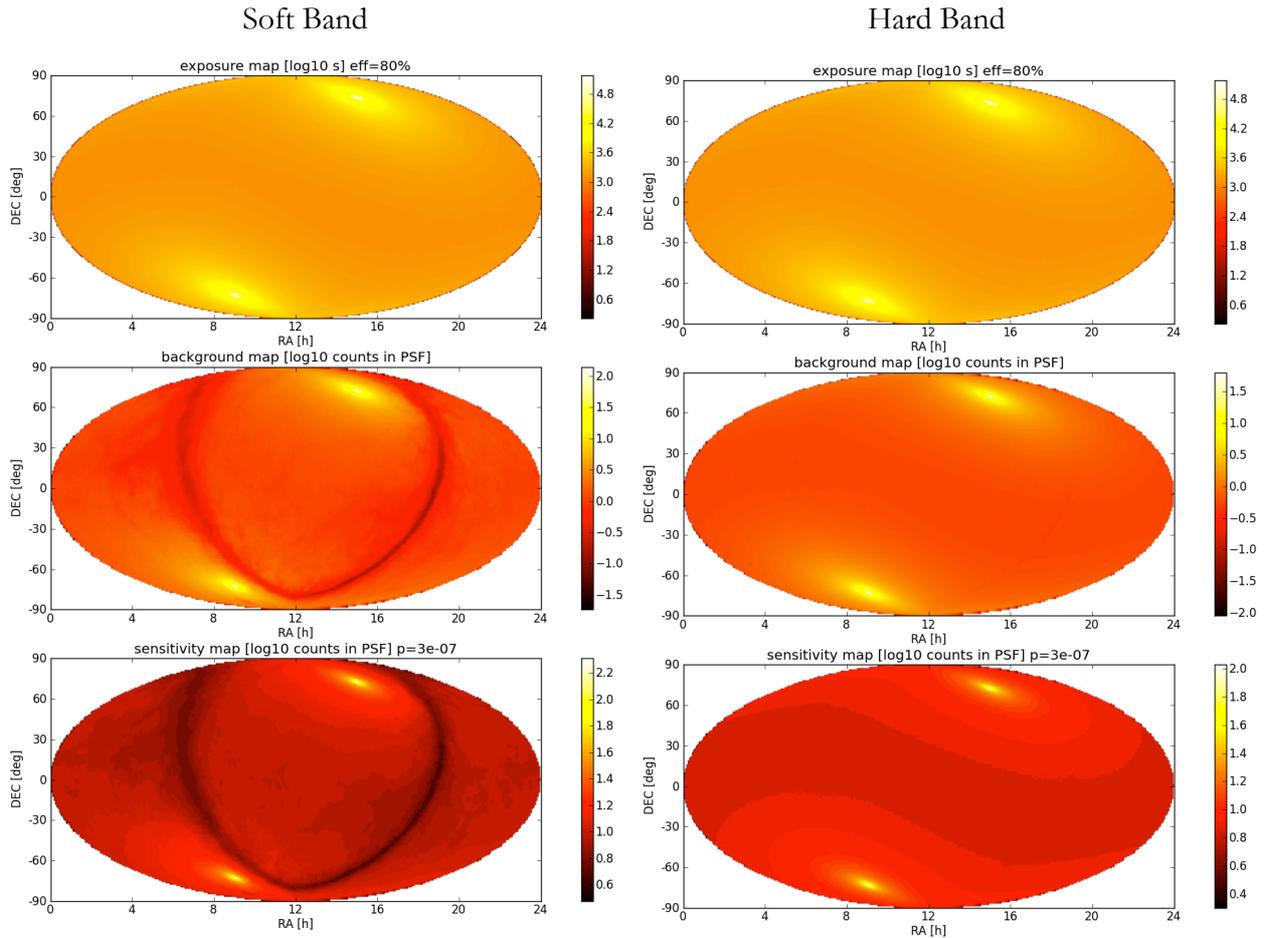

*Fig. 3.1.2 Exposure map, Background map and sensitivity map calculated for the eROSITA soft band (0.5-2 keV; left column) and hard band (2-10 keV; right column) assuming sun-pointing scan axis, a 80% survey efficiency, the background model of section 4.2 below, and a foreground $N_H$ map from Kalberla et al. (2005), known as the Leiden/Argentine/Bonn (LAB) survey. Images courtesy of J. Buchner.*

## 3.2 Survey exposure

After the completion of the four year all-sky survey, an average exposure of 2,548 sec[1] will be reached, assuming an observing efficiency of 100%. This number is independent of the precise scanning law. The approximate exposure at a specific sky location can be computed by the following formula, again assuming 100% observing efficiency (lat is ecliptic latitude): $T_{exp}$ ~ 1627/cos(lat) sec (for -84 deg < lat < 84 deg) and $T_{exp}$ ~ 17,500 sec within 6 degrees of each ecliptic pole, depending on the details of the scanning law.

For more accurate figures, that include a more realistic 80% observational efficiency, Figure 3.1.2 shows the exposure map (top panel) calculated assuming an inclined L2 halo orbit (axis length: 0.96/0.3 Million km, i=30 deg, P=0.5 yr), as well as the background (middle) and sensitivity (bottom) maps in the 0.5-2 and 2-10 keV bands, computed from the background model described in section 4.2 below and a neutral Hydrogen map from Kalberla et al. (2005; LAB survey).

Due to the scanning strategy, the highest *eROSITA* exposure is achieved close to the ecliptic poles and the lowest exposure will be reached at the ecliptic equator. However, the exact exposure law close to the ecliptic poles depends on the orientation of the scan axis throughout the survey period. Relatively high, uniform exposure in an extended area of 100 deg² or more around each pole can be accomplished by not pointing the scan axis directly at the sun but at different locations slightly away from it. A specific scanning scenario currently under investigation calls for pointing the survey scan axis at different, fixed angles away from the sun direction (perpendicular to the plane of the ecliptic), keeping it fixed at each of these locations for one year. This results in the scan poles "precessing" around the ecliptic poles once per year with the selected offset angles. The four angles can be optimized to achieve a widened area of deep exposure around the ecliptic poles.

Different sets of offset angles allow to control the depth of the polar exposure. The increase in size of the deeper exposed area is in the range of a few up to ten percent, depending on the chosen maximum exposure time (see Table 3.1.1). The offsets have positive beta-amplitudes of 0.2, 0.7, 1.2, 1.7 deg (offset scenario #1); 0.3, 1.0, 1.8, 2.5 deg (offset scenario #2) and 1.0, 2.8, 4.8, 7.0 deg (offset scenario #3).

| scan axis offset (degrees) | Sky area above given exposure (deg²) | | | | | Minimal exposure (ks) | Max. exposure at poles (ks) |
|---|---|---|---|---|---|---|---|
| | >10 ks | >15 ks | >20 ks | >25 ks | > 30 ks | | |
| 0 (sun pointing) | 550 | 240 | 140 | 90 | 60 | 1.62 | >100 |
| #1 (0.2, 0.7, 1.2, 1.2) | 560 | 245 | 140 | 95 | 65 | 1.635 | 50 |
| #2 (0.3, 1, 1.8, 2.5) | 565 | 250 | 140 | 90 | 55 | 1.645 | 35 |
| #3 (1, 2.8, 4.8, 7.0) | 580 | 0 | 0 | 0 | 0 | 1.69 | 13 |

*Table 3.1.1: Sky exposure values for the sun- and three offset-pointings described above: the area (in deg²) above fixed given exposure times are given as well as the approximate values for the exposures in the equatorial and polar areas.*

---

[1] The average exposure has been computed in the following way: (*eROSITA* FoV)/41253 deg² ×(four years) = 2548 sec; on the other hand, the minimum exposure is given by: 2×(*eROSITA* FoV)/(360 deg)² ×(four years) = 1622 sec, where the *eROSITA* Field of View (FoV) is equal to 0.833 deg²

# 4. Simulating the *eROSITA* sky

In this section we present a quantitative assessment of the expected technical performances of the *eROSITA* telescope which are relevant for predicting the main characteristics of the planned observations, with particular emphasis on the survey mode. As discussed in Section 1, *eROSITA* has been designed from the very beginning as an instrument capable of exploring the uncharted region of the parameter space where very wide area, sensitive surveys are (see Figure 4.1).

In the following, we describe how a model of the *eROSITA* background can be created, and use this information, in conjunction to that on the telescope effective area (see section 4.1), to calculate the sensitivity as a function of exposure time, which will then critically inform our discussion of the science output in Section 5.

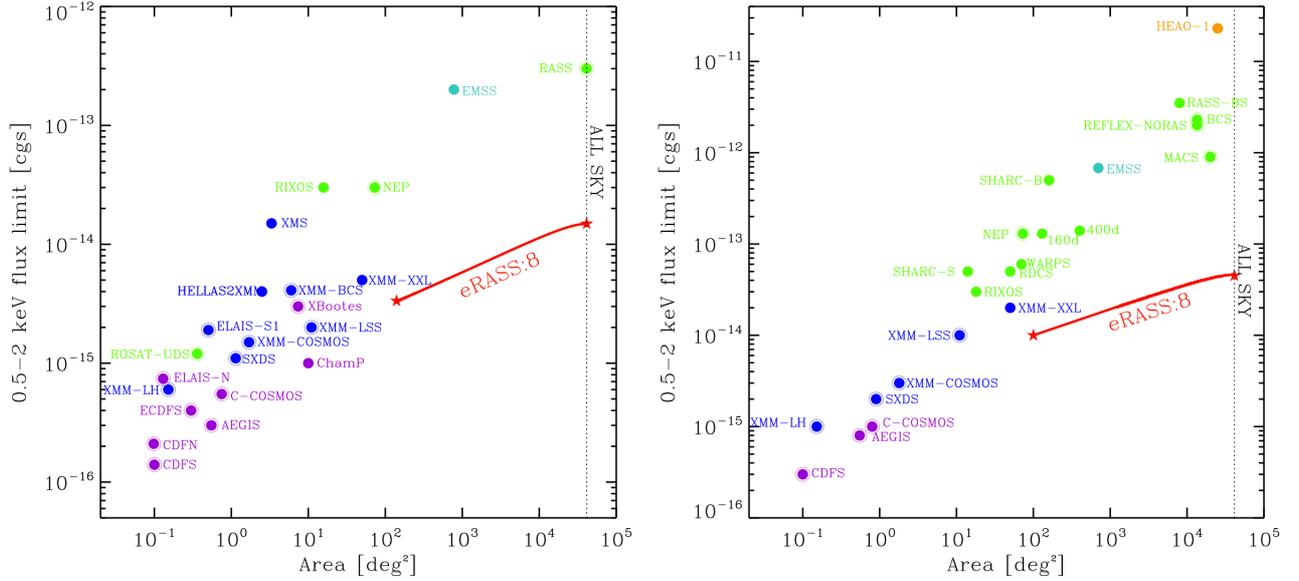

*Figure 4.1 Sensitivity versus area for eROSITA surveys of point-like (e.g. AGNs and stars, left panel) and extended X-ray sources (e.g. clusters, right panel) are shown with thick red solid lines. Exisiting surveys from Einstein (cyan), HEAO-1 (orange), ROSAT (green), XMM-Newton (blue) and Chandra (purple) are shown for reference. Encircled points mark contiguous surveys. eROSITA outperforms in terms of area covered any existing X-ray survey by more than one order of magnitude at the widest areas. The sensitivity calculations are based on the survey strategy presented in section 3 and on the analysis described in sections 4.1-4.3*

## 4.1 *eROSITA* response matrix

The combined effective area (on-axis) of the 7 *eROSITA* telescopes is shown in the left panel of Figure 4.1.1, together with a comparison with *XMM-Newton* EPIC-pn. The right panel of Figure 4.1.1 shows the Field-of-View averaged effective area. Figure 4.1.2 shows instead the "grasp" of the telescope, defined as the product of field of view times (average) effective area. Comparison with *XMM-Newton* PN+MOS and *ROSAT* PSPC clearly highlights the major breakthrough of *eROSITA* in the survey speed and capability over a very wide range of energies.

During the standard data reduction and analysis, the full energy range will be subdivided into five, non-overlapping, energy bands, which will be approximately given by: [$E_{min}$-0.5 keV], [0.5-1 keV], [1-2 keV], [2-4 keV], [4-10 keV], where the minimum energy $E_{min}$ will have to be determined on the basis of the final choice of optical filter and taking into account telemetry constraints. For most of the discussion in the remaining of this document, we will quote numbers relative to two broadband energy ranges, a soft (0.5-2 keV) and a hard (2-10 keV) one, respectively.

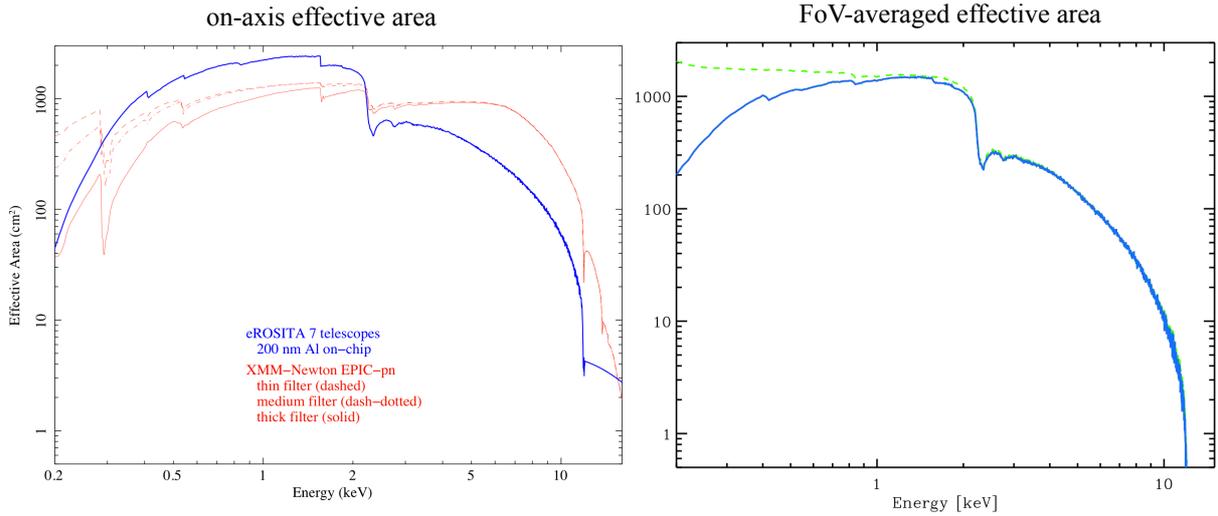

*Figure 4.1.1: Left: Comparison of the effective area (on-axis, in cm²) of the 7 telescopes of eROSITA (blue curves) with XMM-Newton/EPIC-pn for the 3 filters (thick/medium/thin, red curves). Right: Field-of-View averaged effective area for the eROSITA 7-telescopes systems (blue curve), including vignetting, the effects of filters and CCD quantum efficiency; the dashed green line show the effective area for the mirror systems only.*

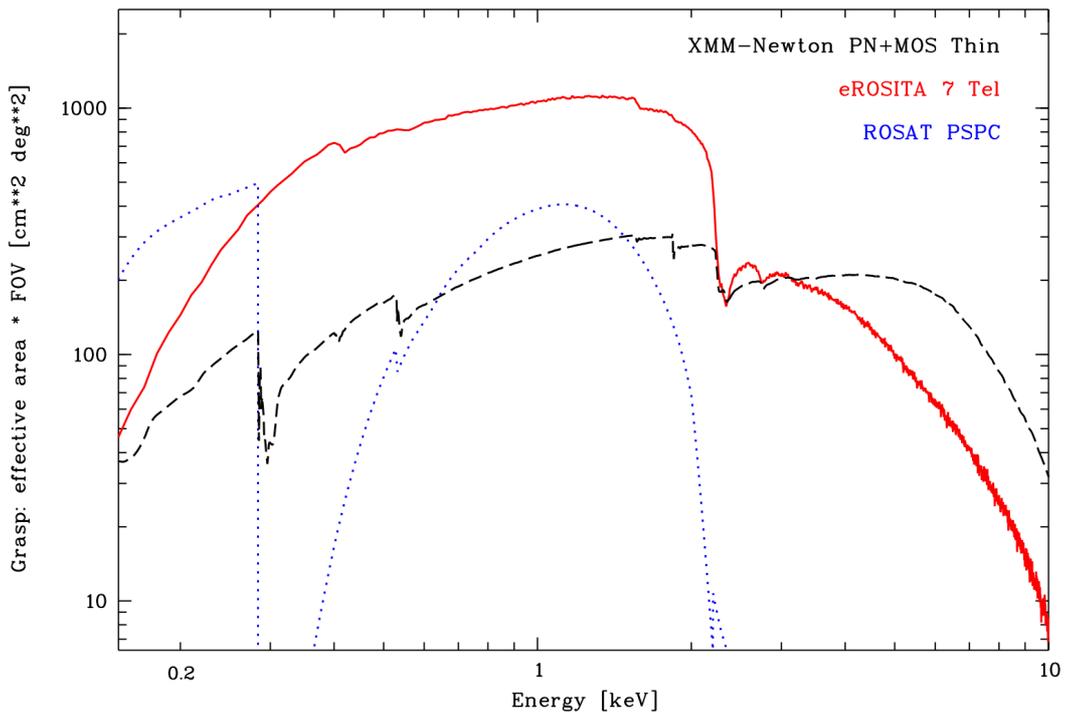

*Figure 4.1.2 eROSITA grasp (red curve), defined as the product of field of view times (average) effective area, as a function of energy. For comparison, the grasp of ROSAT PSPC (blue dotted) and XMM-Newton PN+MOS thin (back dashed) are shown.*

## 4.2 The expected background

The expected *eROSITA* background has been simulated based on photon and high-energy particle spectral components. The cosmic diffuse photon X-ray background has been adopted from the measurements with *XMM-Newton* EPIC camera, as reported in Lumb et al. (2002). The high-energy particle background, which does not interact with the mirror system, has been calculated by way of Geant4 simulations by Tenzer et al. (2010) (see also Perinati et al. 2012). The expected background count rate has been compared with *XMM-Newton* observations, which might provide the best test for the photon background for *eROSITA* around the Lagrangian point L2 before real photon background data will become available. The comparison of the *eROSITA* background model with *XMM-Newton* is described in Section 4.2.1 below. In addition, the count rate plots available at http://xmm.vilspa.esa.es/external/xmm_sw_cal/background/bs_countrate.shtml#14, give an estimate of the scatter of the *eROSITA* background to be expected in 'low background' periods. Deviations from the mean value by factors of about 2 to 3 might occur. Below we discuss in more detail the various spectral components that have been taken into account to produce the *eROSITA* background model.

### 4.2.1 The simulated mean *eROSITA* background model

As in Lumb et al. (2002), the extragalactic unresolved background emission is modeled with a photon index of 1.42 and a normalization[2] of 9.03 photons keV$^{-1}$ cm$^{-2}$ s$^{-1}$ sr-1. The optically thin background emission from the Milky Way is modeled with two mekal models (Mewe et al. 1986; Kaastra 1992; Liedahl et al. 1995) with temperatures of $kT_1$=0.204 keV and $kT_2$=0.074 keV and normalizations of 7.59 and 116 photons keV$^{-1}$ cm$^{-2}$ s$^{-1}$ sr-1, respectively. The values are listed in Table 3 of Lumb et al. (2002). A Galactic foreground absorption of $1.7 \times 10^{20}$ cm$^{-2}$ has been assumed. The high energy particle background, modeled by Tenzer et al. (2010) has a normalization of 1151 counts keV$^{-1}$ s$^{-1}$ sr-1 and a flat spectral energy distribution.

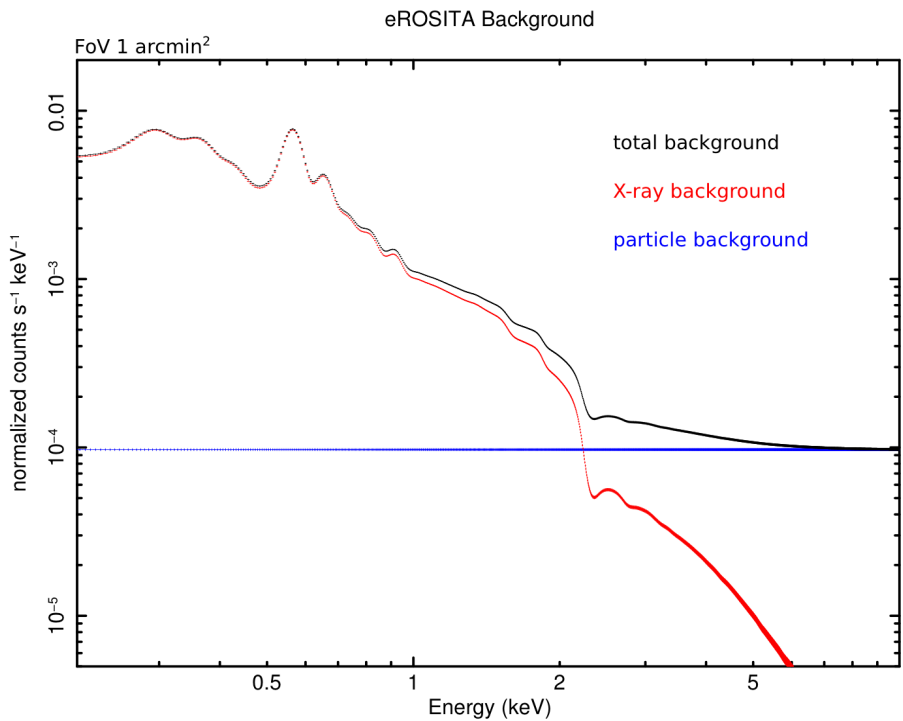

*Fig 4.2.1: Simulated (XSPEC) eROSITA background (black) and spectral components: in red the photon X-ray background (galactic and extragalactic), in blue the particle background and in black the total. Courtesy of K. Borm.*

Based on the spectral model parameters described above, the mean *eROSITA* background count rate model is shown in Figure 4.2.1. In Table 4.2.1 the *eROSITA* background count rates values in various energy bands are listed.

---

[2] According to Georgakakis et al. (2008), from the *logN – logS* curve at the flux limit of $1 \times 10^{-14}$ erg cm$^{-2}$ s$^{-1}$, and assuming that about 30 per cent of the CXRB is resolved, one expects a flux of $1.3 \times 10^{-14}$ erg cm$^{-2}$ s$^{-1}$ in a detection cell of 3.14 arcmin$^2$ in the 2-10 keV band, roughly consistent with the normalization quoted above.

| Energy band [keV] | 0.2-0.5 | 0.5-1 | 1-2 | 0.5-2 | 2-4 | 4-8 | 2-7 | 2-10 |
|---|---|---|---|---|---|---|---|---|
| simulated photon+particle ct. rate [$10^{-3}$ cts s$^{-1}$ arcmin$^{-2}$] | 1.6 | 1.46 | 0.68 | 2.14 | 0.31 | 0.42 | 0.62 | 0.92 |

*Table 4.2.1: eROSITA photon and particle background count rates for different energy bands.*

### 4.2.2 Comparison of the *eROSITA* background model with *XMM-Newton*

The *eROSITA* background model has been folded through the EPIC pn medium full-frame (PN fm) response files for a comparison with the measured values of the *XMM-Newton* photon (without particle) background (cf. Read and Ponman, 2003, Table 7). In Table 4.2.2 we show the comparison between the measured *XMM-Newton* photon background and the simulated photon *eROSITA* background count rates. While the overall diffuse cosmic photon X-ray emission from the model is in good agreement with the observations up to energies of about 5 keV, the model under-predicts the observed photon background count rates above 5 keV published by Read and Ponman (2003). It appears most likely that some of the high-energy particle background is still included in the *XMM-Newton* observations (D. Lumb, private communication). Moreover, at energies above 5 keV the high-energy particle background dominates the spectral energy distribution.

| Energy band [keV] | 0.2-0.5 | 0.5-2 | 2-4.5 | 4.5-7.5 | 7.5-12 |
|---|---|---|---|---|---|
| PN fm observed count rate [$10^{-3}$ cts s$^{-1}$ arcmin$^2$] | 1.13±0.50 | 2.04±0.94 | 0.72±0.36 | 0.64±0.36 | 0.68±0.48 |
| simulated particle count rate [$10^{-3}$ cts s$^{-1}$ arcmin$^2$] | 0.028 | 0.15 | 0.24 | 0.29 | 0.44 |
| simulated photon+particle count rate [$10^{-3}$ cts s$^{-1}$ arcmin$^2$] | 1.6 | 2.14 | 0.36 | 0.31 | 0.44 |

*Table 4.2.2: Comparison of the XMM-Newton photon background with the eROSITA photon background.*

## 4.3 *eROSITA* sensitivity

### 4.3.1 Point sources

For point sources, we first estimated the total net counts needed to securely identify a point source as a function of exposure time (top panel of figure 4.3.1). We did this by estimating the full Poisson probability of spurious detection ($P_{null}$), given the number of background ($n_B$) and total detected counts ($n_T$) (Weisskopf et al. 2007):

$$P_{\text{null}} = 1 - \sum_{m=0}^{n_T - 1} \frac{n_B^m}{m!} e^{-n_B}$$

We have assumed an extraction region of 60" diameter (2×HEW averaged over the FoV), and a background count rate of 7.70 cts/s/deg$^2$ and 3.31 cts/s/deg$^2$ in the soft and hard band, respectively (see Table 4.2.2), and requiring a false detection probability $P_{null}<2.87\times10^{-7}$ (i.e. the Gaussian upper-tail probability for 5σ). The PSF characterization of the *eROSITA* telescope suggests that about 20% of the total point sources flux is expected to lie outside the 60" diameter extraction region, thus a 1.2 multiplicative correction factor has been applied. The overall flux limits in the two bands are then finally computed by multiplying the count rates by the appropriate conversion factors, calculated assuming a power-law spectrum with photon index $\Gamma$=1.8 and column density $N_H$=3×10$^{20}$ cm$^{-2}$. These are given by $1.42\times10^{-12}$ and $2.81\times10^{-11}$ erg/s/cm$^2$/count_rate for the soft (0.5-2 keV) and hard (2-10 keV) bands, respectively. The bottom panel of Figure 4.3.1 shows the predicted sensitivity as a function of exposure time for both the soft and hard bands.

By combining the sensitivity expectation discussed above with the exposure map (Sec. 3.1), the known AGN number counts in the soft and hard bands, and taking into account the effects of the galactic foreground extinction, we can calculate the total number of point-like extragalactic sources to be detected by *eROSITA* at about 3×10$^6$ over the entire sky (see Kolodzig et al., in prep.; see Sec. 5.2).

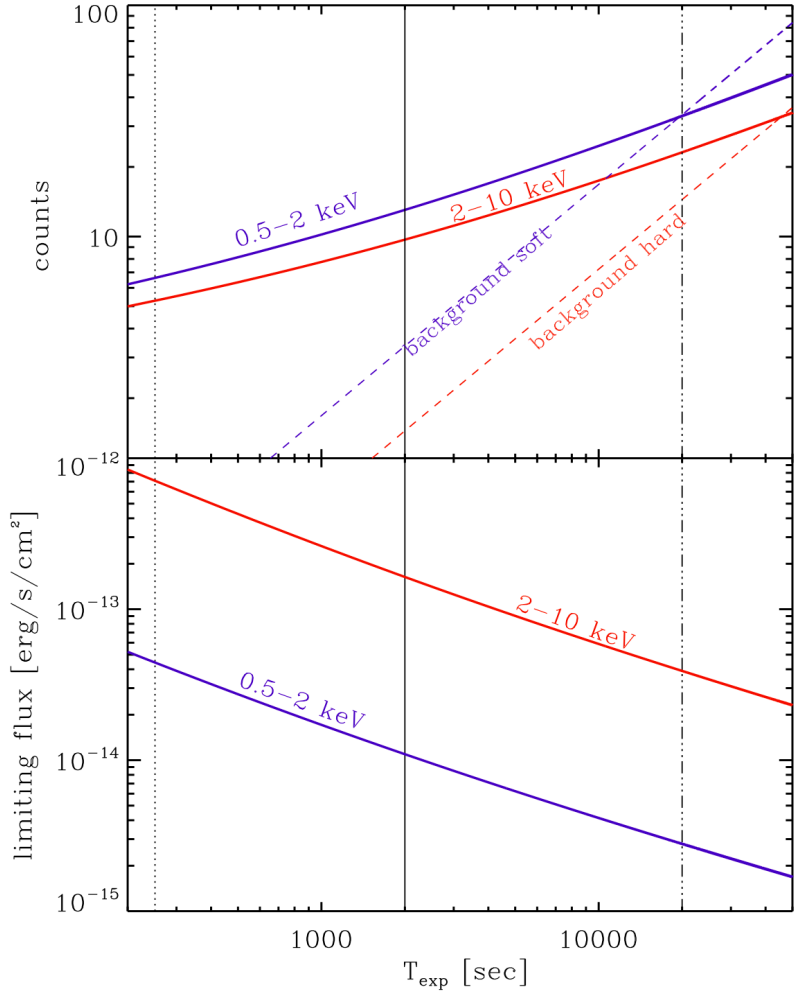

*Figure 4.3.1: Top panel: Minimum number of net counts needed to securely identify a point source in survey mode (HEW~30") as a function of exposure time for soft (0.5-2 keV, blue solid line) and hard (2-10 keV, red solid line) band respectively. The corresponding dashed lines are the expected number of background counts in a 1' diameter extraction region. Bottom: Sensitivity plot for AGN (power-law with photon index $\Gamma$=1.8 and $N_H$=3 x $10^{20}$), e.g. point sources limiting flux versus exposure time. Three vertical lines are shown, marking the average exposure times for one all-sky survey (eRASS:1, 6 months; 250 s, dotted) the final 4-years all-sky survey (eRASS:8, ~2 ks, solid) and the 4-years deep exposure at the ecliptic poles (~20 ks, dot-dashed), where a survey efficiency of 80% has been assumed.*

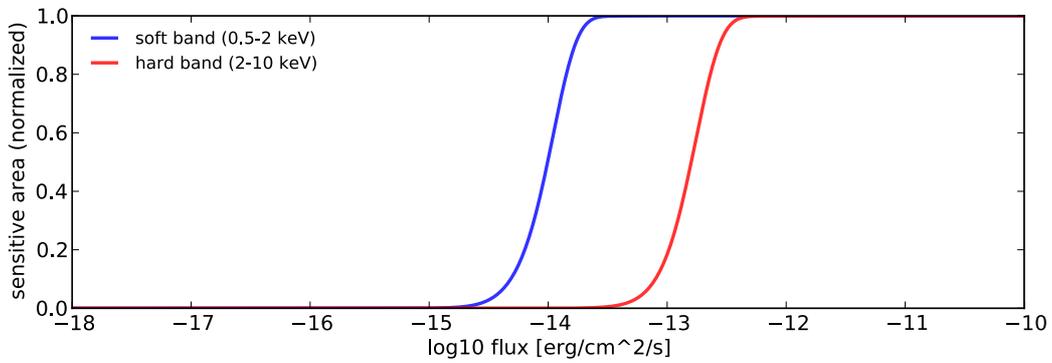

*Figure 4.3.2: Sensitivity curves for the full 4-years eROSITA survey: the normalized sensitive area is plotted as a function of the limiting flux for point source detection for both soft (blue) and hard (red) band. The computations are based on the exposure map and background model of Fig. 3.1.2*

## 4.3.2 Extended sources

For extended sources, the source detection process is more complicated and depends in detail on the source properties. The most numerous and important set of extended sources in the *eROSITA* Survey are galaxy clusters, which are generally characterized by a compact appearance of the source with a bright central region and a steep decline of the surface brightness to larger radii, which can be described with some approximation by a beta-model (Cavaliere & Fusco-Femiano 1976). Fig. 4.3.3 shows a simulation of an *eROSITA* sky area of about 1.9×1.2 degrees in size, with exposures of 2.2 ks (approximately at the average exposure of the all sky survey) and 30 ks (~deepest exposure in the ecliptic pole region) from Mühlegger (2011). The most prominent extended cluster sources are easily noted.

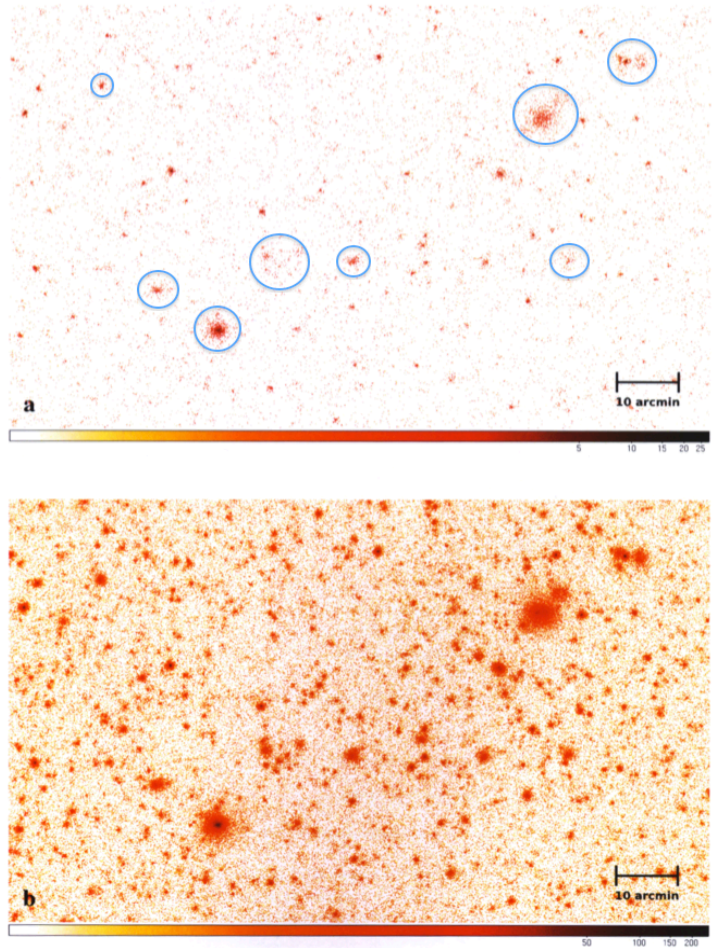

*Figure 4.3.3: Simulated region of the eROSITA survey including point sources, extended cluster sources, and diffuse background. The upper panel shows a region of 1.9x1.2 deg² as would be observed in the main survey with an exposure of 2.2 ks, while the lower panel shows the same region as would be observed at the ecliptic poles with an exposure of 30 ks (from Mühlegger 2011). Extended structures are highlighted in the top panel.*

The detection threshold for extended sources has to be determined from detailed simulations, possibly including realistic structural properties of clusters, such as cool-cores and disturbed morphologies, as well as the evolution of their incidence as a function of cosmic time. Here we present, as an illustration, simple estimates based on analytic calculations which accounts for all known expected instrumental effects. There are two types of thresholds that are of interest for our purpose: the detection threshold as an X-ray source and the threshold for characterizing the source as clearly extended.

As far as detection threshold, we compute this again using the full bimodal probability (see above), computed for a 0.5-2 keV background within an extraction region of 3' diameter. In analogy with the point-source case, in Figure 4.3.4 we show both the net counts (top panel) and the limiting 0.5-2 keV flux (bottom panel) as a function of exposure time. In this case, however, we have set a more stringent limit for spurious detection ($P_{null} < 2.56 \times 10^{-12}$ i.e. the Gaussian upper-tail probability for 7σ). For a rough estimate of how such a threshold needs to be raised when we seek for a characterization of the detected sources, we plot in the same figure (dot-dashed lines) the limiting flux and net counts obtained by requiring a minimal signal-to-noise ratio S/N=7, where S/N=$n_S/(n_T)^{1/2}$ with $n_S$ denoting the net (source) and $n_T$ the total (source plus background) counts, respectively.

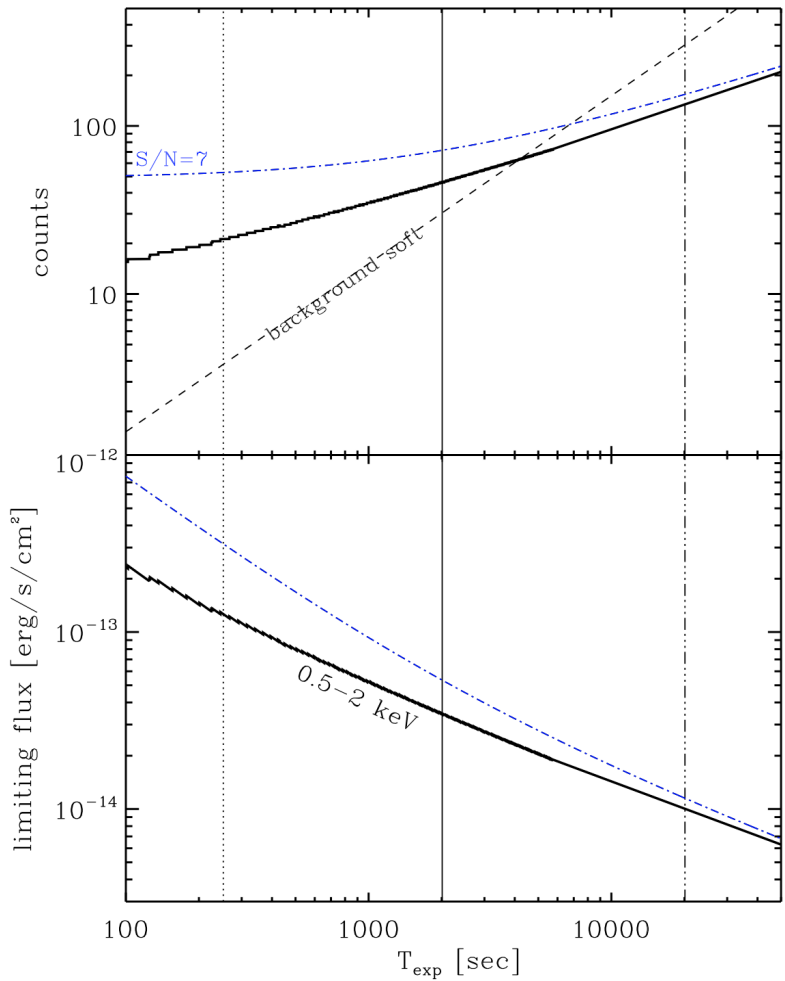

*Figure 4.3.4: Top panel: Minimum number of net counts needed to securely identify an extended source in survey mode as a function of exposure time in the 0.5-2 keV band. Thick solid line is for Poissonian probability threshold, while dot-dashed line mark a fixed S/N=7 (see text for details). The corresponding dashed lines are the expected number of background counts in a 3' diameter extraction region. Bottom: Sensitivity plot for clusters (APEC model in XSPEC assumed at z=0.2) with T=2keV, metallicity of 0.3 solar and $N_H=3 \times 10^{20}$). Three vertical lines are shown, marking the average exposure times for one all-sky survey (eRASS:1, 6 months; 250 s, dotted) the final 4-years all-sky survey (eRASS:8, ~2 ks, solid) and the 4-years deep exposure at the ecliptic poles (~20 ks, triple-dot-dashed).*

In the following section, we estimate the number of galaxy clusters *eROSITA* will detect as a function of the mass and redshift, with a conservative assumption that at least 100 (or 50, see section 5.1.1) source counts are needed to detect an extended X-ray source. In the *ROSAT* survey, with a low background, the majority of the cluster sources were easily detected and marked as extended sources on the bases of 30 source counts (e.g. Böhringer et al. 2004). While the better angular resolution of *eROSITA* will be advantageous for the detection of cluster sources, the higher background and more severe source confusion will have to dealt with great care. More detailed simulations will provide us with a precise cluster selection function, partly dependent on the surface brightness of clusters near the detection threshold. As a guideline, we can adopt a safe detection threshold for X-ray galaxy clusters of 100 counts (in survey mode). This also leads to a high probability for the source to be clearly classified as extended. Based on these results, we will discuss in greater detail in Section 5.1 the expected scientific impact of the *eROSITA* surveys on our understanding of clusters and groups of galaxies in the Universe.

## 4.4 Summary

Table 4.4.1 shows a summary of the expected characteristics of the *eROSITA* telescope which are most relevant for the assessment of the scientific impact of its all-sky surveys. Sensitivity limits correspond to the flux limits at the detection thresholds discussed in section 4.3 above. The "Average All-sky" correspond to a net exposure of 2 ks, while the "Poles" are for 20 ks.

*Table 4.4.1: Summary of expected characteristics of the eROSITA telescope and of its survey performances*

| | | Energy Range | |
|---|---|---|---|
| | | Soft band [0.5-2 keV] | Hard band [2-10 keV] |
| FoV-averaged Effective Area [cm$^2$] | | 1,365† | 139‡ |
| Total Background [$10^{-3}$ cts/s/arcmin$^2$] | | 2.14 | 0.92 |
| Sensitivity, eRASS:1 Average All-sky [erg/s/cm$^2$] | Point sources | $4.4 \times 10^{-14}$ | $7.1 \times 10^{-13}$ |
| | Extended sources | $1.1 \times 10^{-13}$ | - |
| Sensitivity, eRASS:8 Average All-sky [erg/s/cm$^2$] | Point sources | $1.1 \times 10^{-14}$ | $1.6 \times 10^{-13}$ |
| | Extended sources | $3.4 \times 10^{-14}$ | - |
| Sensitivity, eRASS:8 Poles [~140 deg$^2$] [erg/s/cm$^2$] | Point sources | $2.9 \times 10^{-15}$ | $3.9 \times 10^{-14}$ |
| | Extended sources | $1 \times 10^{-14}$ | - |

† At 1 keV
‡ At 5 keV

# 5. Scientific goals

## 5.1 Clusters of galaxies and cosmology

Galaxy clusters are unique large scale astrophysical laboratories and important probes for the study of the structure of our Universe. X-rays offer an extremely powerful way of detecting clusters as gravitationally bound and evolved entities, and to obtain a picture of their structure and dynamical state. In most cases, the X-ray emitting intra-cluster plasma sits in the gravitational potential well of the cluster in nearly hydrostatic fashion. Thus, the gas density and its projection into the X-ray surface brightness trace the iso-potential surfaces of the cluster's gravitational potential. The X-ray images provide therefore a very clear outline of the cluster's structure. In addition, the X-ray emitting plasma contains important information on the energy input into the intergalactic medium by various processes, such as star-formation-driven galactic winds and relativistic jets from accreting supermassive black holes in galactic nuclei. The chemical enrichment of the intra-cluster medium (ICM) by heavy elements – easily detected in the X-ray spectra of galaxy clusters – provides important information on the nucleosynthesis activity of different kind of supernovae over cosmic times. Last but not least, galaxy clusters form an integral part of the cosmic large scale structure, highlighting the density distribution of the dark matter as powerful cosmic lighthouses. They are thus ideal tracers to assess the statistics of the large-scale matter distribution in the Universe and can provide unique tests of cosmological models (see Rosati et al. 2002; Allen et al. 2011 for comprehensive reviews).

Due to their relative low surface density in the sky, very large area surveys are mandatory to amass large enough samples capable of producing competitive cosmological constraints, as we briefly discussed in the first Section of this document. Indeed, the *eROSITA* telescope has been designed with the very purpose of detecting a galaxy cluster sample of about $10^5$ objects. Given our current knowledge of the expected performance of the telescope instrumentation, together with the details of the mission planning strategy that we have presented in the previous sections, we can now provide robust estimates of the overall number of clusters that will be detected by *eROSITA* during its first, surveying, mission phase. We adopt a flat cosmological model with the parameters, $\Omega_M=0.3$, $H_0=70$ km/s/Mpc and $\sigma_8=0.8$. We use a primordial power spectrum with a slope, $n=1$, and use a transfer function which includes all the effects of baryonic acoustic oscillations. We then use the halo mass function of Tinker et al. (2008) to calculate the cluster mass function as a function of redshift, and apply the X-ray scaling relations found in the study of the REXCESS sample (Böhringer et al. 2007, Pratt et al. 2009), and assuming an evolution thereof as in Reichert et al. (2011) and Vikhlinin et al. (2009), to obtain the cluster X-ray luminosities. With the help of the preliminary response matrix for *eROSITA* (section 4.1), we derive the expected photon counts for each cluster depending on its sky position (with given exposure and interstellar absorption column density) to evaluate if the cluster source counts are above the detection threshold.

From these calculations we obtain the results shown in Fig. 5.1.1 for the total number of clusters detected as a function of redshift and for the typical mass of the clusters detected at the threshold for different redshifts. In its four years survey *eROSITA* will indeed be able to detect more than $10^5$ galaxy clusters, increasing our census of this population by at least one order of magnitude. The information contained in these cluster data will therefore enable a very broad range of interesting astrophysical and cosmological studies, and we can only give a very brief overview of them in the following. In particular:

- We will **discover all massive galaxy clusters** in the observable Universe away from the Galactic plane;
- We will **constrain cosmological parameters** with 1-2 orders of magnitude better statistics than currently available X-ray cluster samples;
- We will be able to probe the statistics of the very **large scale matter density distribution** (power spectra and correlation function) in the Universe on scales exceeding 1 Gpc;
- We will **study the detailed (thermo-)dynamical structure of nearby clusters** with the high spectral resolution of *eROSITA* over their entire large projected sizes;
- We will use large sample of clusters detected with high number of counts to study the **evolution of the thermal structure and chemical enrichment with redshift**;
- We will take advantage of the all-sky nature of the *eROSITA* surveys to search for the rarest and most extreme clusters, such as **massive objects at high redshift**, that could constitute crucial tests to cosmological and structure formation models.

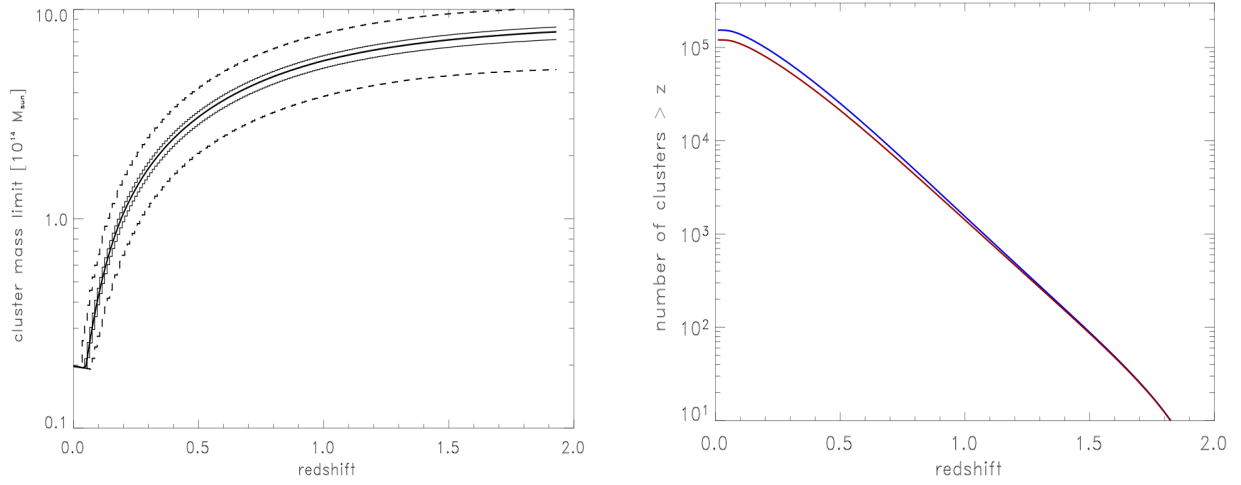

*Figure 5.1.1: Left Panel: cluster mass limit as a function of redshift for the final (4 years) all-sky eROSITA survey (eRASS:8); the mass is defined as $M_{200}$, the mass inside the radius in which the cluster has a mean density 200 times the critical density of the Universe. The lines shown, from top to bottom, are 10% and 30% limits, median, 70% and 90% limits. Right Panel: cumulative number of clusters of galaxies detected in the full all-sky survey by eROSITA at redshift larger than z, as a function of z. Blue line is for the whole sky, red for the extragalactic one only. A conservative limit of 100 photons have been considered in order to guarantee a secure detection of a cluster.*

### 5.1.1 Testing cosmological models

Dark energy makes up about three quarters of the energy content of our Universe and determines its ultimate fate. Yet, we do not understand at all the nature of this enigmatic component that currently accelerates the expansion of the Universe. While *eROSITA* will revolutionize all of X-ray survey astronomy, studying dark energy by discovering all massive galaxy clusters in the observable Universe is its primary science driver. Indeed, *eROSITA* will likely be the first "Stage IV" dark energy probe to be realized (according to the DETF report categories, Albrecht et al. 2006).

There are two major routes to constrain dark energy and most other important cosmological parameters: absolute distance or volume ("geometry") measurements and growth of structure experiments. Distance measurements mostly rely on so-called standard candles or standard rulers; i.e., one assumes the luminosity or physical size of an astrophysical object class or feature to be known and then measures the apparent brightness or size of such objects. This yields the luminosity distance or angular diameter distance, respectively. Comparison between these directly measured distances and the distances predicted based on redshifts, $z$, and assumed values for cosmological parameters leads to constraints on all parameters entering the Friedmann equation

$$H(z) = H_0\, E(z), \qquad E^2(z) = \Omega_r(1+z)^4 + \Omega_M(1+z)^3 + \Omega_k(1+z)^2 + \Omega_{DE}\, e^{3\int [1+w(z)]\, d\ln(1+z)},$$

where $H_0$ is the Hubble constant, and $\Omega_r$, $\Omega_M$, $\Omega_k$, and $\Omega_{DE}$, are the normalised (dimensionless) mean density parameters for relativistic particles and radiation, matter, space curvature, and dark energy, respectively. $w(z)$ is the equation of state parameter of dark energy.

As stressed in the DETF report, geometric methods alone will not be able to differentiate between two competing approaches to dark energy: modification of general relativity and/or of the laws of gravity and the introduction of a new particle or field that dominates the Universe today. Therefore, growth of structure experiments need to be pursued, which are also sensitive to changes in gravity: the growth of structure within standard gravity can be described by a differential equation for the matter density contrast, $\delta = \Delta\varrho/\varrho$,

$$\delta'' + 2H\delta' = 4\pi G \rho \delta,$$

where $G$ is the gravitational constant. One notes that the Hubble parameter, $H(z)$, which describes the expansion of the Universe, acts to suppress the growth of density fluctuations while the gravitational attraction of matter, $\varrho$, in the Universe accelerates it. By testing this equation observationally, all parameters entering $H(z)$ can be constrained again. Additionally, if gravity laws indeed need to be extended, then this equation will also need to be modified, e.g., by adding extra terms. Whether such a modification is required to explain dark energy can *only* be determined with growth of structure measurements.

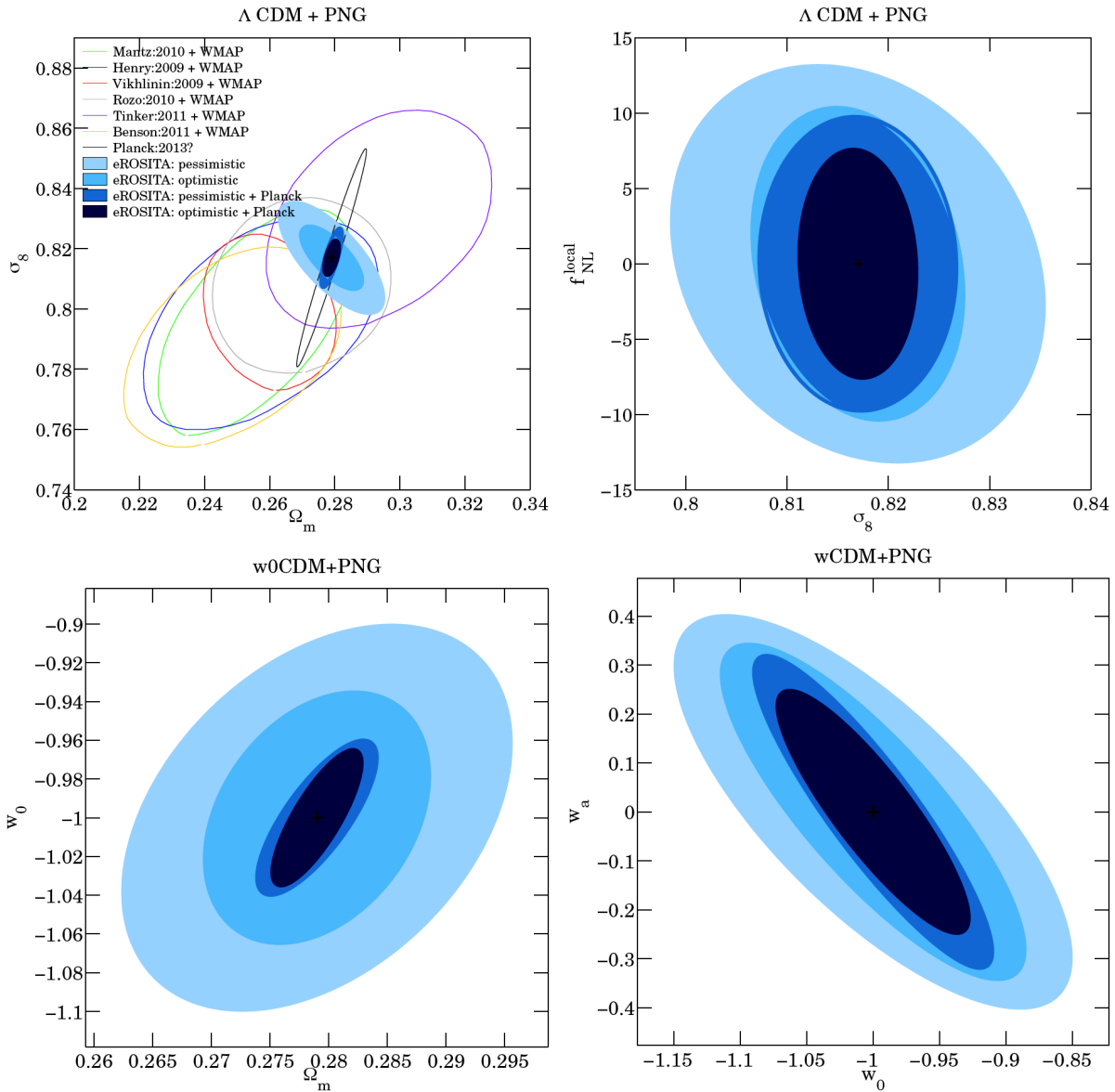

*Figure 5.1.2: Examples for expected cosmological constraints from eROSITA. The pessimistic cases assume current knowledge on the $L_X$-M relation and photometric redshifts while the optimistic ones assume that uncertainties in the $L_X$-M relation parameters improve by a factor of two in the coming years and spectroscopic redshifts can be obtained from planned dedicated follow-up surveys (e.g., SPIDERS and 4MOST, Section 6.2). In the upper left plot, several existing cluster+WMAP constraints are also shown for comparison. From Pillepich et al., in prep.*

Several partially independent cosmological tests are possible with galaxy cluster samples. They include both geometric and growth of structure tests, e.g., the baryon fraction in clusters, the apparent evolution of the intracluster gas mass fraction, the combination of X-ray and Sunyaev–Zeldovich effect observations, the temperature–size relation, the baryon acoustic oscillation feature in the power spectrum, redshift space distortions (e.g., Allen, Evrard, Mantz 2011, for a review); even individual merging clusters can be used to constrain dark matter properties (e.g., Clowe et al. 2006). Two of the most important cosmological tests are the evolution of the cluster mass function ("cluster counts") and the evolution of the cluster power spectrum ("angular clustering") described in the next two sections. Employing independent tests simultaneously allows us to cross check results, tighten constraints, and relax some of the underlying assumptions.

Besides dark matter and dark energy another mysterious ingredient in our standard cosmological model is required: an inflationary period of accelerated expansion very shortly after the big bang. While inflation was introduced to solve the so-called flatness and horizon problems (e.g., Guth 1981; Linde 1982), it has been realised that it could also be the source of the initial tiny matter density fluctuations required for the gravitational growth of galaxies and galaxy clusters observed today (e.g., Hawking 1982; Bardeen, Steinhardt, Turner 1983). The "standard" inflation model implies a nearly Gaussian initial density fluctuation spectrum but a number of interesting alternatives have been suggested (e.g., Bartolo et al. 2004, Chen 2010, for reviews). Observations of the evolution of large scale structure, e.g., of the distribution of galaxy clusters in space and time, should allow us to put tight constraints on the level of primordial non-Gaussianity and thereby to rule out standard inflation, or many of its alternatives, if the constraints exclude a Gaussian statistics.

Figure 5.1.2 and Table 5.1.1 show quantitative forecasts for constraints on some cosmological parameters from expected *eROSITA* cluster counts and angular clustering (Pillepich et al., in prep.; see Pillepich, Porciani, Reiprich 2012 for details on methodology) assuming use of the full extragalactic sky. The combination of these two methods results in tight constraints of $\Delta\Omega_M=0.0026$, $\Delta\sigma_8=0.006$, $\Delta f_{NL}=5.7$, $\Delta w_0=0.023$ (for $w_a=0$), and $\Delta w_a=0.16$ in the optimistic scenario and with *Planck* priors, where $\sigma_8$ represents the amplitude of linear matter density fluctuations, $f_{NL}$ is a non-Gaussianity parameter, $w_0$ and $w_a$ are dark energy equation-of-state parameters. In all cases shown ("PNG"), $f_{NL}$ is left free to vary. Assuming instead $f_{NL}=0$ would slightly improve the constraints, e.g., the dark energy Figure of Merit (FoM) to about ~300.

It is worth noting that, in practice, the cosmological analysis will be more involved than our simplified approach for the forecasts, e.g., the cosmological parameters will not only need to be fit simultaneously with the scaling relation parameters (which we do) but also with the cluster selection function (see section 5.1.4; for the forecasts shown here we employ a simple photon count cut). However, the constraints shown here are also conservative in the sense that we ignore the 1,000s of galaxy clusters observed with *eROSITA* that will have mass proxies better then $L_X$, e.g., the gas temperature (Borm et al., in prep.), directly measured from the survey as well as from the dedicated *eROSITA* pointed follow-up (see below).

The good complementarity of *eROSITA* and *Planck* also becomes apparent in the upper left plot in Figure 5.1.2 since *eROSITA*+Planck will significantly improve upon constraints from *Planck* alone. As a Stage IV dark energy probe, *eROSITA* will improve upon Stage III experiments currently being built or conducted. As an example, expected constraints from a survey similar to the Dark Energy Survey (DES) are included in Table 5.1.1., including weak lensing and galaxy clustering. For instance, the improvement is about a factor of 1.5 each for $w_0$ and $w_a$. Several years after *eROSITA*, *Euclid* is expected to be launched – another powerful Stage IV mission. Expected constraints from a *Euclid*-like survey are also included in Table 5.1.1, using the same probes as for the DES-like survey. This shows that *Euclid* and *eROSITA* will yield comparable precision on the parameters considered here. Moreover, the *eROSITA* data will be combined with DES and other Stage III experiments to make further progress before the *Euclid* launch.

The expected constraints on $f_{NL}$ are particularly exciting in view of their complementarity with CMB studies. Using the bispectrum of temperature anisotropies, the *Planck* satellite will constrain the shape and amplitude of primordial non-Gaussianity at a similar level as *eROSITA* (with *Planck* priors from the CMB power spectrum) but on different scales and with different systematics (e.g. Komatsu & Spergel 2001; Sefusatti et al. 2009). A robust detection of primordial non-Gaussianity would rule out the standard (single-field, slow roll) inflationary scenario. The combination of CMB data on large scales with the *eROSITA* analysis will tell us about the scale dependence of the non-linearity parameter and thus put tight constraints on multi-field inflation (e.g. Byrnes et al. 2010). Stage III galaxy surveys like DES will have slightly less constraining power on $f_{NL}$ than *eROSITA* (Giannantonio et al. 2012), mainly due to the lower clustering amplitude of galaxies with respect to clusters (galaxies are less biased tracers of the mass distribution). Additional constraints could be derived from the galaxy bispectrum (Liguori et al. 2010) which, however, being an intrinsically non-linear quantity, requires sophisticated modeling of gravitational dynamics and galaxy biasing (Pollack, Smith & Porciani 2012) and thus is more prone to systematic effects.

For these forecasts, clusters with masses roughly below $5\times10^{13}\ h^{-1}M_\odot$ were excluded (this cut was incorporated in the analysis as a redshift dependent *eROSITA* photon count cut, so no mass information is actually required). Why is this conservative step done? Why throw away the information from more than half of all the detected clusters? The reason is the current uncertainty in scaling relations for galaxy groups. While several studies of groups are available (e.g., Gastaldello et al. 2007; Sun et al. 2009), a study of a *complete* group sample is still lacking, which would allow us to correct for selection effects. More importantly, it is clear that the scatter in most scaling relations increases for groups (i.e., the scatter then depends on cluster mass, Eckmiller, Hudson &

Reiprich 2011) and especially its uncertainty is larger. More observations of galaxy groups help to further improve cosmological constraints from *eROSITA* as much as understanding the most distant clusters will do.

| Data | Redshifts | Prior Scenario | Model | $\Delta f_{NL}^{local}$ | $\Delta\sigma_8$ | $\Delta\Omega_m$ | $\Delta w_0$ | $\Delta w_a$ | $FoM^{DETF,1\sigma}$ |
|---|---|---|---|---|---|---|---|---|---|
| eROSITA | photo-z | Pessimistic | LCDM+PNG | 8.1 | 0.012 | 0.0101 | - | - | - |
| eROSITA | spectro-z | Optimistic | LCDM+PNG | 6.4 | 0.007 | 0.0060 | - | - | - |
| eROSITA + Planck | photo-z | Pessimistic | LCDM+PNG | 6.5 | 0.006 | 0.0021 | - | - | - |
| eROSITA + Planck | spectro-z | Optimistic | LCDM+PNG | 5.0 | 0.004 | 0.0015 | - | - | - |
| eROSITA | photo-z | Pessimistic | w0CDM+PNG | 8.2 | 0.016 | 0.0109 | 0.066 | - | - |
| eROSITA | spectro-z | Optimistic | w0CDM+PNG | 6.6 | 0.009 | 0.0063 | 0.043 | - | - |
| eROSITA + Planck | photo-z | Pessimistic | w0CDM+PNG | 6.9 | 0.007 | 0.0034 | 0.026 | - | - |
| eROSITA + Planck | spectro-z | Optimistic | w0CDM+PNG | 5.6 | 0.005 | 0.0025 | 0.023 | - | - |
| eROSITA | photo-z | Pessimistic | wCDM+PNG | 8.2 | 0.018 | 0.0120 | 0.098 | 0.27 | 57.4 |
| eROSITA | spectro-z | Optimistic | wCDM+PNG | 6.6 | 0.011 | 0.0066 | 0.075 | 0.23 | 103.1 |
| eROSITA + Planck | photo-z | Pessimistic | wCDM+PNG | 7.0 | 0.007 | 0.0036 | 0.059 | 0.21 | 179.4 |
| eROSITA + Planck | spectro-z | Optimistic | wCDM+PNG | 5.7 | 0.006 | 0.0026 | 0.048 | 0.16 | 263.3 |
| DES | photo-z | WL+2D photometric | wCDM+PNG | 8.6 | 0.009 | 0.0082 | 0.093 | 0.61 | - |
| DES + Planck | photo-z | WL+2D photometric | wCDM+PNG | 8.2 | 0.009 | 0.0074 | 0.090 | 0.35 | - |
| Euclid | photo-z | WL+2D photometric | wCDM + PNG | 4.7 | 0.005 | 0.0048 | 0.054 | 0.32 | - |
| Euclid | spectro-z | WL+2D spectroscopic | wCDM + PNG | 5.7 | 0.005 | 0.0051 | 0.051 | 0.35 | - |
| Euclid + Planck | photo-z | WL+2D photometric | wCDM + PNG | 4.5 | 0.005 | 0.0044 | 0.052 | 0.20 | - |
| Euclid + Planck | spectro-z | WL+2D spectroscopic | wCDM + PNG | 5.3 | 0.005 | 0.0037 | 0.035 | 0.15 | - |

*Table 5.1.1 FoM is the dark energy figure-of-merit as defined in the DETF report using 1-σ uncertainties. Parameters not shown are marginalized over. From Pillepich et al., in prep.; expected DES- and Euclid-like constraints are taken from Giannantonio et al. (2012).*

In general, external constraints on scaling relations will be vital. Along with multi-wavelength follow-up (e.g., weak lensing and velocity dispersion measurements for independent mass estimates, or radio observations to characterize the dynamical state, see section 5.1.9) another crucial aspect is X-ray follow up. Luckily, *Chandra*, *XMM-Newton*, *Suzaku* and possibly *NuSTAR*, will likely still be operating. In addition, *Astro-H* will be launched around the same time as *eROSITA*, a coincidence that carries great synergy potential, possibly similar to the *ROSAT* and *ASCA* era. Moreover, *eROSITA* will follow up itself. After the survey, it is planned that *eROSITA* performs pointed observations for about three years. A fraction of this time is dedicated to follow up galaxy clusters discovered with *eROSITA* in order to measure mass proxies that are more precise than $L_X$ (e.g., gas mass and temperature). It has been shown that such follow-up observations can significantly improve cosmological constraints from a survey (e.g., Majumdar and Mohr, 2004; Wu et al. 2010; Mantz et al. 2010). The use of the very same instrument for this ensures that systematic instrumental calibration uncertainties cancel out. The pole regions will have deeper exposure then the average assumed for the constraints in Figure 5.1.2 (see Section 3). This will be taken into account in the optimization of the follow-up strategy in terms of redshift range, mass range, and exposure time.

## 5.1.2 Studying the evolution of the galaxy cluster mass function

Galaxy clusters form from overdense peaks in the matter density fluctuation field originating in the early Universe. While the density fluctuations grow by the action of the gravitational field, larger and larger regions above a certain density threshold decouple from the Hubble flow and collapse to virialized objects. The largest collapsed objects we find today have masses in the range from $10^{14}$ to few $10^{15}$ solar masses: the galaxy clusters. With the growth of structures, the galaxy cluster mass function also grows, with the largest effect found at the high mass end of the cluster mass function, where the cluster number counts are exponentially sensitive to the growth of structures, itself dependent on the dark matter and dark energy content of the Universe. This, in a nutshell, makes galaxy clusters sensitive probes for the test of cosmological models, as demonstrated, e.g., by the study of Vikhlinin et al. (2009), where the comparison of sample of only 37 moderate redshift clusters with a nearby cluster sample provides an important cosmological test.

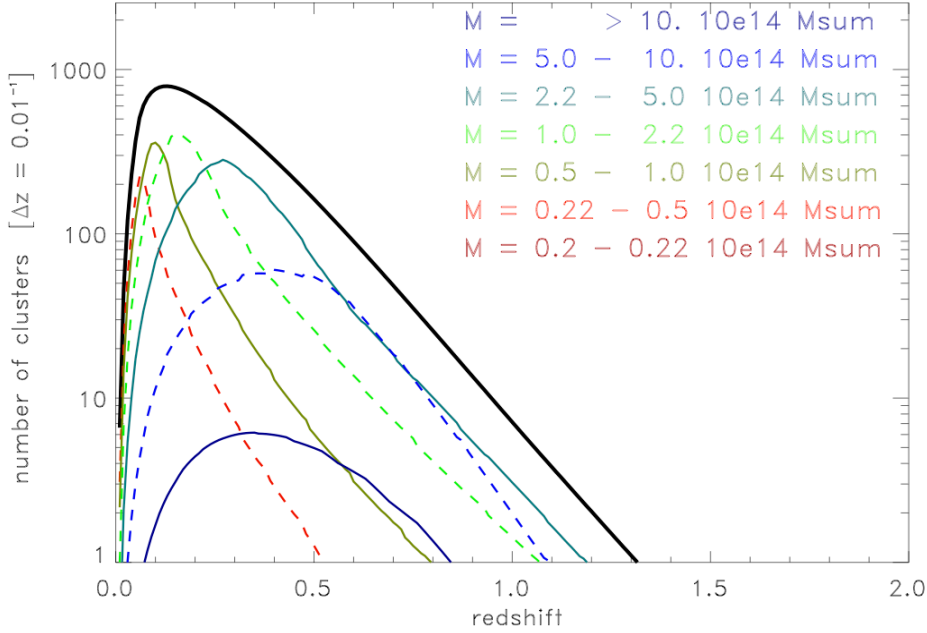

*Figure 5.1.3: Mass - redshift distribution of the detected clusters in the full 4-years eROSITA all-sky survey, assuming 100 photons to secure a detection. Black solid line is for the total, with lines of different colors corresponding to different mass bins, as in the inset. The cluster numbers are given for redshift bins of $\Delta z = 0.01$*

The assessment of the evolution of the galaxy cluster X-ray luminosity function (with $L_X$ as proxy for the galaxy cluster mass) for the all-sky survey up to redshifts of $z\sim1$ and in the scanning poles up to redshifts of $z\sim1.5$, as well as that of the cluster temperature for the brightest ~5,000 galaxy clusters in the survey is therefore a prime task of the survey analysis. Figure 5.1.3 shows the expected Mass-redshift distribution of *eROSITA* detected clusters. For the major part of the cluster sample, where during the survey only photometric redshifts are available, galaxy cluster densities will be determined in bins of about $\Delta z\sim0.1$ with a redshift accuracy per cluster of ≤0.03. In regions with sufficient redshift information (see Section 6.2), the evolution can be characterized more precisely with accurate redshifts.

Besides, a self-consistent model of clusters detected by *eROSITA* and relying on X-ray observables (e.g. count-rates in selected energy bands) can favorably make use of the full population and provide constraints on both the evolution of the mass function and scaling relations between X-ray quantities and mass (Clerc et al. 2012).

## 5.1.3 Clusters and the large-scale structure

The evolution of the galaxy cluster number densities is probing the growth of structures on Lagrangian comoving scales of the order of 10 Mpc. Studying the distribution of the galaxy clusters above a given lower mass cut, the large scale matter distribution can be assessed statistically e.g. by the two-point correlation function or the density fluctuation power spectrum. The amplitude of the power spectrum, or correlation function, for the cluster distribution follows the fluctuations in the overall matter density distribution in a proportional and amplified way (an effect called biasing). This behaviour has been well studied theoretically, supported by N-body simulations, and it has been successfully tested with observations of galaxy clusters in the RASS (Schuecker et al. 2003, Balaguera-Antolinez 2010).

The analysis of the galaxy cluster distribution in the *eROSITA* survey will reveal the statistical properties of the matter density distribution in the Universe on scales exceeding 1 Gpc, only rivaled by the largest redshift surveys planned for the future (and the cosmic microwave background studies in the early Universe). Interesting effects on the large-scale power spectrum are predicted by some of the exotic cosmological models. The exploration of the very large scales accessible by the large *eROSITA* survey volume are therefore particularly exciting.

The large-number of well defined galaxy clusters from the *eROSITA* Survey, together with optical surveys providing information on the spatial distribution of galaxies and AGN, will also allow the first large statistical determination of the cross-correlation between AGN and clusters (e.g. Cappelluti et al. 2007) and that of galaxies and clusters (Sanchez et al. 2005). While simple statistical models predict a biased cross-correlation function with a biasing strength just given by the mean of the galaxy and cluster autocorrelation function, deviations from this simple behavior are expected to reveal interesting environmental effects of the triggering of AGN and star formation activity in the surroundings of galaxy clusters and in the large-scale structure network (e.g. Braglia et al. 2007).

The dense sampling of galaxy clusters in the *eROSITA* survey also enables to study the cluster properties as a function of their environment. In a study of RASS galaxy clusters Schuecker et al. (2001) found a first indication that galaxy clusters with substructure are more often found in regions of high cluster density. Similarly, cluster ellipticities and elongations may show correlations with their environment. *eROSITA* will allow such studies, which currently tantalizing evidences, with much better precision.

Galaxy clusters are expected to be preferentially located at the intersection of the filaments of the large-scale structure network of the matter distribution. It is extremely difficult to detect X-ray emission from the filaments, even though they are expected to contain hot gas. But the gas temperatures are lower than for the clusters themselves and the gas densities are also very low (e.g. Werner et al. 2010). A promising way to find filamentary emission is thus the stacking of regions connecting neighboring clusters. *eROSITA* will provide by far the largest statistical sample of close cluster pairs with reasonably deep X-ray observations to improve much over the previous less successful searches for X-ray emission from cosmic filaments (see e.g. Briel and Henry 1995).

Finally, the large number of clusters of galaxies to be detected by *eROSITA* will enable us to construct catalogs of superclusters of galaxy clusters (Einasto et al. 2008, Nowak 2005) with large enough statistics to draw conclusions on the nature of superclusters. One of the prime goals, besides characterizing the superclusters themselves, is a comparison of the cluster properties in the superclusters to those in the field.

## 5.1.4 Survey selection function and survey simulations

The work described in the previous chapters concerning the construction of the cluster X-ray luminosity function, the assessment of the large-scale structure, and the test of cosmological models requires a precise knowledge of the properties of the cluster selection function and observational biases. Given the higher X-ray background in the *eROSITA* survey (see Section 4.2) compared to the *ROSAT* All-Sky Survey, we will face a more complex task, as the selection function will depend on many factors: the nature of the instrumental and X-ray sky background, the X-ray morphology of the clusters, and the used detection algorithm. This complex behavior of the cluster selection can be properly evaluated by means of dedicated simulations of the cluster detection and characterization process, as we have sketched in Section 4.3.

In this section we describe in greater detail the various approaches that have been (and will be) implemented to tackle this issue and allow an accurate determination of the *eROSITA* clusters survey selection function. Diverse approaches are indeed necessary, for two main reasons: (i) different questions concerning the detection properties are best answered by specific methods, and (ii) not all ingredients required for a faithful simulations of the cluster sky are known a priori, such as the distribution of morphological characteristics of clusters as a function of redshift. The latter information can be "bootstrapped" from the survey itself, or taken from cosmological N-body/hydrodynamical simulations. The different survey simulations performed so far comprise: (i) simulations with parametrized cluster morphologies, where the detection probabilities are determined for a grid of parameter values (e.g. Pacaud et al. 2006), (ii) simulations with templates taken from observations of representative cluster samples from *XMM-Newton* or *Chandra*, (iii) simulations with clusters taken from cosmological N-body/hydrodynamical simulations. All these are currently run with an artificial X-ray background, but may alternatively be performed into actually observed survey fields, once the survey has started. The later simulations have the advantage to feature the exact properties of the real X-ray background, but they cannot be used to determined the degree of contamination by false sources among the detections, since this evaluation would require an a priori knowledge of the sources coinciding with real clusters (see e.g. Mühlegger 2011). A proper characterization of the cluster selection function will require simulations of tens of Millions of galaxy clusters to cover a realistic grid of cluster properties with sufficient statistics, a challenge that will be met by the efforts currently prepared.

The test of cosmological models requires two further complementary approaches: (i) a precise census of the cluster population in form of e.g. the X-ray luminosity function as a function of redshift or the cluster density power spectrum, both based on a predetermined selection function and (ii) the cosmological modeling of the observations which requires a translation of the cluster masses into observational properties. The latter is generally achieved by the construction of scaling relations, e.g. the scaling relations of mass and X-ray luminosity. Such scaling relations are usually determined from cluster samples obtained in surveys featuring specific selection functions and are in general affected by Malmquist and Eddington bias. It is important to model these bias effects properly. The *eROSITA* survey by itself will be an ideal source of galaxy clusters useful for the determination of scaling relations between various X-ray parameters and between these parameters and cluster mass.

The modeling of the bias effects requires again a perfect knowledge of the selection function and of the statistical properties of the scatter of the scaling relations. *eROSITA*, with its enormous statistics of cluster detections, will be the best source to provide us with a good characterization of the scatter in scaling relations. Attempts to model these effects have been made for the much lower statistics of the *ROSAT* surveys (e.g. Stanek et al. 2006, Balaguera-Antolinez et al. 2011, see also Allen et al. 2011). A special role is played by the scaling relations of X-ray observables and cluster mass. Since, for the vast majority of the clusters, the mass cannot be determined from the *eROSITA* data alone, an external calibration of these relations is required with, e.g., a gravitational lensing analysis. For a calibration with sufficient precision a large number of lensing measurements are necessary, which will come from optical sky surveys like DES, KIDS and others (see Section 6.1), improving the current statistics of small samples by several orders of magnitude. With such a statistics, we expect to calibrate the biases in the mass determination to a few percent.

### 5.1.5 Structure and astrophysics of nearby galaxy clusters and groups

The closest galaxy clusters like Coma, Perseus and, in particular, the Virgo cluster, cover regions of the sky larger than one square degree and therefore have only been fully mapped in the *ROSAT* All-Sky Survey. Due to their proximity, we can study many details there which are simply not detectable in more distant clusters. The Virgo cluster, for example, has a very complex morphology, but shows a very close correspondence between the structure of the X-ray emitting gas and the galaxy population, with at least four dynamically independent components around the massive elliptical galaxies M87, M86, M60 and M49 (Böhringer et al. 1994). The X-ray emitting gas follows also very closely the complex distribution of early type galaxies and early type dwarfs (Schindler et al. 1999). While the RASS provided a first complete X-ray image of this nearby cluster over a sky region of about 5 by 8 degrees, *eROSITA* will enable us to study the structure of this cluster and its ICM by X-ray spectroscopy over this large region. Similarly, other larger structures can be studied in their full extent with *eROSITA*, like e.g. the Shapley supercluster, of which we only have a partial mosaic by *XMM-Newton* and Chandra observations and for which the RASS only provides an indication with its very shallow exposure.

A large number of nearby clusters will be detected with several thousand photons, allowing detailed studies of their structure. This will enable a systematic study of the fraction of dynamically young clusters and clusters featuring a major merger. One can obtain a good account of the temperature and metallicity distribution in those clusters pushing results to higher precision by grouping clusters of similar properties, and stacking the data, which is only possible for *eROSITA* due to the large number of objects at hand. Finally, we will have observations of a very large number of galaxy groups (characterized by ICM temperatures below 2 keV), for which we have at present only sparse results with poor statistics compared to our data on galaxy clusters. In addition, the (small) samples of galaxy groups studied in X-rays so far have been selected in very different ways and they are far from forming a homogeneous data set. *eROSITA* will provide for the first time a large, homogeneously X-ray selected sample of these objects for many detailed studies.

More than 5,000 galaxy clusters in the eRASS:8 catalogue (in the extragalactic sky) will have more than 1,000 photons. This is about twice the total number of presently known X-ray galaxy clusters, for most of which we just have a detection in X-rays. *eROSITA* will allow us to study in some detail the morphology of this large number of galaxy clusters, which are mostly nearby (at $z < 0.3$). Among the most interesting morphological studies is the characterization of the cluster cool cores, which are observed in about half of the galaxy clusters in the present day Universe (e.g. Peres et al. 1998, Hudson et al. 2010) and the statistical characterization of substructure by different statistical measures. While the observational analysis provides at first just an inventory of cluster morphologies, these studies gain extreme importance when they are compared to numerical simulations of galaxy clusters. As shown e.g. in Böhringer et al. (2010), a detailed comparison of the observed statistics of clusters morphologies with those seen in simulations provides a stringent test of the physics treated in the simulations themselves. Reproducing the observed occurrence of cool cores, for example, is still one of the most difficult unsolved challenges for cosmological hydro/N-body simulations. The very good statistics that will be provided by the *eROSITA* survey, which is better by at least one order of magnitude over our present-day knowledge, will put the comparison of observations and simulations on a completely new footing. It is also worth noting that the large *eROSITA* samples will be complementary to those that will be amassed in large optical surveys, that probe less massive and/or less evolved systems.

## 5.1.6 Galaxy cluster scaling relations and mass-observable calibration

Galaxy clusters have been shown to be a nearly self-similar family of objects, with e.g. total mass used as the scaling parameter (e.g. Stanek et al. 2010, Reichert et al. 2011, Böhringer et al. 2012). Scaling relations, and the scatter about them, are therefore very useful tools for a statistical description of the properties of the overall cluster population. Comparisons of the observed relations with theoretical predictions for different assumed cosmological models are also essential for the cosmological modeling of the cluster population. In this context, it is the mass-observable scaling that plays the most critical role.

Indeed, *eROSITA* will provide an unprecedented large number of galaxy clusters, allowing for a very precise determination of scaling relations. It will also allow, for the first time, the intrinsic scatter to be determined accurately by contrasting scaling relations obtained with various selection techniques. One of the most interesting and directly accessible scaling relation is that of X-ray luminosity and temperature. For more than 5,000 galaxy clusters detected with more than 1,000 counts we will be able to measure temperatures accurately, as well as obtain good estimates of the cluster's gas mass from the surface brightness profile fitting. Both parameters are good mass proxies (e.g. Kravtsov et al. 2006, Okabe et al. 2011) and important for the cosmological modeling. As an example, Figure 5.1.4 shows the simulated spectrum of a cluster with a ICM temperature of 4 keV at $z=0.2$ detected with 1,000 source counts in a 2 ksec exposure. Background modeling and subtraction was included in the simulation. In general, we find that with 1,000 source counts we can measure cluster temperatures to an accuracy of 10 - 35%, depending on the cluster temperature (ranging from 1-2 to 10 keV). In the group regime, where the current knowledge of the scaling relations is very poor, we will be able to obtain a good temperature measurement for objects with just about 500 counts (since at low temperatures the spectrum is richer in features).

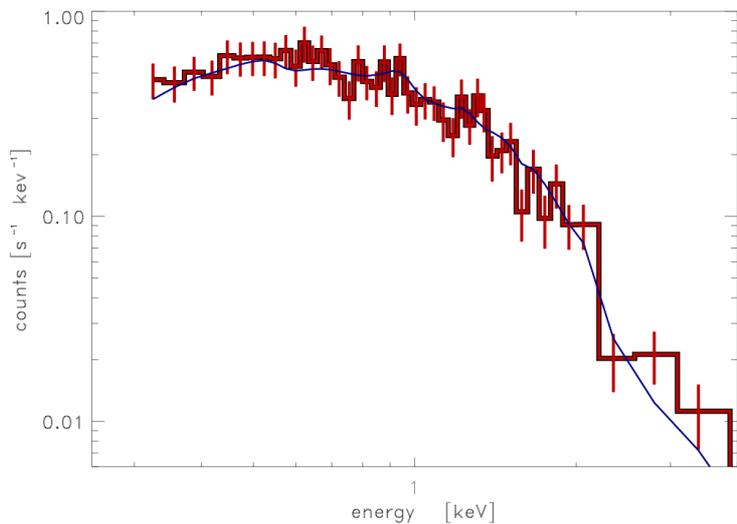

*Figure 5.1.4: A simulated spectrum of a cluster with a ICM temperature of 4 keV at $z=0.2$ detected with 1000 source counts by eROSITA in 2000 sec exposure. An extraction radius of 3 arcmin was assumed for the simulation.*

For cosmological applications it will be important to calibrate the mass-observable relations for the *eROSITA* clusters. For the majority of the clusters, the X-ray luminosity is the prime observable parameter. With the above described detailed observations, we will get a precise statistics of the relation between X-ray luminosity and the best mass proxies (temperature and gas mass), which have a scatter about a factor of two smaller than X-ray luminosity (e.g. Pratt et al. 2009) in the redshift range $0<z<0.6$. To further calibrate the temperature and gas mass to total mass relationship, we will use: (i) the deepest observations on the survey, (ii) previous *XMM-Newton* and *Chandra* observations for masses based on the hydrostatic approximation (known to be biased low by about 10 – 15%, e.g. Mahdavi et al. 2008, Zhang et al. 2010), (iii) the increasing number of weak lensing results becoming available, and (iv) the statistics of the galaxy velocity dispersions (e.g. from SDSS), which will be obtained through spectroscopic follow-up campaigns (Section 6.2 below). For the interpretation of the comparison of the different data sets on galaxy cluster masses, detailed and large cosmological simulations including hydrodynamics will be crucial.

## 5.1.7 The evolution of the clusters' thermal structure and chemical enrichment

The ICM keeps an interesting record of astrophysical processes related to galaxy formation and AGN activity: the energy released by these processes increases the internal energy and entropy of the ICM and nucleosynthesis associated to star formation, with the subsequent triggering of galactic winds, leads to the chemical enrichment of the ICM with heavy elements (metals) in addition to primordial H and He.

Both effects can easily be measured from spectroscopic X-ray observations of galaxy clusters. Recently the trend of the redshift evolution of the luminosity-temperature relation was established for the first time with sufficient statistics, and early energy input into the ICM (at redshift significantly higher than one) is supported by the observations, consistent with the predictions of "pre-heating" models (Reichert et al. 2011). But the constraints are still very weak. The good statistics of the *eROSITA* survey and dedicated pointed cluster follow-up data will allow us to put much stronger constraints on these feedback models. In this context, the close comparison of the simulations and observations for these effects will very much improve the degree of realism of the simulations and will thus make them more trustworthy calibrators for the cosmological modeling of cluster observations.

The ICM of rich clusters contains more metal mass than the galaxies. Therefore most of the nucleosynthesis yields of star formation have left the galaxies and can only be detected in X-ray observations of the ICM. So far we have only an indication of an evolution of the metallicity in the ICM with redshift (Balestra et al. 2007) and an approximate model for what types of supernovae are responsible for the enrichment by oxygen and heavier elements (de Plaa et al. 2007). The large amount of data supplied by *eROSITA* will enable us to provide a clear picture of the metal enrichment as a function of the galaxy population in the galaxy clusters, as a function of cluster mass and morphology, and as a function of redshift. While the exposures of individual clusters are low compared to the studies with *XMM-Newton* and *Chandra*, the advantage of *eROSITA* is the very large number of objects that will be studied. Therefore the breakthrough results will come from the analysis of stacked spectra of galaxy clusters grouped by their properties, and from the systematic *eROSITA* pointed phase cluster follow-up.

## 5.1.8 *eROSITA* and the highest redshift clusters

Observations of surprisingly massive galaxy clusters at redshifts $z>0.8$ (Menanteau et al. 2012; Foley et al. 2011; Jee et al. 2009; Santos et al. 2011) have recently attracted significant attention as a new tool to test predictions of the standard cosmological $\Lambda$CDM model for the growth of structure (Mortonson et al. 2011; Hoyle et al. 2011; Williamson et al. 2011; Baldi & Pettorino 2011; Waizmann et al. 2012; Harrison & Coles 2012).

For a given survey area, $\Lambda$CDM structure growth models can predict the mass limits as a function of redshift above which no clusters should be found (at, say, 95% confidence limit). Figure 5.1.5 shows such a prediction in a concordance cosmology (from Mortonson et al. 2011). The discovery of more massive distant objects than predicted, with sufficiently reliable and tight mass constraints, would point towards a significant level of non-Gaussianity in the primordial density field (see e.g. section 5.1.1; Pillepich et al. 2012; Sartoris et al. 2010) or exotic forms of Dark Energy, such as coupled DE scenarios (Baldi & Pettorino 2011).

The currently known massive and distant test clusters observed in the first half of cosmic time (e.g. at $z>0.8$) were detected either in deep X-ray surveys with moderate survey areas ($\leq$100 deg$^2$) or SZE surveys of a few thousands square degrees (see lower panel of Fig. 5.1.5). For realistic estimates of survey areas, the currently observed distant and massive cluster population appears not to be in significant tension with the $\Lambda$CDM predictions (Williamson et al. 2011; Menanteau et al. 2012).

*eROSITA,* with its all-sky coverage and sensitivity sufficient to detect the most massive clusters out to at least $z\sim1.6$, will constitute a unique tool for the discovery of the most massive distant objects in the Universe. The 4-year survey (eRASS:8) will enable a dense sampling of the most relevant mass-redshift plane (green box in top panel of Fig. 5.1.5) with dozens of test objects and will hence allow tight parameter constraints on non-standard cosmologies. The pre-requirements for such a challenging test up to the highest cluster redshifts accessible to *eROSITA* will be very robust detection algorithms, efficient follow-up identification techniques, and accurately calibrated mass proxies.

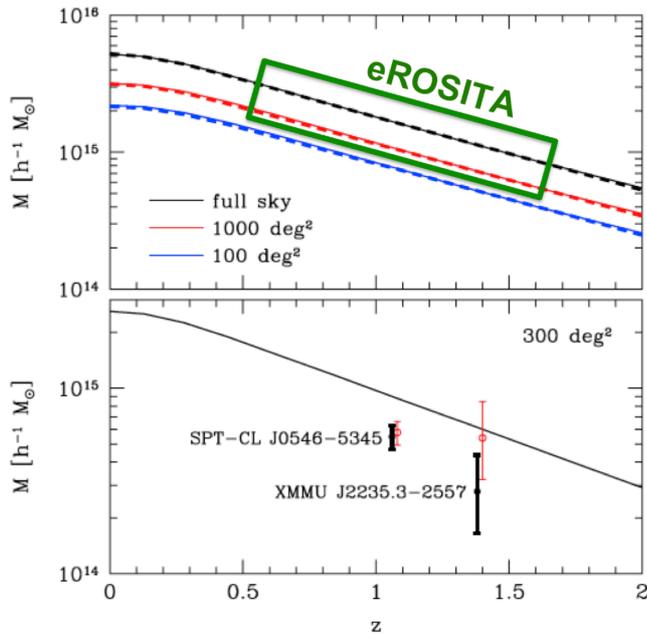

Figure 5.1.5: ΛCDM exclusion curves (95% CL) from Mortonson et al. (2011) for different survey areas (top panel) and including the locations of two prominent detected distant galaxy clusters (bottom panel). The discovery of a single galaxy cluster above the black line in the top panel would not be consistent with standard model predictions. eROSITA will be able to probe the most massive clusters in the full sky all the way to $z\sim 1.6$ (green box).

### 5.1.9 Comparison of the cluster X-ray properties with data from other wavelengths

The much deeper and more detailed observations of *eROSITA* on galaxy clusters in the entire sky compared to the RASS, will provide much more leverage in the comparison of cluster properties obtained with sky surveys at other wavelengths. The very interesting work of the correlation of SDSS and RASS results on the galaxy populations and X-ray bulk parameters of galaxy clusters (e.g. Popesso et al. 2006, Rykoff et al. 2008) gives only a first impression of the importance of such X-ray/optical studies. Two important aspects of these results are the clear trend that galaxy formation is more efficient in less massive cluster systems and the very good success in calibrating the optical richness-mass relation with the help of X-ray detections to enable a successful cosmological modeling of the optical surveys (Rozo et al. 2010). The large statistics of X-ray properties of galaxy clusters from *eROSITA* will provide very precise relations of X-ray versus optical properties and help to calibrate the selection function of clusters in optical and X-ray surveys and will allow for firm conclusions on the galaxy formation efficiency in clusters as a function of mass and redshift.

With the possibility of characterizing the morphology of thousands of clusters we can also shed more light on the claims, that the galaxy population in clusters has some dependence on the cluster morphological state. The Butcher-Oemler effect, that is the increases with redshift of the ratio of blue to red galaxies in a cluster (Butcher & Oemler 1984), may depend on the dynamical age of the cluster. It has also been reported that there are more post-starburst (E+a) galaxies in dynamically young or merging clusters (Poggianti et al. 2006). While these findings have always been based on small number statistics, and have always been debated, the *eROSITA* Survey in combination with photometric and spectroscopic optical surveys has the potential to finally answer these questions.

Similarly important will be the correlation of the *eROSITA* X-ray detections of galaxy clusters with the Sunyaev-Zeldovich effect signal from *Planck* and the *SPT* and *ACT* surveys (e.g. Planck Collaboration 2011; Carlstrom et al. 2011; Marriage et al. 2011), and the polarization sensitive *SPTpol* and *ACTpol*. Also the *LOFAR* radio array and the *ASKAP*/EMU survey will be fully operational during the *eROSITA* survey. First estimates lead to expectations of the detections of thousands of galaxy cluster radio halos (Ensslin & Roettgering, 2002; Norris et al. 2011). The possibility to study the occurrence and structure of the radio halos in connection with information on the morphology and bulk parameters of the clusters from *eROSITA* will surely provide new insights into the formation and astrophysics of cluster radio halos and the cosmic ray population and magnetic fields in galaxy clusters. Since many/most of the radio-detected (through halos, relics, wide-angle-tail radio galaxies, etc.) clusters will be dynamically active, a cross-comparison with *eROSITA* data could serve as a possibility to "flag" merging clusters and treat them differently in the cosmological interpretation (e.g., use of a different scaling relation).

There is an especially large synergy between the *eROSITA* Survey and the *Euclid* ESA mission concerning galaxy cluster science and cosmology. The large number of galaxy redshifts that will be obtained with *Euclid* over a large area of the sky will also provide redshifts for cluster galaxies comprising most of the X-ray detected clusters and will most importantly cover the more distant galaxy clusters not covered by the ground based spectroscopic surveys. For the cosmological model testing, the calibration of the X-ray luminosity mass relation with the help of *Euclid* will be particularly important. The lensing analysis of the high angular resolution optical/NIR images obtained by *Euclid* will allow us to determine galaxy cluster masses by weak gravitational lensing modeling. Since we know that there is a substantial uncertainty in the weak lensing mass determination due to the intervening structure in the line of sight of at least 30% (e.g. Hoekstra 2003) whereas the mean signal is expected to be unbiased, large statistics is a prerequisite for a precise mass calibration. As we expect a lensing signal to be detected for the majority of the *eROSITA* clusters (having masses larger than about $2\times10^{14}$ solar masses) we will have sufficient statistics for mass calibrations to a few percent for many bins in mass and redshift. Obviously, extensive, and accurate, photometric redshift information is needed for a proper evaluation of the lensing data. We discuss in section 6.1 below the landscape of wide area optical/NIR imaging surveys that will be used towards this end.

## 5.2 Active Galactic Nuclei and normal galaxies

The emergence of the three-dimensional structure of the cosmic web over the history of the Universe displays very distinctive features when observed in X-rays. X-ray emission is a signpost of accretion of matter onto the supermassive black holes that seed the whole population of galaxies and may strongly influence their formation and subsequent evolution. A complete and homogeneous inventory of X-ray sources, if complemented with accurate redshifts, can therefore provide a wealth of crucial information on the nature of the large-scale structure itself, as well as on the history of accretion in the Universe and the role of AGN in galaxy formation and cosmology. Indeed, the *eROSITA* X-ray sky will be dominated, numerically, by the population of Active Galactic Nuclei (AGN) that together make up the vast majority of the Cosmic X-ray Background (CXRB) light.

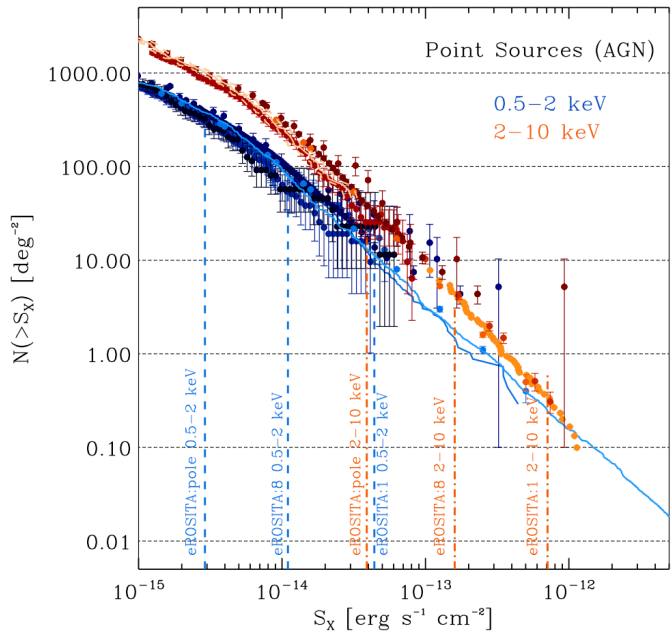

*Figure 5.2.1: Compilation of cumulative X-ray source counts (number of objects brighter than a given flux per square degree) in the soft (0.5-2 keV, cyan-blue colors) and hard (2-10 keV, red-orange colors) bands. The observational data are taken from Cappelluti et al. (2009) (see references therein) and Mateos et al. (2008). Vertical lines show the expected flux limits of three different eROSITA all sky surveys: the first one after six months (eRASS:1) and the one completed after the planned eight passages (eRASS:8) for soft (blue) and hard (orange) X-ray bands, respectively.*

The presence of Super-massive Black Holes (SMBH) in the nuclei of virtually all local massive galaxies establishes a link between high-energy X-ray astronomy and galaxy evolution. Despite the observational evidence that the formation and growth of SMBHs and their host galaxies are related processes (i.e. Ferrarese & Merritt 2000; Gebhardt et al. 2000), it is still unclear which physical mechanisms are responsible for their coupling. In order to discriminate among the many and different model predictions, sizable samples of AGN over different phases of their evolution are needed. X-ray selected AGN samples are extremely valuable, as they are less biased by obscuration effects than optical ones; they also tend to be more complete at the low end of the AGN luminosity function, where optical selection is affected by host galaxy light dilution. Up to now, the samples of X-ray selected, spectroscopically identified AGN are small compared to the ones available from large area optical surveys (Brandt & Hasinger 2005). The *eROSITA* all-sky survey will deliver about 3 Millions X-ray selected AGN and QSOs, thus providing the statistics needed to assemble all the different populations caught at different evolutionary phases, and study their clustering properties. In the following subsections, we describe in detail the main goals of the study of AGN with *eROSITA*. They can be summarized as follows:

- We will determine the **evolution of the accretion history onto supermassive black holes** by studying with unprecedented detail the X-ray AGN Luminosity Function;
- We will study the **clustering properties of X-ray selected AGN** and the relationship between BH growth in galactic nuclei and the Large Scale Structure of the Universe, at least up to z~2;
- We will amass a statistically significant sample of tens of thousands **obscured AGN** which will enable accurate studies of the relationship between accretion on the nuclear black hole and host galaxy properties (mass, star formation, metallicity, etc.);
- We will study **rare AGN sub-populations**, such as high-redshift objects, highly obscured nuclei, etc.;
- We will be able to study how the overall **Spectral Energy Distribution (SED) of X-ray selected AGN**, as well as the detailed spectral feature in the *eROSITA* band (~0.3-10 keV) changes with redshift, luminosity, host galaxy properties, by compiling highly accurate stacked template spectra with large number of objects.

## 5.2.1 Evolution of the AGN population

To determine the evolution of the AGN population with cosmic time we need powerful deep surveys sampling the highest redshift at any particular luminosity, but we also need powerful wide-field surveys which sample the rare high-luminosity objects and give a local anchor to the luminosity function evolution.

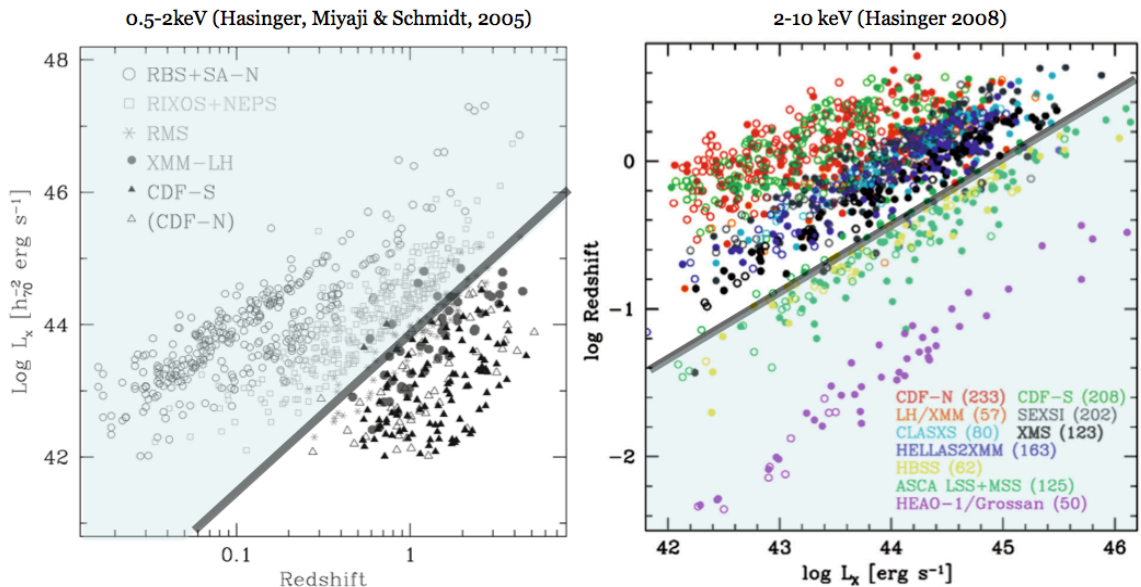

*Figure 5.2.2: Luminosity redshift plane location of 0.5-2 keV (left) and 2-10 keV (right) selected AGN from existing surveys (colored points, from Hasinger et al. 2005 and Hasinger 2008, respectively). The parameter space covered by the all-sky eROSITA survey (eRASS:8, average exposure) is shown as a yellow shaded area in both planes.*

Figure 5.2.2 shows the samples that have been used for the highest quality studies with the largest numbers of identified X-ray sources in the 0.5-2 keV and the 2-10 keV band, respectively. In the soft X-ray band the combination of the *ROSAT* All-Sky-Survey with the deep *ROSAT* and later *Chandra* plus *XMM-Newton* surveys could convincingly establish a luminosity-dependent density evolution for the AGN population with a peak in space density that moves from higher to lower redshifts with decreasing X-ray luminosity (e.g. Ueda et al. 2003; Hasinger et al., 2005). This downsizing behaviour of the AGN population has in the meantime been confirmed for all other electromagnetic bands (e.g. Hopkins et al. 2007). The right hand side of the figure shows some of the best samples currently available in the 2-10 keV band (Hasinger 2008). As one can see, while there is an excellent coverage in the luminosity-redshift plane from the deep *Chandra* and *XMM-Newton* surveys, there is a glaring gap at brighter fluxes. Bright sources are rare and large cosmological volumes are required for good statistics. Aird et al. (2010), for example, resorted to *ASCA* Medium Deep Survey (Ueda et al. 2001, Akiyama et al. 2003) to improve the statistics at the bright end of the XLF and complement deep *Chandra* surveys, and still had just a few tens of QSO with Log $L_X$ > 45. For the 2-10 keV selected samples, the luminosity function still has to be anchored by the 30 years old HEAO-1 survey (or extrapolated from the more recent harder band *Swift* BAT data). eRASS will perfectly fill this gap in a similar way as the RASS has done in the soft band (see Table 5.2.1).

We have assessed the AGN content of the *eROSITA* survey after 4 years of operation (eRASS:8), taking into account the survey scanning strategy and the galactic foregrounds (see Section 3), based on the AGN population of the Gilli et al. (2007) model for the CXRB. Figure 5.2.3 shows the expected redshift and luminosity distributions of both obscured and unobscured AGN, for a 0.5-2 keV *eROSITA* selected sample, including the important effect of luminosity-dependent obscured AGN fraction (see e.g. Hasinger 2008). The overall population, largely dominated by un-obscured objects, will peak at redshift slightly lower than unity and luminosities of about $L_X \sim 10^{44}$ erg/s.

The reconstruction of the black holes accretion history is the first step toward constraining the evolution of the black-hole mass function. That will serve as a benchmark to test competing models for the growth of SMBH in the Universe. Once the XLF is well described over wide accretion luminosity and redshift baselines the most robust constraint on the SMBH mass function at any redshift can be obtained by integrating the continuity equation backwards in time, using the local mass function as a boundary condition (Merloni & Heinz 2008). A byproduct of this approach is the accretion rate distribution of AGN as a function of redshift and luminosity. This provides additional constraints to different models for the growth of SMBH which make specific testable predictions on how the Eddington ratio of AGN should evolve with epoch and accretion luminosity (e.g. Hopkins et al. 2006, Fanidakis et al. 2011).

|  |  | Log X-ray (0.5-2 keV) Luminosity [erg/s] | | |
| --- | --- | --- | --- | --- |
|  |  | 44-45 | 45-46 | 46-47 |
| **Redshift range** | 0-1 | $2.20 \times 10^5$ ($4.90 \times 10^4$) | $2.02 \times 10^3$ ($8.15 \times 10^2$) | 12 (7) |
|  | 1-2 | $1.00 \times 10^6$ ($1.13 \times 10^5$) | $4.14 \times 10^4$ ($1.29 \times 10^4$) | 355 (200) |
|  | 2-3 | $1.81 \times 10^5$ ($2.37 \times 10^4$) | $7.90 \times 10^4$ ($2.32 \times 10^4$) | 765 (400) |
|  | >3 | $3.20 \times 10^3$ ($3.32 \times 10^2$) | $2.14 \times 10^4$ ($5.50 \times 10^3$) | 472 (3) |

*Table 5.2.1: Number of AGN that will be detected in eRASS:8 (All-sky survey, 4 years) over a total extragalactic area of 30,000 deg² in various luminosity and redshift bins. The numbers in parenthesis are for obscured AGN (Log $N_H$ >21).*

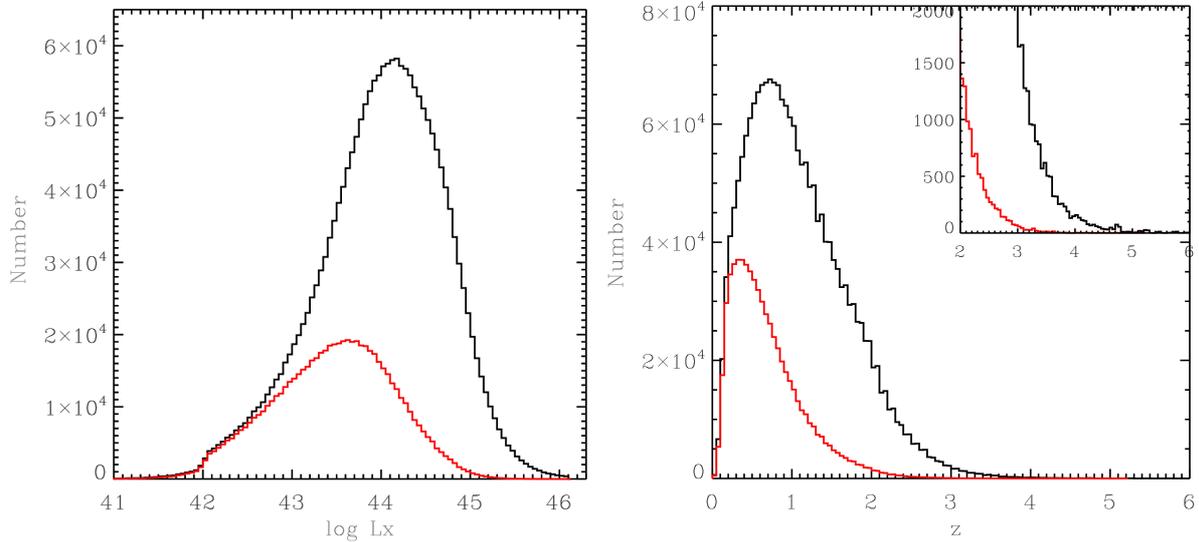

*Figure 5.2.3: Left: Observed 0.5-2 keV luminosity distribution of soft X-ray selected AGN in eRASS:8. Right: Redshift distribution of eROSITA AGN (0.5-2 keV selected). The inset shows a zoom into the high redshift part of the population. In both panels, black lines are for the total population, red for that of obscured AGN.*

## 5.2.2 Relation of black holes' growth to the large scale environment

The environment of AGN, i.e. the large scale dark matter halo structures in which they live, has recently emerged as a powerful diagnostic of the conditions under which SMBH grow across cosmic time. The triggering mechanism of the nuclear activity (e.g. major mergers, disk instabilities, cooling flows), the accretion mode onto the SMBH (e.g. hot vs. cold gas accretion) and the interplay between SMBH growth and host galaxy, imprint detectable signatures on the clustering of AGN. Indeed, recent cosmological simulations predict very different dependences of the AGN dark matter halo mass on accretion luminosity and redshift, as a direct result of the physics adopted to grow SMBH (e.g. Marulli et al. 2009, Fanidakis et al. 2011, Allevato et al. 2011; see Figure 5.2.4).

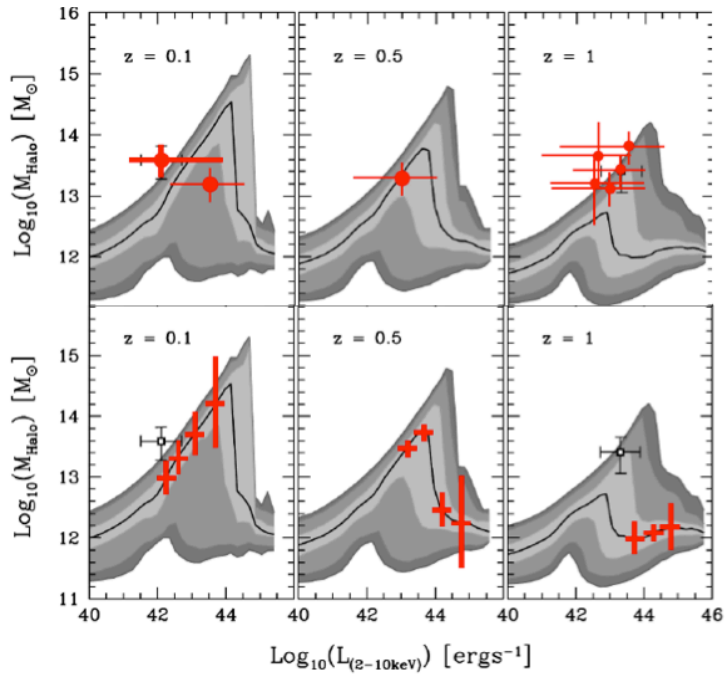

*Figure 5.2.4: Top: Current constraints on the dark matter halo mass hosting AGN of a given luminosity, as deduced from clustering analysis (red points). Bottom: prediction of the quality of constraints possible with eROSITA, where the available AGN from the all-sky survey can be binned in a large number of L-z independent intervals. In both panels the solid lines and grey shaded area represent the prediction from the Fanidakis et al. (2011) semi-analytic models of AGN-galaxy co-evolution. The solid lines correspond to the mean DM halo mass at a given $L_X$. The different shadings correspond to the 68, 95 and 97.3 percentiles around the median.*

*eROSITA* will open up a new region of the parameter space to environmental investigations by detecting, for the first time, large numbers of powerful sources, $L_X \gtrsim 10^{44}$ erg/s. Although such bright AGN dominate the SMBH growth in the Universe (Aird et al. 2010), they are very rare objects. The number of such sources in current X-ray surveys, which typically cover limited areas on the sky, is therefore too small to determine their clustering. This is a major limitation as it does not allow constraints on the physics that drive the bulk of the SMBH growth in the Universe. The alternative approach of tackling this problem by studying the properties of AGN hosts (e.g. morphology, star-formation history, stellar mass) is also problematic, as contamination of the host galaxy light by emission from the central engine is close to maximal at bright accretion luminosities. This is the regime that clustering studies can provide the most robust constraints on the physical processes responsible for the growth of SMBH.

We estimate the expected *eROSITA* AGN density (per square degree) at different redshifts from the logN-logS AGN distribution (see figure 5.2.1 above). Using constant redshift bins of $\Delta z$=0.2 we can compare the AGN volume density (co-moving) at different redshift with the volume density of SDSS Luminous Red Galaxies, which have been extensively used for clustering measurement to study the distribution of these massive galaxies within each of the dark matter halos and to detect the first Baryonic Acoustic Oscillation (BAO) signal ever (Zehavi et al. 2005; Zheng et al. 2009; Eisenstein et al. 2005; Sanchez et al. 2012). Simulations show that the auto-correlation function of the *eROSITA* AGN in 10,000 deg² will have uncertainties of approximately 3% in redshift bins of $\Delta z$=0.2 in the range z=0.4-1.6, which is similar to the high accuracy clustering measurements achieved for LRGs. Consequently, we will be able to place stringent constraints on the typical mass of dark matter halos hosting AGN of different luminosities, thus providing unique tests of current models of AGN activation and AGN-galaxy co-evolution, as shown in Figure 5.2.4. Even at $z$~3 we still expect to achieve uncertainties of 20-25% which is comparable to SDSS based AGN clustering measurements with the same size of the redshift bins. Furthermore, AGN clustering measurements will be able to detect the BAO signal at $z$~1 with uncertainties similar to the BAO measurements with LRGs (Klodzig et al., in prep). A major challenge for comprehensive AGN clustering studies is the need for uniform spectroscopic redshift information for such a large sample. We discuss current plans and opportunities for massive spectroscopic follow-up campaigns for *eROSITA* sources in Section 6.2.

### 5.2.3 The host galaxies of obscured AGN

The obscured AGN population will represent, numerically, a small fraction of the entire *eROSITA* sample. They will be, however, highly complementary to the dominant unobscured QSOs population, and will enable key studies of the host galaxies of AGN over a crucial redshift range. Recent studies in the local Universe ($z$<0.2) from the SDSS survey pointed out the apparent existence of two distinct modes in the Eddington ratio distribution for different host galaxy stellar population ages (Kauffmann & Heckman 2009), and the correlation between the bolometric AGN luminosity and the star formation luminosity. The spectroscopic follow-up campaigns of *eROSITA* selected AGN (see Section 6.2 below) will therefore be ideally suited to study the same characteristics over a wider, and more uniformly sampled, redshift interval. Spectral features, e.g. emission line intensities and the depth of the Balmer break, measured in the optical spectra will be used to constrain the star-formation rates, the reddening by dust, the stellar population ages and metallicities of the AGN host galaxies; hence, it allows to test the luminosity and redshift dependence of the dichotomy on the accretion rate distribution, and the metallicity history in AGN host galaxies. Moreover, *eROSITA* all-sky survey will deliver the first, sizable sample of obscured, luminous quasars caught in the most critical phase of AGN-galaxy evolution at $z$=0.3-2 (about $1.7 \times 10^5$, see Table 5.2.1 above). With such a sample, we will extend the knowledge of detailed physical properties of type-2 AGN hosts with high statistical significance, well beyond the local Universe and into the epoch of the most vigorous SMBH growth.

### 5.2.4 Rare AGN populations

Our current knowledge of the properties of very high redshifts ($z$>6) QSOs comes from optical surveys, such as the SDSS and CFHT (with a total of ~40 QSO currently available, Willott et al. 2009, Jiang et al. 2009). The optical surveys only sample the high-luminosity tail of the QSO population and therefore they are likely not representative of the overall population at those redshifts. In addition, most models for the co-evolution of AGN and galaxies predict that, in the early Universe, powerful AGN may also be obscured. X-ray selection is less sensitive to obscuration effects and is be more efficient at picking up the bulk of QSO population at lower, more representative luminosities. However, the space density and evolution of X-ray selected QSOs and AGN at $z$>6 is today completely unconstrained, due mainly to the lack of area coverage at substantial depth (see, e.g. Brusa et al. 2010): the highest redshift, spectroscopically confirmed, X-ray selected sources in medium-deep Chandra fields are at $z$~5.2-5.4 (Barger et al. 2003; Steffen et al. 2004; Civano et al. 2011), and only a few candidates at higher $z$ have been claimed using the photometric redshift technique (Salvato et al., 2009, 2011). The systematic search for luminous, X-ray selected AGN at $z$>3, and the precise constraint of their luminosity (and mass) functions is an important goal of the *eROSITA* survey. *eROSITA* will bring, for the first time, the statistics of high-redshift X-ray selected QSO at the same level of SDSS QSOs (see Figure 5.2.5); as an example, (conservative) extrapolations from lower-redshift luminosity functions predict few tens of $z$>6 QSOs in the *eROSITA* surveys, and the real number may sum up to about a thousand, if the most optimistic extrapolations are considered. Given that the total number of AGN expected from the *eROSITA* surveys is ~3 millions, it is essential to develop a powerful selection technique to efficiently isolate such an elusive population of sources at high-$z$. In particular, an efficient culling technique is required to discriminate among the many faint counterparts in the relatively large *eROSITA* error circle. Deep NIR surveys in combination with a dropout selection technique will be likely necessary ingredients of any attempt to identify the most distant *eROSITA* AGN.

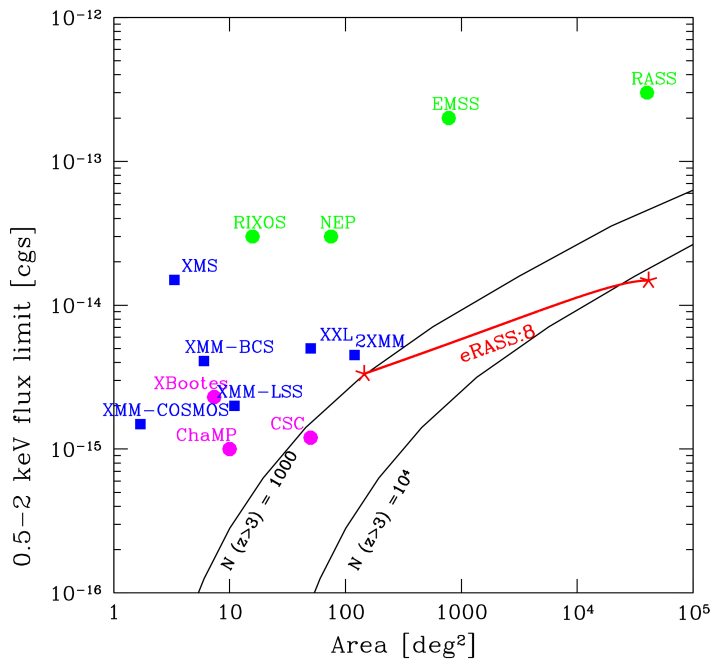

*Figure 5.2.5: The loci of constant expected numbers of high redshift ($z$>3) AGN (solid lines) are plotted in the Area-Flux limit (0.5-2 keV energy range) plane. Lines marking 1000 and $10^4$ are plotted. Past X-ray surveys from Chandra (purple dots), XMM-Newton (blue squares) and ROSAT (green circles) are shown alongside the predictions from the 4 years eROSITA surveys (eRASS:8, red curve).*

Current models for the evolution of the AGN population rely heavily on the results of wide and deep X-ray surveys with *Chandra* and *XMM-Newton*. Those, however, are essentially unable to detect the most obscured Compton Thick (CT) AGN (those with absorbing column densities of $N_H$>$10^{24}$ cm$^{-2}$), even if a significant fraction of those sources (up to 50% of the obscured AGN population) have been postulated to explain the shape and overall normalization of the diffuse Cosmic X-ray background (Gilli et al. 2007). Given that the observed population of Compton Thin AGN is able to explain the observed background at lower X-ray energies (<4-5 keV; Worsley et al. 2005), CT AGN remain the last "missing" population in our census of growing black holes. Therefore, the search for these heavily obscured sources using a variety of multi-wavelength selection techniques has become a prolific industry. Indeed, expectations on the true size of the missing Compton Thick population may vary wildly. Models that synthesize the XRB spectrum have a number of input parameters, such as the exact shape of the AGN X-ray spectra (reflection component, cutoff energy, power-law exponent) or the level of contribution from radio-loud AGN. Small changes of those parameters to plausible values have a large impact on the number of CT sources required to explain the shape of the CXRB (Treister et al. 2009, Draper & Ballantyne 2010). As a result of this uncertainty, the most robust constraints on the space density of CT AGN come from direct detections of such sources at different flux limits (e.g. Georgantopoulos et al. 2008, Tueller et al. 2008). Although the *eROSITA* energy response is not geared toward heavily obscured and Compton Thick AGN, because of the sheer volume of the Universe surveyed, it will detect many AGN in the Compton Thick regime, providing a constraint on the number of CT QSOs at the yet unexplored flux limit of $\sim 10^{-14}$ erg s$^{-1}$ cm$^{-2}$.

## 5.2.5 *eROSITA* potential for X-ray spectroscopy of AGN

As far as X-ray spectra are concerned, we still lack a deep understanding of the exact mechanism responsible for AGN X-ray emission and its physical location. As a very general diagnostic, the "X-ray loudness", usually characterized by the $\alpha_{ox}$ parameter, i.e. the slope of the spectrum between 2500 Å= 5 eV and 2 keV: $\alpha_{ox} = 0.3838 \log(F_{2keV}/F_{2500A})$ can be used to characterize the fraction of bolometric light carried away by high-energy X-ray photons. Recent studies of large samples of both X-ray and optical selected AGN have demonstrated that $\alpha_{ox}$ is itself a function of UV luminosity (see e.g. Steffen et al. 2006; Young et al. 2009). However, no redshift evolution can be discerned in the data. Moreover, large collecting-area X-ray telescopes allow a more precise determination of the Xray spectra of AGN, which are usually characterized by a power-law, upon which emission lines and absorption features are superimposed. Up to the highest redshift where reliable spectral analysis of AGN can be performed, no clear sign of evolution in the X-ray spectral slope has been detected (Young et al. 2009). Similarly, while the narrow iron Kα emission line, the most prominent feature in AGN X-ray spectra, is clearly dependent on luminosity (the so-called Iwasawa-Taniguchi effect: Iwasawa & Taniguchi 1993; Bianchi et al. 2007), it shows no sign of evolution in its equivalent width with redshift, at least up to $z\approx1.2$ (Chaudhary et al. 2010). Finally, a fundamental open question for X-ray spectroscopy of AGN is the origin, incidence, and possible evolution of the broad, relativistically skewed iron Kα line (Fabian et al. 2000). Since the first unambiguous detection of relativistically broadened Fe K emission in the X-ray spectrum of MCG-6-30-15 observed with the *ASCA* satellite (Tanaka et al. 1995), sincere efforts have been made to test for the presence of broad Fe K emission lines in AGNs. *Chandra*, *XMM-Newton* and *Suzaku* have revealed that statistically significant broad Fe lines can be detected in relatively deep observations (relativistic lines are detected in ≈50% of nearby *XMM-Newton* AGN when the "well-exposed" spectra with >10,000 net counts in the 2–10 keV band are considered; Nandra et al. 2007; de la Calle et al. 2010). Very little is known, however, about the properties of the Fe Kα broad emission lines at higher redshift. Using rest-frame stacking procedures to compute the mean Fe Kα profile in a sample of ~250 AGN from the 2XMM survey (at $0<z<2$), Chaudhary et al. (2012) have recently shown that the average Fe Kα line profile in that sample is best represented by a combination of a narrow and a broad line, with the equivalent widths of the narrow and broad components are ~30 eV and ~100 eV, respectively. Such low measured values of EW, in particular for the broad line component, imply that very good photon counting statistics (and good detector calibration, to better constrain the underlying continuum) are needed to make further progress in the field. What will the contribution of *eROSITA* be in this field?

Only a small fraction of individual AGN in the *eROSITA* sky will be detected with enough X-ray counts to ensure accurate spectral analysis. Nonetheless, the sheer number of objects means that the survey is endowed with tremendous potential for the study of the physical conditions of the accreting gas nearby supermassive black holes across cosmic time. This can be harnessed provided sufficiently complete and extensive follow-up campaigns allow at least a photometric redshift determination (Salvato et al. 2011) for the detected sources (see section 6 below), by stacking the observed spectra in different bins in the luminosity-redshift plane. This will make it possible to reach spectral accuracy in each selected bin high enough to study in detail the possible evolution with redshift and luminosity of the absorption column density, of the average spectral slope and X-ray loudness parameter, as well as of individual spectral features, such as the narrow and broad, relativistic components of the iron line, or possible absorption features due to fast outflowing material in powerful accretion disc winds (Pounds et al. 2003; Tombesi et al. 2010). To illustrate this point, we show in the top panels of Fig. 5.2.6 the total number of counts to be detected in eRASS:8 from AGN as a function of their redshift and Luminosity. A high level of redshift completeness could allow the construction of a large number of stacked X-ray spectral templates of very high quality.

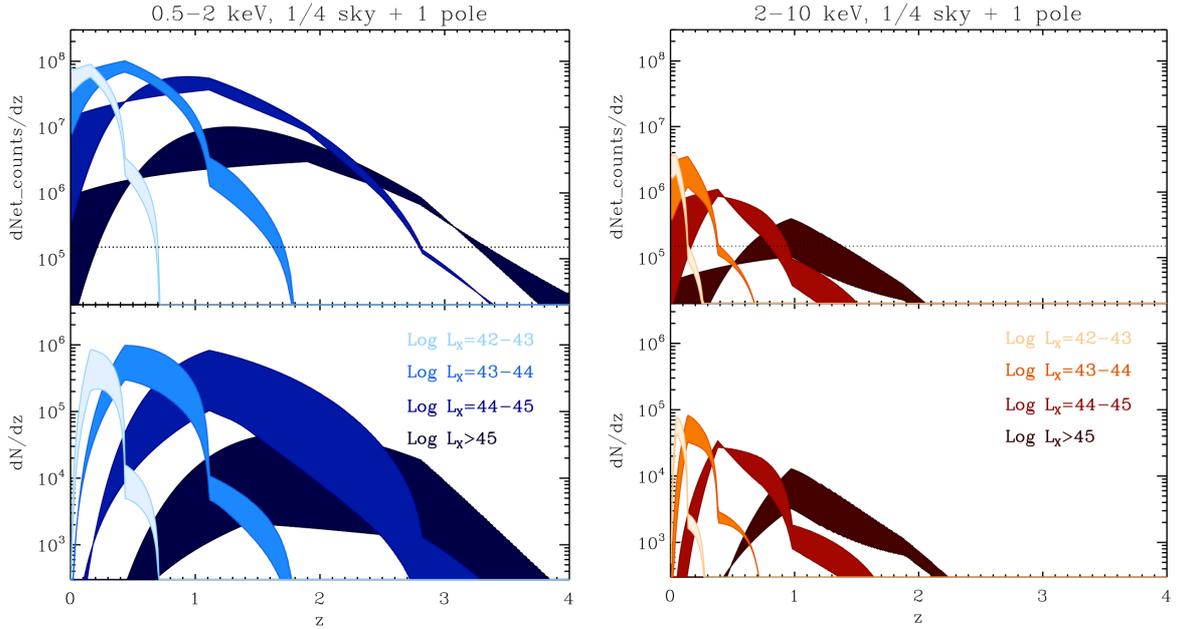

*Figure 5.2.6: An illustration of the potential of eROSITA for X-ray AGN spectroscopy. The bottom panels (left for soft X-ray selected AGN, right for hard X-ray selected ones) show the differential number counts as a function of redshift, split in different luminosity bins, as per legend. The top panels show the corresponding total X-ray counts (calculated assuming the conversion factors and sensitivity discussed in section 4.3) that could be accumulated by stacking spectra of sources with similar X-ray luminosity and redshift. In all panels, the numbers refer to all AGN visible over ~10,000 deg² of the eRASS:8 all-sky survey, plus one of the two deeply exposed ecliptic poles. The shaded areas measure the uncertainty due to the choice of X-ray luminosity functions (we have used Ebrero et al. 2009 and Hasinger et al. 2005 for the 0.5-2 keV XLF, and Ebrero et al. 2009 and Silverman et al. 2008 for the 2-10 keV one). The horizontal dotted line in the top panels mark the empirical count limit above which a meaningful determination of broad Fe K$\alpha$ lines properties in local AGN is possible (Guainazzi et al. 2006).*

## 5.2.6 Normal galaxies in the *eROSITA* survey

We close this section with a brief outline of the potential of *eROSITA* to study X-ray emitting galaxies where the nuclear SMBH emission represents a subdominant component of the observed radiation. The X-ray emission of such galaxies is, to a large extent, dominated by the collective emission of their X-ray binary populations. X-ray binaries are divided into Low-Mass X-ray Binaries (LMXBs) and High-Mass X-ray Binaries (HMXBs) according to the mass of the optical companion, resulting in drastically different evolutionary time scales and different relations of their luminosity functions with the present star formation of the parent galaxy. The fast evolution of HMXBs makes them a good tracer of the very recent star formation history (see Antoniou et al. 2010 for the case of the SMC) while from the slow evolution of LMXBs no relation to the present star formation rate is expected. Instead, the LMXB population reflects the total stellar content of a galaxy formed during its lifetime (see e.g. Grimm et al. 2002). Grimm et al. (2003) conclude that the luminosity distribution of HMXBs in a galaxy can be approximately described by a universal luminosity function with a normalization which is proportional to the star formation rate.

About 15,000 to 20,000 normal galaxies will be detected in eRASS:8 (Prokopenko & Gilfanov 2009) from which about 8,400 are of early type and 7,000 to 10,000 of late type. Galaxies up to distances of 50 – 70 Mpc will be detected as extended sources, while for galaxies within 20 Mpc the brightest sources can be resolved. This applies in particular to the class of ultra-luminous X-ray sources (ULXs) with luminosities larger than $10^{39}$ erg s$^{-1}$. This source class was first identified by the Einstein Observatory (Fabbiano 1989) and is discussed to consist of intermediate mass black holes with masses between $10^2$ and $10^4$ solar masses (e.g. Miller & Colbert 2004). ULXs with luminosities of $10^{40}$ erg s$^{-1}$ at a distance of 35 Mpc will have fluxes of $2\times10^{-14}$ erg s$^{-1}$ cm$^{-2}$, about the sensitivity limit of the *eROSITA* survey. About 100 ULXs are expected within a distance of 35 Mpc.

## 5.3 Stellar mass compact objects

Galactic compact X-ray sources are single or binary stars harboring white dwarfs, neutron stars or stellar mass black holes. They form a very heterogeneous class of X-ray emitters, that may be powered by accretion, thermo-nuclear explosion, magnetic field decay, stellar spin-down, collapse, internal energy and other less common mechanisms. Accretion as the most common form of powering happens via Roche-lobe overflow and via the Bondi-process either from a stellar wind or from the interstellar medium. Galactic compact X-ray emitters hold the key to understand evolutionary channels both of single and binary stars. As individuals, they offer rich diagnostic power via timing and spectroscopy; as a class, they provide key information on population synthesis models and are essential to unveil the nature of the diffuse galactic ridge X-ray emission (GRXE) observed in the Milky Way and in other galaxies.

The scientific impact of the *eROSITA* survey in the field of compact objects in our (and nearby galaxies) is governed by its discovery and its monitoring potential, since galactic compact objects are among the most variable objects in the sky.

The discovery power lies in the *eROSITA* ability to detect many new point-like X-ray sources identifying galactic compact candidates (or compact objects in the Magellanic Clouds) among them. The latter is a very challenging task which will in most cases require follow-up X-ray or multi-wavelength observations (see Chapter 6). However, a number of identifications will already be possible on the basis of the *eROSITA* data alone, thanks to the instrument's spectral (up to several keV) and timing resolution which were very limited in previous all-sky X-ray surveys.

The monitoring potential of *eROSITA* is related to the observational strategy during the survey, which allows multiple observations of a considerable portion of the sky around ecliptic poles. Many X-ray sources located in those regions will therefore be repeatedly observed with an "XMM-class" instrument for the first time, which will allow an unprecedented study of their long-term flux, spectral and timing variability (see Section 5.7 below).

With *eROSITA* we will discover and study several 10,000 galactic compact objects of all kinds and we will:

> - Measure the **space density, galactic scale height and luminosity function** of the various kinds of white dwarf accretors (*cataclysmic variables, double degenerates, Super-soft sources, Recurrent Novae, and symbiotic binaries*);
> - Determine the **X-ray luminosity function of X-ray binaries** (neutron star accretors) in the Milky Way;
> - Determine the contribution of high- and low-luminosity objects to the **Galactic Ridge X-ray emission** and ultimately determine the composition of its resolved phase;
> - Uncover the population of **isolated neutron stars**, within 2 kpc and constrain the evolutionary links between various known sub-classes of the neutron stars family (XDINS, AXPs, SGRs, RRATs).

### 5.3.1 Cataclysmic Variables

Cataclysmic variables are binary systems containing either magnetic or non-magnetic white dwarfs (CVs and MCVs). Some subclass is expected to contain the long-sought progenitor of the Supernova type Ia explosions, candidates being the double degenerates (DD, Schaefer & Pagnotta 2012), the Recurrent Novae (RN), the hard X-ray emitting Symbiotic Binaries (Kennea et al. 2009), and the Supersoft X-ray Sources (SSS). Although almost 1,000 CVs are known, we are far from establishing a coherent picture about their evolutionary role and their contribution to the X-ray active Milky Way. Main reasons are the highly biased sample composition and the uncertainties in the distance estimates. Cataclysmic binaries were detected in various ways, due to their strong variability, their blue color or due to pronounced soft X-ray emission, and each method left largely unknown imprint on the observed population. Complete samples that were used to determine space density comprise typically one or two handful of objects and estimated space densities vary by up to two orders of magnitude (Hertz et al. 1990, Schwope et al. 2002, Pretorius et al. 2007, Pretorius & Knigge 2012).

The situation will substantially change since the *eROSITA* survey will unravel the zoo of compact binaries for the first time within ~1kpc radius down to a luminosity of $10^{30}$ erg s$^{-1}$, the minimum luminosity of CVs in the *ROSAT* Bright Survey (RBS, Schwope et al. 2002) and the *ROSAT* NEP (North Ecliptic Pole survey, Pretorius et al. 2007). Assuming the current best estimates of the population parameters (Pretorius & Knigge 2012) one may expect a cumulative number of CVs as a function of the Galactic co-latitude as shown in Figure 5.3.1 for scale heights of 150 pc and 250 pc, respectively, representing a moderately young and old population. In the calculation, a mid-plane space density $\varrho_0 = 0.6\times10^{-5}$ pc$^{-3}$ was assumed, which is uncertain by a factor of two or more. Hence, *eROSITA* is expected to detect several thousands non-magnetic CVs in a flux-limited sample with radius 900 pc. On the other hand, predictions were made that the mid-plane space density of CVs might be as high as $2\times10^{-4}$ pc$^{-3}$ (Kolb 1993). Should such a population exist, it would have luminosities below $2\times10^{29}$ erg s$^{-1}$ (Pretorius et al. 2007) and about 5,000 of those low luminosity sources would populate the *eROSITA* X-ray sky.

Most CVs will be discovered in the soft band between 0.5 and 2 keV. Recent surveys at higher energies with *Swift*, *INTEGRAL*, and *RXTE* have revealed a significant number of magnetic CVs (Intermediate Polars, IP, and asynchronous Polars) that might be sufficiently abundant to synthesize the GRXE. Adopting a minimum IP luminosity in the 2-10 keV band of $10^{32}$ erg s$^{-1}$ and a 10% IP fraction among the CVs, *eROSITA* will yield a flux-limited IP sample with radius 1.7 kpc that contains 9,000 IPs. Hence, the whole *eROSITA* sky will contain tens of thousands CVs of all flavors.

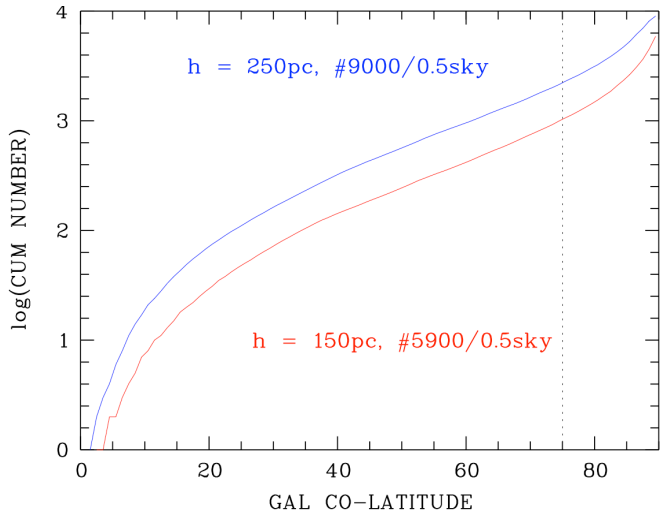

*Figure 5.3.1: Cumulative number counts as a function of Galactic zenithal distance of non-magnetic CVs predicted in each eROSITA hemisphere based on a luminosity function of ROSAT-detected CVs (Pretorius & Knigge 2012).*

Distance determination with *GAIA* and optical identification with spectroscopic facilities are essential to proceed further. With this unique data set we will (a) uncover the parent population of CVs free of selection and detection bias from flux-limited samples comprising about $10^3$ objects. For the first time the true composition of the CV population with magnetic and non-magnetic systems will be uncovered to constrain the effect of magnetic fields on close binary evolution. We will furthermore (b) probe the existence of the putative large population of low-luminosity CVs with $L_X=10^{29}$ erg s$^{-1}$, predicted by binary population synthesis, and determine their local CV space density with 10% accuracy or better. This number will have profound impact on theoretical models of close binary evolution (strength of angular momentum loss) and on CV birth rates. And finally we will (c) measure the galactic scale height and luminosity functions of both magnetic and non-magnetic CVs. The population parameters will be determined in a local volume to synthesize the galactic ridge X-ray emission with high fidelity. We will finally solve the decades-long debate about the true nature and composition of the GRXE by fully characterizing the local sample (radius 1 to 2 kpc) and extrapolating into the central regions of the Milky Way.

5.3.2 X-ray binaries in the Milky Way

Over 300 Galactic XRBs containing NSs are known (114 HMXBs and 187 LMXBs, over 100 of them are X-ray pulsars; Liu et al. 2006, 2007), and yet this may be just the tip of an iceberg. They exhibit amazing diversity of properties, often variable on timescales ranging from ms to years. Many of XRBs are heavily absorbed, which makes it difficult to detect and identify them in existing soft X-ray surveys. As a matter of fact, the Galactic XRB luminosity function in the 2-10 keV energy range is poorly constrained for fluxes below $10^{-10}$ erg/s, while the majority of sources are expected to be fainter (Grimm et al, 2002). On the other hand, many objects are dim/soft enough to escape existing hard X-ray surveys. *eROSITA* is expected to improve sensitivity in the 2-10 keV energy range by a factor of $\sim10^3$, and to exceed by far the sensitivity of existing hard X-ray all sky surveys (about $4\times10^{-12}$ erg/s for *INTEGRAL*, Krivonos et al. 2010).

From the observed luminosity function and current population synthesis studies one might expect about 3,000 new XRBs (mostly LMXBs) to emerge from the *eROSITA* survey (Grimm et al. 2002; Belczynski et al. 2004). And in fact this population already started to emerge in pointed observations (Muno et al. 2006). *eROSITA* will provide a complete census of the Galactic population of the accretion powered X-ray binaries down to luminosities of about $10^{33...34}$ erg/s. Many fainter sources as well as new transient sources will be discovered not only in the Galaxy but also in the LMC and SMC. The *eROSITA* survey will provide a robust estimate of the galactic XRB X-ray luminosity function and provide constrains for population synthesis models. This will help to understand the origin and evolution of X-ray binaries in the Galaxy.

### 5.3.3 Isolated Neutron Stars

The observed population of neutron stars is dominated by radio pulsars. In recent years, however, different subclasses of isolated neutron stars (INSs), characterized by peculiar properties and not yet understood physics, have been discovered: magnetars (AXPs, SGRs), thermally emitting INSs (a.k.a. the "Magnificent Seven", or XDINS), central compact objects in supernova remnants and rotating radio transients (RRATs). While currently few of them are known, they might represent a considerable fraction of the neutron stars in the Galaxy. The few sources known have already had a deep impact on our understanding of the physics of matter at extreme conditions of gravity and magnetic field.

Understanding evolutionary relations between different subgroups and establishing a comprehensive picture of neutron stars birth and evolution in the Milky Way requires larger sample than known today. Highly sensitive optical surveys (such as SDSS, VISTA, etc.) coupled to the >20-fold increase of the *eROSITA* sensitivity with respect to *ROSAT*, will enable the mission to discover and identify many new INS. In particular, XDINS constitute a homogeneous group of seven nearby, cooling, middle-aged INSs discovered by *ROSAT*, which display unique properties. Their proximity and the combination of strong thermal radiation and absence of significant magnetospheric activity make them ideal targets for testing NS atmosphere models, deriving radii and constraining the equation of state of neutron star interior. It is remarkable that a group of very similar sources, displaying at the same time unique properties that are so different from ordinary radio pulsars, are all detected in the very local Solar vicinity. Is this fact an anomaly caused by the Sun's current location near regions of active star formation of the Gould Belt or is it really signaling that radio surveys do miss a large population of INSs, at least as large as that of standard radio pulsars? Answering these questions will be possible with the unprecedented survey efficiency of *eROSITA*. The eRASS will allow the detection of an estimated number of 60 to 100 new X-ray thermally emitting INSs, thus increasing the population by an order of magnitude. A major statistical and observational challenge will be the identification of the new XDINS in suitably tailored optical follow-up programs. Before *ROSAT*, predictions were made to discover thousands of isolated neutron stars reheated by accretion from the interstellar medium (e.g. Treves & Colpi 1991; Blaes & Madau 1993) but none were found. The much larger sensitivity of *eROSITA* will shed new light on the Bondi accretion efficiency and the NS velocity distribution.

### 5.3.4 Unique science opportunities with eROSITA surveys

The CV surveys will definitely uncover rare and/or interesting objects of great importance for astronomy and fundamental physics. Examples are the hard X-ray emitting Symbiotic Binaries, the Double Degenerates and the Galactic SuperSoft X-ray sources, all of them being regarded as SNIa progenitor candidates. Together with them, we describe here a number of interesting classes of objects for which we expect significant progresses to be made thanks to the exploitation of eRASS data. These includes isolated white dwarfs and black holes, as well as quiescent LMXB.

- *Double Degenerates* (UCBs and AM CVn stars) exist in ultra-compact configurations with orbital periods ranging from 65 min down to 5.4 min (see Solheim 2010 for a recent review). They are important laboratories for binary stellar evolution theory, in particular to elucidate the elusive common-envelope phase, and can potentially produce rare, sub-luminous, SN Ia-like explosions (e.g. Bildsten et al. 2007, Nelemans et al. 2001, Podsiadlowski et al. 2003). They are the strongest known sources predicted to emit gravitational-wave below the detection threshold of space-based gravitational wave detectors (e.g *LISA* or *NGO*; see Nelemans et al. 2004, Stroeer et al. 2005, Roelofs et al. 2007). Although $10^4$ and $10^5$ systems are predicted to be in the Milky Way, only 27 have been discovered so far. Of those, nine have been detected as soft X-ray emitters with unabsorbed fluxes of $\sim 10^{-13..-14}$ erg cm$^{-2}$ s$^{-1}$ corresponding to luminosities of $\sim 10^{30..33}$ erg s$^{-1}$. In this case, and assuming the recent estimated density of about $(1-3) \times 10^{-6}$ pc$^{-3}$, *eROSITA* is expected to uncover roughly 1,300 Double Degenerates.

- *Galactic Novae and Super Soft Sources (SSS)*: The outbursts of classical novae are caused by the explosive hydrogen burning on the WDs hosted in CVs. After sufficient H-rich material is transferred to the WD, ignition in degenerate conditions takes place in the accreted envelope and a thermonuclear runaway is initiated. As a consequence, the envelope expands and causes the brightness of the star to increase to maximum luminosities up to $10^5$ L$_\odot$. A fraction of the envelope is ejected, while a part of it remains in steady nuclear burning on the WD surface. This powers a bright supersoft X-ray source (SSS) radiating at about the Eddington limit which can be observed as soon as the expanding ejected envelope becomes optically thin to soft X-rays (Gallagher & Starrfield 1978). The duration of the SSS phase is inversely related to the WD mass, while the turn-on of the SSS is determined by the mass ejected in the outburst. The SSS phase can last from less than a month to more than ten years (Pietsch et al. 2007). In addition, shocks may form in the ejecta giving rise to a lower luminosity hard X-ray source (V382 Vel, Orio et al. 2001). Henze et al. (2011) derived correlations between several optical, X-ray and physical parameters of a sample of novae in M 31. In the Galaxy, to date about eight novae are detected every year (Pietsch 2010). Most of these novae have been followed up in X-rays with the help of monitoring campaigns with the *Swift* satellite (Ness et al. 2007) and with *Chandra* and *XMM-Newton* observations. They showed a diversity of time variability patterns and partly complicated variable spectra with emission and absorption lines.

Indeed, the small population of currently known SSS is very inhomogeneous and heavily biased towards unabsorbed, high luminosity sources, which do not undergo temporal variations. Assuming that the known sources are representative of the entire population, it appears that the initial theoretical expectations of ~1,000 sources per galaxy (Di Stefano & Rappaport 1994) are highly overestimated, a finding which is partially confirmed by the deficiency of integrated soft X-ray flux from several elliptical galaxies (Gilfanov & Bogdan 2010). There is however the possibility that a large fraction of SSS is predominantly in a state where the high energy flux is shifted into the unobservable UV (as observed in the transient sources CAL 83 or RXJ0513.9-695), or hidden by absorbing interstellar material. *eROSITA* has the great potential of uncovering such hidden population in our galaxy by detecting X-rays not from the shell burning white dwarf, but from the accretion process itself.

During its four year all sky survey, *eROSITA* will allow us to get a homogeneous census of the X-ray behavior of all Galactic (and Magellanic Cloud see below) novae with snapshots every half a year. While we may miss novae with a shorter SSS phase, we will efficiently follow the development of novae with long SSS phases. We will be able to investigate X-ray spectra and light curves and also trigger follow-up X-ray monitoring campaigns with shorter observation intervals and/or observations with X-ray instruments providing higher spectral resolution. During the survey we also may detect SSS from novae where the outburst has been missed in optical observations. This was the case for XMMU J115113.3-623730, which was detected during an *XMM-Newton* slew maneuver and later identified as a nova (see Greiner et al. 2010). Detailed modeling will allow us to constrain the number of novae showing a SSS phase and to extend correlations between nova parameters and compare them to those derived for M 31.

- *Isolated White Dwarfs* are thermal soft X-ray emitters if they are sufficiently hot, i.e. with effective temperature in excess of $T_{eff}$~20,000 K. Since their X-ray spectra can be modeled to a great levl of detail (Werner et al. 2004) they are invaluable for the calibration of spectrally resolving X-ray instruments. The hydrogen-rich DA-type white dwarfs HZ 43 A ($T_{eff}$ = 51,100 K) and Sirius B ($T_{eff}$ = 24,900 K) in particular were used to establish soft X-ray standards allowing inter-calibration between X-ray observatories (Beuermann et al. 2006) and will be used as ideal calibration targets for *eROSITA* as well.

- Stellar-mass black holes have so far only been identified in binary systems. However, stellar-population-synthesis models and chemical enrichment models indicate that also a large number of *isolated stellar-mass black holes* (~hundreds of Millions to a few Billions) should reside in the Galaxy. These sources could in principle be detected through their X-ray emission arising from Bondi-Hoyle accretion from the interstellar medium. Rate predictions for *eROSITA* depend on a large number of uncertain input parameters. While black holes should be less common than isolated neutron stars, their higher masses and on average lower space velocities should lead to ~$10^3$ times higher Bondi-Hoyle accretion rates. However, due to the lack of a hard surface, the accretion on a black hole should have significantly less efficiency than accretion on a neutron star. Depending on the exact value of the efficiency (could be between $10^{-2}$ to $10^{-4}$, detections of isolated black holes may be more or less frequent than detections of isolated neutron stars (Algol & Kamionkowski 2002). Most isolated black holes should be persistent accretors, and predominantly located in high-density regions such as molecular clouds. This complicates the distinction from e.g., X-ray emitting hot corona of massive stars. *eROSITA*'s spectral coverage and time capability will be important to identify these systems.

- *Transient LMXBs in quiescence: eROSITA* will monitor state transitions in known LMXBs and observe many thermonuclear bursts with high time resolution (several per each of the almost 50 known bursters). These observations, as discussed by Poutanen et al. (2010), will most likely provide new constraints on the equation of state of neutron stars. New high-quality data taken with *RXTE*, *INTEGRAL*, and *Suzaku* on a number of transient and persistent X-ray pulsars allowed a detailed study of their spectral properties as a function of luminosity, i.e. of the mass accretion rate. The data reveal the existence of at least two different modes of spectral-flux dependencies, which are most probably due to different accretion regimes realized in different sources depending on the averaged accretion rate (Klochkov et al. 2011). The emerging diversity of the accretion modes, which is of key importance for understanding of the physics and configuration of the accretion flow close to the neutron star surface is currently studied on a relatively small sample of X-ray pulsars. The monitoring capability of *eROSITA* will allow to extend this analysis to a larger sample of sources by measuring the slope of the X-ray continuum at different luminosity states of transient sources. The Liu et al. (2006) catalog of XRB contains 28 transient accreting pulsars which will be included in the sample using *eROSITA* data.

### 5.3.5 X-ray sources in nearby galaxies

The study of X-ray source populations and diffuse X-ray emission in nearby galaxies is important in understanding the X-ray output of more distant galaxies as well as learning about processes that occur on interstellar scales within our own Galaxy. Because we live within the Galaxy, our view is badly compromised by obscuring clouds of dust and gas, a lack of reliable distance indicators, and a confusing welter of overlapping features along each line of sight. To observe and ultimately understand stellar birth and death, the recycling of matter from interstellar gas to stars and back into space we must look beyond the Milky Way (MW). Existing X-ray surveys of Local Group galaxies indicate very different compositions of their X-ray source populations. The large numbers of supernova remnants (SNRs) and HMXBs in the Magellanic Clouds indicate high star formation rates over the last tens of millions of years, while much older populations (low mass X-ray binaries and classical novae) dominate in M31. Due to the vicinity of the Magellanic Clouds to the southern ecliptic pole the eRASS will in particular cover the Large Magellanic Cloud with high exposure. The higher sensitivity, broader energy band pass and the better energy resolution of the *eROSITA* survey compared to *ROSAT* will allow us to detect external galaxies to larger distances and investigate them in much greater detail.

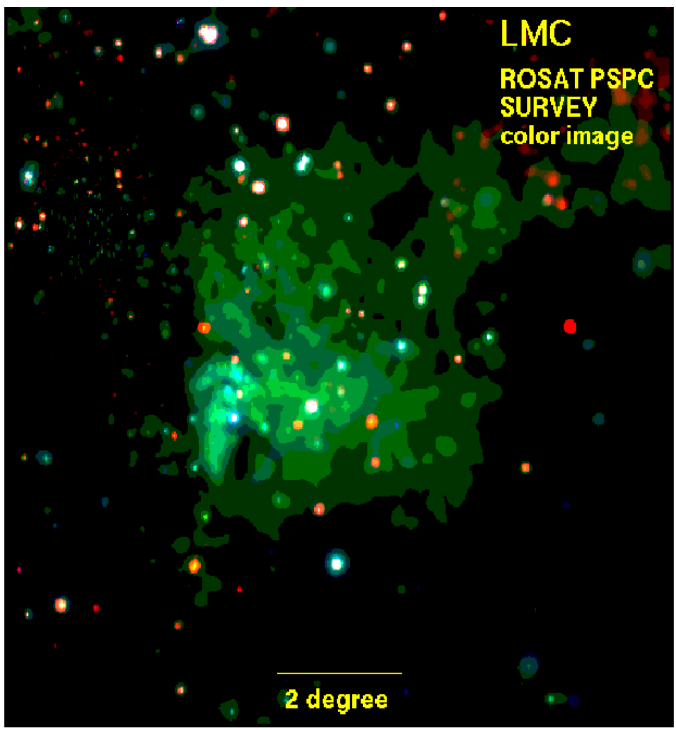

*Figure 5.3.2: ROSAT all-sky survey image of the LMC in the 0.4 – 2.4 keV energy band. Besides the hard point sources structured diffuse emission is clearly visible.*

A.**The Magellanic Clouds**: The first soft X-ray survey of the LMC with imaging instruments was performed with the Einstein observatory (Long et al. 1981; Wang et al. 1991). More than 100 discrete X-ray sources were detected and in addition large-scale diffuse emission originating from hot gas with temperatures of several Million K was revealed (see Figure 5.3.2). About 50% of the point-like sources were identified with objects in the LMC, while the remainder was assigned to Galactic foreground stars and background AGN. Due to the vicinity to the South ecliptic pole the LMC was observed with high sensitivity by the *ROSAT* PSPC during the *ROSAT* all-sky survey. Pietsch & Kahabka (1993) analyzed an area of 13 degrees by 13 degrees centered on the LMC and found more than 500 X-ray sources. Due to the large area, a complete coverage of the LMC was so far only achieved as part of the *ROSAT* all-sky survey. The *eROSITA* survey will be the only opportunity to repeat this with much higher sensitivity, wider energy band and multiple coverage within the 4 year survey.

The deep and homogeneous exposure across the LMC is of particular importance for the investigation of the large-scale diffuse emission from the hot interstellar medium (Sasaki et al. 2002, see section 5.5.3). The complete coverage of the LMC will not be possible with any other current or planned X-ray mission (the 2 Ms very large *XMM-Newton* project of the LMC survey with a mosaic of pointed observations covers with ~10 square degrees only less than ~15% of the whole galaxy). On the other hand, the vicinity to the south ecliptic pole also provides a unique coverage of the LMC sources. The overlapping scans will allow to monitor variable sources like X-ray binaries for up to ~20 (12) days in the northern (southern) part of the LMC, repeated during every half year survey. This will provide important information about the duty cycles of X-ray transients like Be/X-ray binaries in the LMC, as demonstrated by the 40 day light curve of A0538-66 during the *ROSAT* all-sky survey (Mavromatakis & Haberl 1993).

The smaller size of the SMC allowed to cover it with imaging instruments by a mosaic of multiple pointings (Einstein Observatory, Wang & Wu 1992; *ASCA*, Yokogawa et al. 2003; *ROSAT*, Kahabka et al. 1999; *XMM-Newton*, Haberl & Pietsch 2008; Haberl et al. 2012). The large field of view of the *ROSAT* PSPC provided a comprehensive catalogue of discrete X-ray sources and revealed the existence of a hot thin plasma in the interstellar medium of the SMC with temperatures between $10^6$ and $10^7$ K (Sasaki et al. 2002). From the *XMM-Newton* EPIC observations a catalogue of more than 3,000 sources from an area of 5.5 square degrees was compiled (Sturm et al. 2012).

From the X-ray surveys of the SMC it has emerged that there is a substantial population of HMXBs in this irregular galaxy, comparable in number to the Galactic population (e.g. Haberl et al. 2008 and references therein). Unlike in the Milky Way, all except one (the supergiant system SMC X-1) of the HMXBs in the SMC are Be/X-ray binaries. In these systems a compact object, in most cases a neutron star, revolves around a Be star in a generally wide and eccentric orbit. Many Be/X-ray binaries show strong outbursts, associated with enhanced accretion from the Be circumstellar disc.

The strong variability and transient behaviour of most of the SSSs and HMXBs easily will allow us to discover new such sources in the *eROSITA* survey data of the Magellanic Clouds. HMXBs in outburst can reach X-ray luminosities of $10^{37}$ erg s$^{-1}$ or more, resulting in *eROSITA* count rates of ~3 counts s$^{-1}$ yielding a spectrum with 600 counts within 200 s of exposure, sufficient to differentiate HMXBs from AGN. The *eROSITA* survey will not only cover the LMC and SMC, but the whole Magellanic system. This includes the Magellanic Bridge, which is thought to be the product of tidal interaction between the two Clouds. It contains both gas and a young stellar population, which has likely formed in situ. From *ROSAT* observations one HMXB has been found in the western part of the Bridge (Kahabka & Hilker 2005) and *INTEGRAL* observations of the SMC region have resulted in the serendipitous detection of two further hard X-ray sources in the Magellanic Bridge which were identified as HMXBs (McBride et al. 2010). The *eROSITA* survey will allow for the first time to constrain the population of HMXB in the Magellanic Bridge.

B. **The spiral galaxies M 31 and M 33**: A large number of X-ray sources in M 31 and M 33 shows high variability or is of transient nature (e.g. Stiele et al. 2008). In the *ROSAT* and *XMM-Newton* surveys (mosaic pointings) of M 31 (Supper et al. 1997, 2001; Stiele et al. 2011) and M 33 (Haberl & Pietsch 2001; Pietsch et al. 2004) many super-soft and hard X-ray transients were detected. Interestingly, the majority of the SSSs in M 31 was identified with optical novae (Pietsch et al. 2005, Henze et al. 2010, 2011), which enter a SSS state some time after optical outburst with onset and duration of the SSS state varying strongly from source to source. A source in M 31 or M 33 will be scanned several times during one day, repeated every half year during the *eROSITA* survey. This leads to a sensitivity limit of $2\times10^{-13}$ erg cm$^{-2}$ s$^{-1}$ (0.5-2 keV) for this one-day scanning period (with about 150 s net exposure). For a distance of 800 kpc this corresponds to a luminosity of $1.5\times10^{37}$ erg s$^{-1}$. The expected number of sources detected in a half year survey in M 31 and M 33 are listed in Table 5.3.1. A very small fraction of the classical novae in M 31 with low absorption and relatively high temperatures (above 70 eV) will be detectable in their SSS state during the *eROSITA* survey.

| Limiting flux [$10^{-13}$ erg/s/cm$^2$] | # of sources M31 | # of sources M33 |
|---|---|---|
| >25 | 3 | 1 |
| 16-25 | - | 1 |
| 10-16 | 8 | 2 |
| 6.3-10 | 8 | 1 |
| 4-6.3 | 12 | 2 |
| 2.5-4 | 17 | - |
| 1.6-2.5 | 15 | 2 |

*Table 5.3.1: Expected number of sources detected in M31 and M33 in a single all-sky scan for different flux levels.*

## 5.4 Stars and stellar systems

The largest fraction of galactic X-ray sources that will be detected with *eROSITA* are stars and the eRASS data will allow us to study them with a sample of unprecedented size. Stellar X-ray emission is found in virtually all regions of the Hertzsprung-Russell Diagram (HRD), as shown in Figure 5.4.1. However, different X-ray generating mechanisms operate in the various types of stars and X-ray luminosities span an extremely large range, going from $L_X \approx 10^{26}$ erg s$^{-1}$ for very low mass stars (late M dwarfs) up to $L_X \approx 10^{33}$ erg s$^{-1}$ for massive O-type stars. Cool late-type stars generate X-rays from magnetic activity that is powered by a stellar dynamo; in these stars the energy output in X-rays strongly declines (by roughly four orders of magnitude) with increasing age, i.e. Log $L_X/L_{bol} \approx -3 \ldots -7$. In contrast, the X-ray emission of hot, massive early-type stars is generated in wind-shocks and a fairly constant fraction of the total luminosity is emitted in X-rays, i.e. Log $L_X/L_{bol} \approx -7$, independent of age. For an overview on stellar X-ray properties see e.g. the review articles by Güdel (2004) and Güdel & Naze (2009).

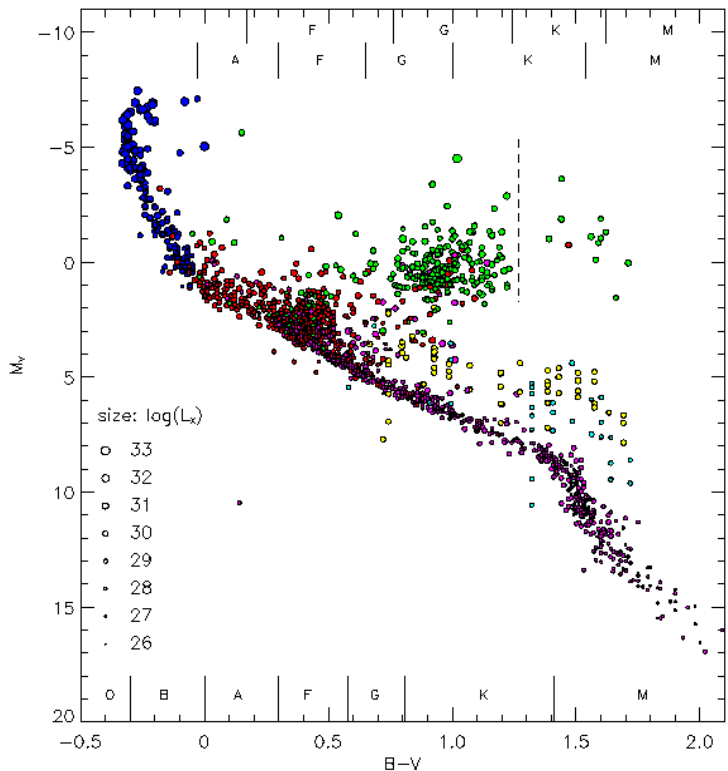

*Figure 5.4.1: The X-ray HRD compiled from the ROSAT survey and pointing data (Güdel 2004). Spectral types for main-sequence stars are indicated in the bottom of the figure, at the top those for giants and supergiants.*

The stellar X-ray sky will be dominated numerically by magnetically active low mass stars and stellar systems. Thus, most of the *eROSITA* stars will be coronal sources, i.e. late-type stars with spectral types F-M. Since these stars are most active in their youth, X-ray surveys are especially well suited to detect and characterize young stellar populations. Further, the energy fraction emitted in X-rays by active stars is a function of rotation, hence age. Thus, for a given range of activity level (or, roughly speaking, stellar age), the X-ray sensitivity limit corresponds to a stellar mass limit, down to which X-ray surveys are complete at a given distance. Relatively young stars can be found everywhere in the extended solar neighborhood, but the youngest ones, with ages of ≲10 Myr are strongly concentrated among pre-main sequence stars in star forming regions (SFRs).

The *eROSITA* sensitivity to stellar sources depends on the emitted stellar spectrum and the absorption along the line of sight. For most stars, however, the dependence is only moderate, apart from highly absorbed objects in SFRs. The majority of the stellar X-ray sources detected in eRASS will be active stars; by using them as a reference template, we can derive an eRASS sensitivity limit for stars of about $1\times10^{-14}$ erg s$^{-1}$ cm$^{-2}$ (cf. section 4.3 above). This flux limit translates to an X-ray horizon for detecting stars of approximately $L_{X,min} \approx 1\times10^{24}\ d^2$(pc) erg s$^{-1}$ with $d$(pc) being the respective stellar distance (in parsecs). These numbers are valid for a typical sky-exposure at equatorial to mid-latitude regions and the complete all-sky survey (eRASS:8). Overall, the eRASS will be at least about 20 times more sensitive than the RASS (*ROSAT* all-sky survey) for typical X-ray stars, i.e. moderately to highly active coronal sources. The eRASS will do even better for hard or absorbed sources and thus it especially improves the sensitivity for very active, more distant or embedded stars; in contrast, a more moderate advancement is achieved for very soft sources like inactive stars.

Using galactic population models combined with X-ray luminosity distributions (Besançon X count model, Guillout et al. 2006), about 0.3 - 0.5 Million stars are expected to be detected with the eRASS. This translates in average stellar surface densities reaching from ~30 per deg$^2$ in the galactic plane down to ~5 per deg$^2$ at the galactic poles.

To illustrate the power of the eRASS in detecting stars, a comparison of the *eROSITA* survey sensitivity with stellar sources detected by *ROSAT* and in several *Chandra* deep fields is shown in Figure 5.4.2. For example a young active solar analog like EK Dra with $L_X \sim 10^{30}$ erg/s will be detected out to distances of 1 kpc in the eRASS, whereas our Sun, a rather inactive star with an average $L_X \sim 10^{27}$ erg/s, will be detected up to distances of about 30 pc.

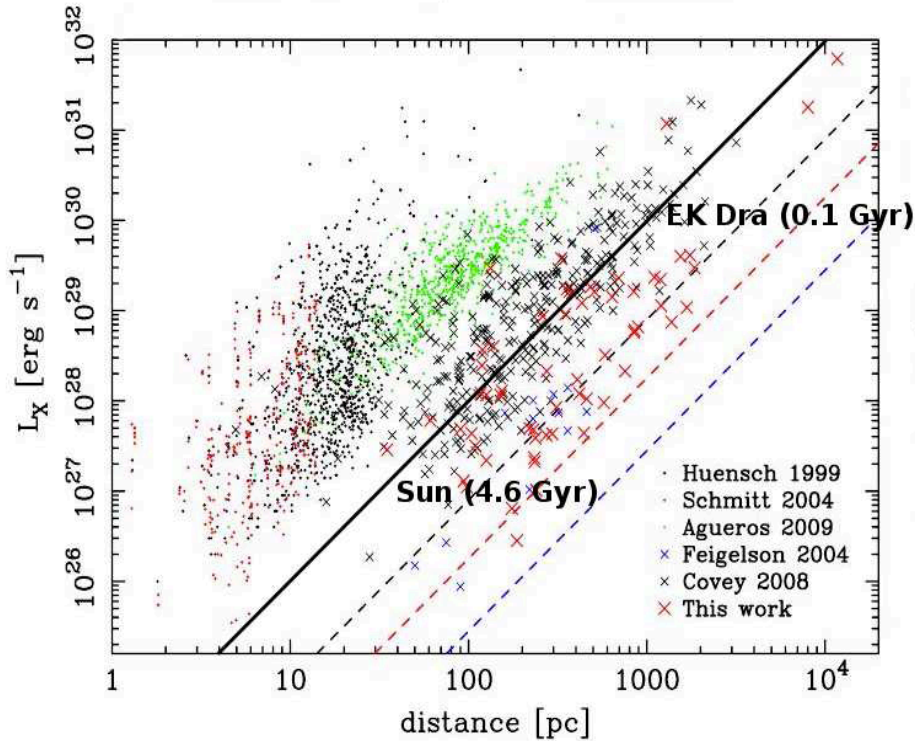

*Figure 5.4.2: eROSITA all-sky survey sensitivity with two Suns (adapted from Wright et al. 2010). eROSITA sensitivity (black line) vs. stellar sources detected by ROSAT: survey + pointings (dots) and in Chandra deep imaging surveys: CDF-N, ChaMP and COSMOS (crosses).*

Compared to the *ROSAT* all-sky survey, the eRASS will not only detect more stars and more photons per star, but also provide more accurate spatial and spectral information on the individual targets. Further, the whole sky is surveyed eight times over four years of observations and each source is observed several times per individual all-sky scan, allowing to study variability on very different timescales (see Section 5.7 below).

Undoubtedly, a large number of astrophysical questions can be addressed with the eRASS data in a more sophisticated manner than allowed by previous studies. The main goals of the *eROSITA* survey can be summarized as follows:

- Determination of **X-ray properties of diverse stellar populations** by obtaining volume complete samples in the solar neighborhood and flux limited sample in the local milky way;
- Identification and study of **young stellar objects** in star forming regions and moving groups that trace recent and ongoing stars formation;
- Study of X-ray emission and its evolution in relation to other stellar parameters (**mass, rotation** etc.);
- Identification and characterization of X-ray emission from **rare stellar objects**;
- Investigation of X-ray generating mechanisms and **time-variable phenomena**.

## 5.4.1 Population studies

The first large field of eRASS science are population studies and local star formation. Since very large regions have to be observed (the study of the complete solar neighborhood obviously requires full sky coverage) these scientific topics are ideal and often exclusive science grounds for an all-sky survey. The study of stellar populations at various ages requires, in its most basic form, only X-ray detections and luminosities, but spectral data will complement for the brighter targets.

A. **Solar neighborhood**: This is of course the best region in the all-sky survey for the study of X-ray faint sources, that will only be detected if located in the vicinity of the Sun. The eRASS is expected to detect virtually all stars of the GJ catalog (~4,000 stars within 25 pc), allowing to perform a complete census of the X-ray properties of nearby stars. The solar neighborhood contains stars with a diverse mix of stellar ages, thus this sample does not only include the more active field population, but also weakly active stars or intrinsically X-ray faint objects, such as very low-mass stars. For these objects the eRASS will provide an unprecedented and unbiased sample in each case. Complete, i.e. volume-, rather than flux-limited, samples are required to determine the statistics of X-ray properties of various classes of stars, and *eROSITA* opens up a much larger space volume for all these studies

B. **Star forming regions:** The eRASS will provide an extensive coverage of young T Tauri and HAeBe stars in all nearby star forming regions (SFRs) (at a distance of about ≲150 pc) like η Cha/Chamaeleon, Taurus-Aurigae, ϱ Oph, Lupus, Sco-Cen or CrA. These low mass SFRs contain a rich population of young (age ≲10 Myr) pre main-sequence stars at low and intermediate masses. Since many of these regions are quite extended (up to 100 deg²), an all-sky survey is optimal to study young stellar populations at different ages and born in different environments. Naturally, also the even wider spread-out, dispersed component will be spatially covered and eRASS is ideal to perform membership studies, since X-ray activity strongly decreases with age. With the cores of many star forming regions, where source confusion and absorption should become a severe issue for *eROSITA*, already covered by deep X-ray observations (*Chandra, XMM-Newton*), the eRASS data provide a valuable and complementary dataset for more complete population studies.

C. **Associations, moving groups & open clusters:** Associations and movings groups contain slightly older stars than typical SFRs. With ages of 10 Myr to 100 Myr these groups, or streams, of young stars, like TW Hya association, β Pic moving group, AB Dor moving group, Tucanae or Pleiades, contain a rich sample of older T Tauri and zero age main-sequence stars. They allow for example to track recent star-formation in the local galaxy including the already more dispersed stars from the above mentioned star forming regions, the so-called Gould-Belt. Also young open clusters belong to this age range, whereas older open clusters are more dispersed and contain stellar populations with ages in the range between 100 Myr and 1 Gyr. Albeit galactic mixing becomes important for these objects, clusters like the Hyades, Praesepe, UMa or IC 4665 do contain a large population of almost coeval stars and allow to study the time evolution of stellar activity up to a maximum age of about 1 Gyr. Further, the IMF (initial mass function) of these groups can be studied in greater detail and scenario of multiple populations will be tested, due to a comprehensive coverage of their population(s).

D. **Massive SFRs, OB stars & associations:** In contrast to the low mass SFRs mentioned above, massive SFRs also harbor O and early B-type stars. These hot stars will probably be the most distant stellar sources detected by eRASS, given an X-ray horizon of a few tens of kpc for the X-ray brightest ones. While the closest of these region, the Orion nebula cloud, contains only a few massive stars, a much larger number of OB stars will be detected in more distant regions. Source confusion is again an issue, but more isolated or dominant X-ray sources should be identifiable in the eRASS data. Beside the study of global X-ray properties of stellar clusters, new insights will be gained from the increased number of X-ray detected hot, massive OB stars. Examples are the X-ray luminosity function ($L_X/L_{bol}$) along the hot star sequence including binary OB stars and Wolf-Rayet stars, their binary fraction, etc.

E. **Peculiar and evolved stars, substellar objects:** The eRASS will also contain an unprecedented number of rare stellar objects, such as peculiar Ap/Bp stars, young brown dwarfs or evolved stellar objects like giants of various spectral types, AGB stars, planetary nebulae, etc. The significantly increased statistics opens up the possibility to study properties of a wide range of very different astrophysical objects. Emission processes, formation scenarios and evolution of these object belong to the large variety of astrophysical questions that can be addressed with X-ray observations.

## 5.4.2 Spectral studies

Combining the high sensitivity of the eRASS with the superb spectral resolution of *eROSITA*, spectroscopic studies will become feasible for many thousand stars. Spectral analysis, beyond hardness ratio determinations, starts with sources at roughly ten times the detection flux level with quality becoming increasingly better for the successively brighter sources. In addition to the RASS energy range, the spectral coverage of *eROSITA* reaches well beyond 2.0 keV and thus allows us to study harder X-ray emission in an all-sky survey for the first time. This spectral window is also very important for stellar sources, in particular when studying hot plasma, fluorescence emission (e.g. Fe K$\alpha$) or embedded young stellar sources. An example of a simulated spectrum of a typical nearby (5 pc) active M dwarfs (AD Leo) seen by *eROSITA* in a 2.5 ks exposure is shown in Figure 5.4.3.

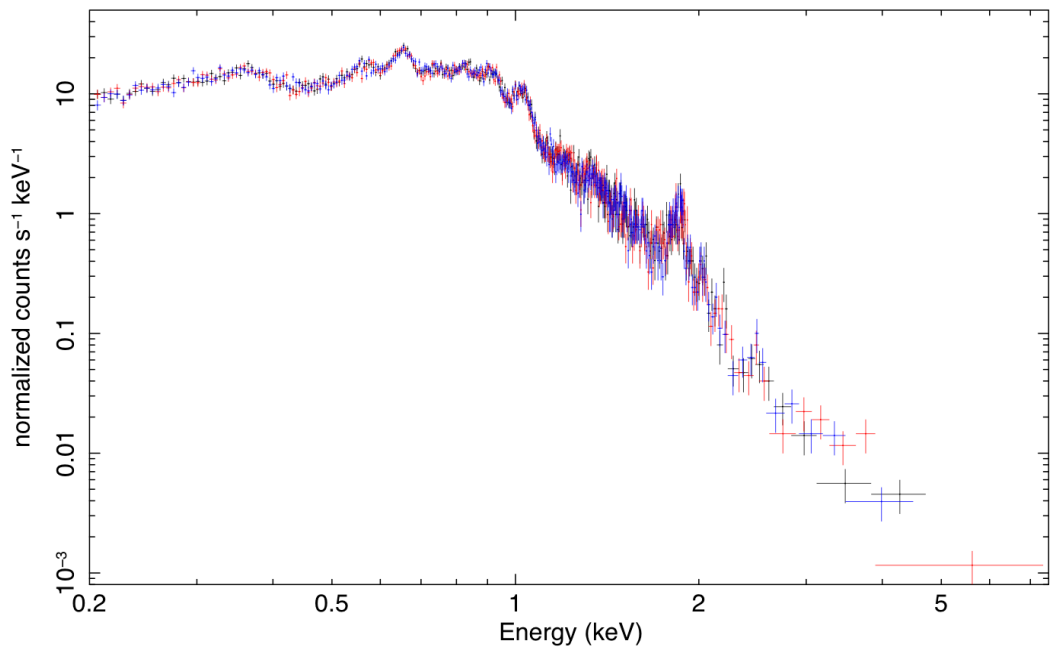

*Figure 5.4.3: The eROSITA all-sky survey will also provide spectra and light curves for many thousand nearby stars, enabling the detailed study of magnetic activity and stellar coronae. Shown is the simulated spectrum of a nearby active M dwarf at a distance of 5 pc with an exposure time of 2.5 ks.*

Spectral analysis will reveal basic X-ray properties, such as coronal temperatures, for many thousand stars. For the brighter, or better exposed, targets this opens up the opportunity to perform time resolved spectroscopy, study abundances or even strong individual emission lines/complexes. Many physical question can be addressed with the spectroscopic information, for example the properties of cool excess plasma produced in accretion shocks or abundance anomalies attributed to the accretion of metal depleted gas as observed in young accreting sources like classical T Tauri stars. Another topic is the search for signatures of magnetically confined wind shock emission in early-type stars and possible effects of stellar evolution. Other hot star topics are the investigation of variations in temperatures and luminosity of massive single stars with age, spectral type, etc. Concerning interstellar matter, with the many lines of sight in the direction of bright stellar X-ray sources, X-ray absorption effects as measured in stellar spectra can be used to map the distribution of interstellar material in the local galactic environment.

## 5.4.3 Time variability

Since stars are variable on all timescales, the eRASS data will be of large scientific value for studying variability and transient phenomena. Time windows reachable by eRASS range from seconds (during individual scans) over hours and days (during consecutive scans) up to months and several years (during consecutive all-sky surveys; see Section 5.7 below). On short timescales (seconds to hours) bursts and flares are the main source of variability. Beside its scientific interest as a phenomenon alone, these events can increase the X-ray brightness of the specific object by up to a factor of hundred and will allow to detect X-rays from deeply embedded pre main-sequence stars or more distant very low mass stars. On intermediate timescales (hours to days), long duration flares or rotational modulation can be studied with the eRASS data, especially for stars at higher ecliptic latitudes that are scanned for several days. On timescales of months and years, activity cycles, long duration trends, accretion outbursts are the main phenomena that can be studied with eRASS. Including additional information, especially from the RASS data, opens the time window of decades. X-ray trends have not been explored for a larger number of stellar sources in this time regime, thus the eRASS data allows for the first time to systematically test the long term stability of coronal X-ray emission in various types of stellar sources.

### 5.4.4 Related astrophysics

As outlined above, X-ray data for various stellar population at different ages with good statistics will be obtained with the eRASS. Therefore the eRASS data is well suited to address possible correlations within the stellar sample and evolutionary trends that link activity with age, rotation or other parameters. Importantly, these studies will be possible with many subsets of the stellar data, allowing to study their dependencies on stellar properties like mass, color or effective temperature. Theses studies have also implications for dynamo theory, allowing to investigate saturation and super-saturation effects, the $L_X/L_{bol}$ evolution or transition effects at the fully convective boundary in mid M dwarfs.

The combined data from young nearby stellar population can be used to trace the local star formation history and map the present day galactic structure as seen in its X-ray active stellar component. Additionally, it will allow to address the effects of high-energy radiation on disk chemistry and the early evolution of planetary systems. Further, the X-ray data of individual SFRs give a better handle on their basic properties like total mass or their IMF. As a sample these data allows for test, if a uniform IMF is present or if it might differ between the individual regions with possible dependencies on mass, size etc. This involves an understanding of the individual star formation history and thus the modes of star formation and possible mixing due to mergers.

Another topic to be addressed is pseudo-diffuse emission. Putting the X-ray emission from all stellar populations together, allows to estimate the stellar contribution to the integrated X-ray flux that is observed in many cases as diffuse X-ray emission, for example in nearby galaxies or even in more distant regions of our own galaxy. A proper estimation of this component is a key issue when investigating the true diffuse emission from hot gas, as present e.g. in the vicinity of massive stars or in supernova bubbles.

### 5.4.5 Synergies

The identification of new young stars in the solar neighborhood and of nearby stellar associations enable deepened insight in the local galactic structure as a whole and even more localized in specific star forming regions. A large data base of optical/NIR observations of SFRs and young stellar clusters is available for cross-correlation with the eRASS X-ray catalogues. As a consequence this allows to study stellar evolution in a broader approach and corresponding relations in a large variety of stellar populations at different ages in an unprecedented fashion.

It is expected that *eROSITA* will increase the number of stellar X-ray detections by up to two orders of magnitude; however it will not only detect stars in a larger space volume, but also detect fainter and thus less massive stars, providing an excellent sample to tackle astrophysical questions. Here missions like *GAIA* or ground based existing and forthcoming surveys will add valuable information. *GAIA* will provide distances (parallax), 3D space motions and stellar identifications for virtually all stars that will be detected in the eRASS and photometric periods become available for larger stellar samples from monitoring. Adding moderate resolution spectroscopic information from ground based optical follow-up observations gives access to lithium absorption lines or Hα, which would provide independent information on stellar age or activity level respectively. Synergistic effects are also expected to work the other way; *eROSITA* X-ray detections of interesting stellar objects will likely trigger observations at other wavelength or deeper pointed X-ray observations.

## 5.5 Studies of diffuse X-ray emission: SNRs, superbubbles and the hot ISM

Supernova remnants (SNRs), interstellar bubbles, and superbubbles, bubbles for short, are generated by one or multiple stellar explosions combined with stellar winds and are driven by the expansion of strong shock waves propagating into an inhomogeneous interstellar medium (ISM). The evolution of SNRs and bubbles can best be studied in soft X-ray line and continuum emission, since these plasmas are very hot ($10^6 - 10^7$ K). Line emission of the thermal plasma, which was not resolved with older X-ray telescopes like *ROSAT*, is now evident in the spectra taken with *XMM-Newton, Chandra,* or *Suzaku*. *eROSITA* will for the first time provide us with detailed spatial and spectral information on nearby, highly extended sources, a complete sample of SNRs and interstellar bubbles in our Galaxy or in the Magellanic Clouds, and the hot interstellar medium in general. This will allow us to better understand the physical state and thus also the evolution history of the interstellar plasma.

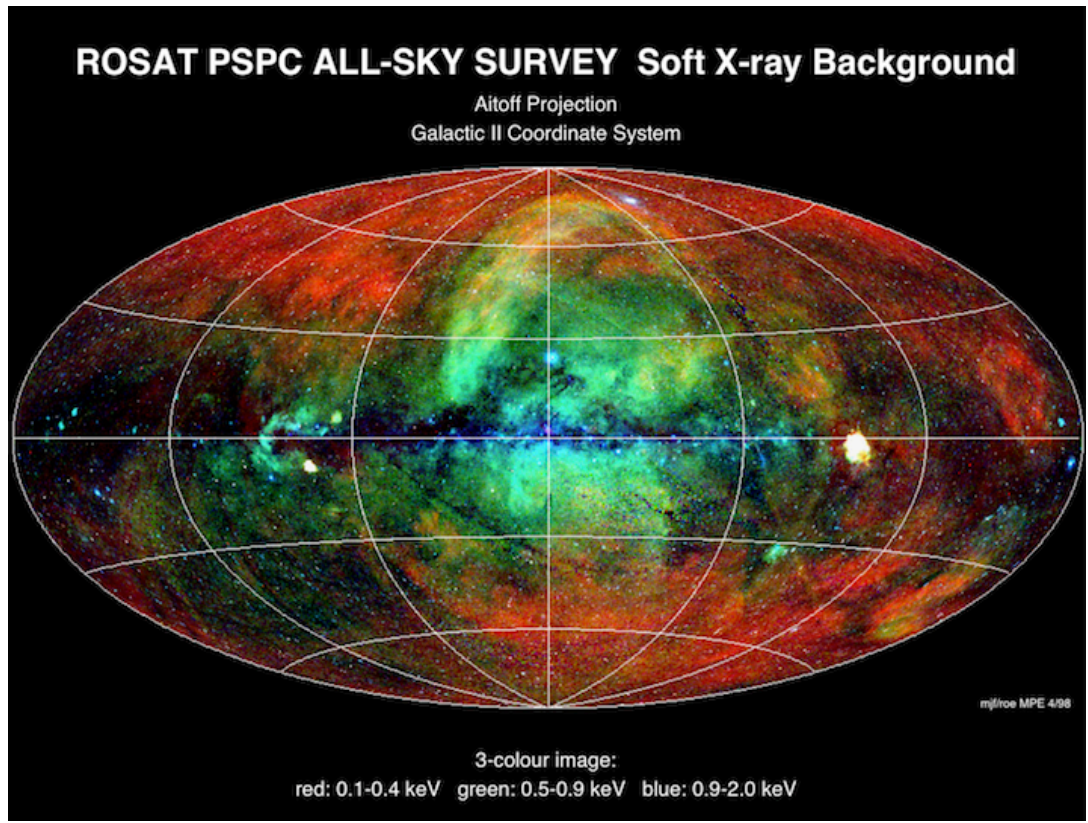

*Figure 5.5.1: 3 colors image of the soft X-ray background detected by ROSAT in the PSPC all-sky survey. Courtesy of M. Freyberg.*

Owing to the coverage of the entire sky and the high sensitivity in the soft X-rays, the key questions and topics that can be addressed with *eROSITA* are:

- What is the **nature of the local ISM**? How much is the contribution of the absorbed emission and how much is that of the Local Hot Bubble? What effect do the solar X-ray background and the solar wind charge exchange have?
- Which are **the nearest SNRs**? What is their age and the physical conditions of their hot plasma? What can the nearby, and thus largely extended SNRs tell us about the interaction of the SNR shocks with the ambient ISM?
- Compilation of a completes sample of **Galactic X-ray SNRs**.
- Global study of the **hot phase** of the interstellar medium in a galaxy, in particular that of the **Large Magellanic Cloud (LMC)**. The LMC is known to undergo active star formation and is thus filled with SNRs, bubbles, superbubbles, and large-scale shells filled with diffuse hot gas.
- General study of the **non-equilibrium ionization (NEI) effects** in the interstellar plasma based on the various initial conditions and ionization history of the plasma at different locations in a galaxy.
- What is the origin of the **Fermi Bubbles**? Did the Galactic center show AGN-like activity in the past?

## 5.5.1 Local Interstellar Medium

Diffuse soft X-ray emission from the Local Interstellar Medium (LISM) has been known since the very beginning of X-ray astronomy. Various models were set up to explain the spatial and spectral distribution of this Soft X-ray Background (SXRB). The observed general anti-correlation of the X-ray intensity with the Galactic neutral hydrogen column density favored an "absorption model", where the emitting regions were located behind the absorbing material. The non-vanishing flux in the Galactic plane could not be explained. "Interspersed" models that mixed the emitting and absorbing material had to clump the absorbing gas by more than allowed from 21 cm measurements. In the "displacement model" the emitting X-ray flux originates in front of the absorbing material in some kind of "Local Hot Bubble" (LHB), where the length of the sight line through the emitting plasma corresponds to the observed X-ray intensity. With the results of the *ROSAT* All-Sky Survey the interpretation became more difficult; detection of X-ray shadows showed that a significant fraction of the SXRB originates beyond the bulk of the absorbing material, and the extent of the local cavity void from neutral hydrogen differed from the extent derived from soft X-ray intensity. The discovery of solar wind charge exchange emission in the vicinity of the Earth (see Section 5.6 below) motivates the question of the "true" zero-level of the SXRB, and about the detailed distance distribution of the components contributing to the observed SXRB.

These possible components (geospheric, heliospheric, at LHB boundary, at Loop I, etc.) can be observed very well with *eROSITA* due to its improved spatial resolution, its significantly enhanced effective area above 0.28 keV, its better spectral resolution especially at low energies, and an orbit at L2 outside Earth's residual atmosphere. The heliosphere shields the Solar System (see next section) from the LISM. At this boundary time-variable effects could be expected and could then be detected with *eROSITA* due to the multiple all-sky coverage. Recent progress in self-consistent modeling of the LISM will also help to interpret the results and to answer the question about the origin and physical state of the LISM.

## 5.5.2 Supernova Remnants

- *Extended Nearby SNRs*: The *Monogem Ring* is a large, old ($t \approx 10^5$ yr, Plucinsky et al. 1996) SNR in our Galaxy located above the Galactic plane close to the anti-center direction. This highly extended object was observed in the HEAO-1 survey (Nousek et al. 1981) and the *ROSAT* All-Sky Survey (Plucinsky et al. 1996). The remnant has a very soft X-ray spectrum with characteristic temperatures of $T = (0.9 - 2.0) \times 10^6$ K. Figure 5.5.2 (left) displays an image of the SNR in the RASS R1+R2 bands (0.100- 0.284 keV, Snowden et al. 1997). The X-ray remnant has a roughly circular morphology with a lower intensity in the center and higher intensity toward the edges.

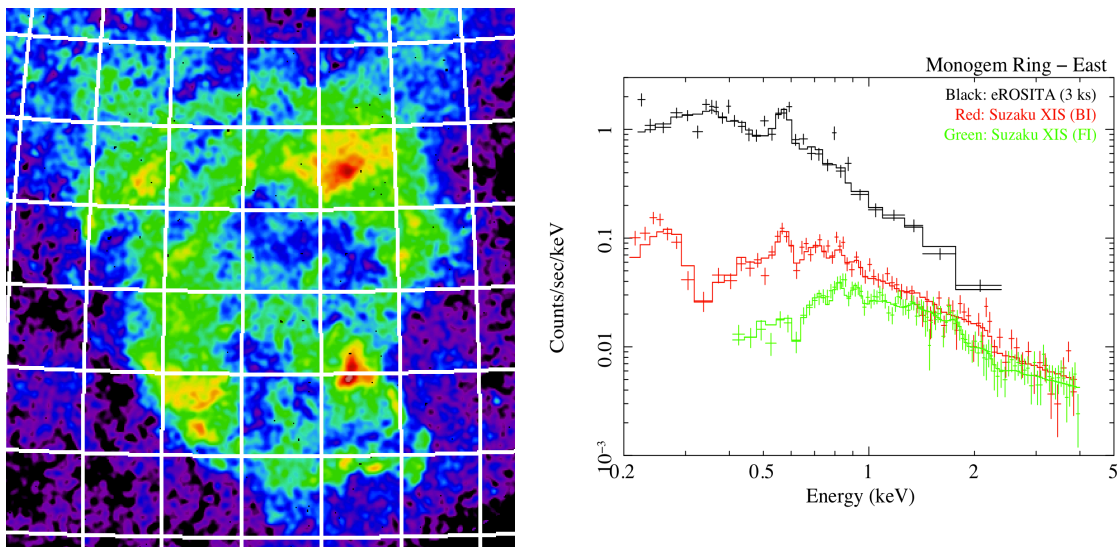

*Figure 5.5.2: Left: ROSAT R1+R2 map of the Monogem Ring in Galactic coordinates. The SNR is about 25 degrees in diameter. Right: Suzaku XIS spectra of the X-ray bright region in the east and best fit CIE model with variable abundances (red: front-illuminated XIS, green: back-illuminated XIS). The simulated eROSITA spectrum assuming the best fit model for the Suzaku spectra and an exposure of 3 ks is shown in black.*

At a distance of $d \sim 290$ pc (Brisken et al. 2003) the Monogem Ring is one of the closest, if not the closest, SNR and has a very low foreground column density $N_H$. It is the oldest SNR observable in soft X-rays to study the evolution of the NEI plasma in the ISM after it has been shocked by an SNR and where mixing of hot and cold gas takes place. The *Suzaku* X-ray Imaging Spectrometer (XIS) spectra of the X-ray bright region in the shell (Plucinski 2009; Figure 5.5.2 right, red and green) indicate that the plasma in the SNR shell is characterized by non-equilibrium ionization with a significantly low ionization timescale of $\tau = n_e t = 1.9 \times 10^{10}$ s cm$^{-3}$ = 600 yr cm$^{-3}$. This number corresponds to a strikingly low density of about $n_e = 0.006$ cm$^{-3}$, assuming the age of the remnant of $t = 10^5$ yr. This observed region is far from the Galactic plane and the SNR should have expanded into a relatively low-density ISM. However the derived $n_e$ is one to two orders of magnitudes lower than what is expected in the Galactic disk. A similar inconsistency between the age of an older remnant (10,000 – 100,000 years) and an apparently too low ionization time scale of the hot plasma of <1000 yr cm$^{-3}$ has also been found for Galactic SNRs as well as SNRs in the Large Magellanic Cloud.

Evolved SNRs like the Monogem Ring or the Cygnus Loop (*Diameter* ≈ 3°, $t = 10^4$ yr, $d = 540$ pc, Blair et al. 2005, see Fig. 5.5.3), present a *unique* opportunity to observe SNRs in the late stage of evolution and its interaction with the ISM, since they are located so nearby and the intervening absorption is low. Figure 5.5.2 also shows a simulated *eROSITA* spectrum of the emission of the Monogem Ring observed with *Suzaku* assuming the best fit model. We assume that the emission fills the entire field of view of the telescope and the exposure time is 3 ks, a typical value for the All-Sky survey. For comparison, the exposure time of the *Suzaku* observation was 54 ks. It is obvious that even with this relatively short exposure time, we will obtain spectra of the Monogem Ring with similarly good statistics as the long *Suzaku* observation. The advantage of *eROSITA* in particular is the high sensitivity in the soft X-ray band of ~0.3 - 2.0 keV, where emission lines of the elements C, N, O, and Ne are expected. Highly ionized species of these elements are tracers of plasma at temperatures of $10^{5-7}$ K and will reveal the ionization structure of the plasma.

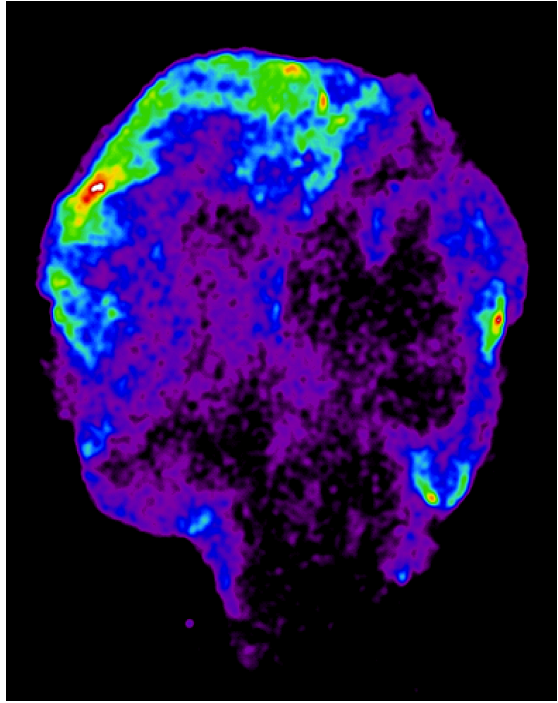

*Figure 5.5.3: Simulated eROSITA all-sky image of the Cygnus Loop SNR. Courtesy C. Schmid.*

- *Complete sample of Galactic X-ray SNRs*: Today, about 275 supernova remnants have been identified in the radio band (Green 2009, and references therein). Although this catalogue represents the result of more than ~50 years of intensive search with the world's largest radio telescopes, it is considered to be strongly biased by selection effects. For a source to be included in the catalogue, it must be: (1) extended, with diameter > 3'; (2) located at low galactic latitude, and (3) have a non-thermal spectrum with spectral index in the centimeter band steeper than -0.3 (Helfand et al. 1989). The Crab, on the other hand, would have a 20 cm flux density of ~3 Jy, a diameter of less than 1', and a spectral index flatter than -0.2 at a distance of ~10 kpc. Similarly distant SNRs can easily fail two of the three standard radio selection criteria, leaving center-filled supernova remnants with pulsar-powered plerions under-represented among radio-identified remnants. Apart from selection effects, other conditions may prevent a remnant from being a strong radio source. Consequently, SNRs have been missed in previous radio searches if:

(i) the supernova remnant shock wave expands within the hot phase of the ISM and reaches a very large diameter before it can sweep up sufficient mass from the low-density gas to have formed a radio shell. Density inhomogeneities in such a large volume will cause distortions in the shell and can make its radio identification as an SNR rather difficult, in particular in the presence of confusing unrelated emission from the same area. The fraction of sources thus undetected depends also on the fraction of the ISM containing the hot component.

(ii) The shock wave expands in a very dense medium. In this case the SNR lifetime is rather short, since material is quickly swept up and decelerated. Such an environment is likely to be relevant, e.g., for OB-associations surrounded by dense molecular and warm gas. Even during their short lifetime, such SNRs are difficult to identify within the strong thermal radio emission from these regions.

(iii) The remnant is old and its radio emission has faded already, thus making it difficult for a faint radio source to be identified by the limited surface-brightness sensitivity of previous sky-surveys (e.g., Reich et al. 1992).

The correlation of *eROSITA* data with radio surveys will allow us further identification studies on more than 140 faint and diffuse X-ray sources, which have been by now classified as SNR candidates in the *ROSAT* all-sky survey (e.g., Prinz & Becker 2012).

## 5.5.3 The hot ISM in the LMC

Bubbles and superbubbles are large structures with extents up to 100 pc or larger and very soft thermal X-ray emission. Due to their sizes and, more crucially, due to the absorption by matter in the Galactic plane, it is difficult to study these objects in our own Galaxy. The Large Magellanic Cloud is an irregular galaxy with indications for spiral structures and one of the closest neighbors of our Galaxy. Its proximity with a distance of 48 kpc and modest extinction in the line of sight (average Galactic foreground $N_H = 0.6 \times 10^{21}$ cm$^{-2}$) make it the ideal laboratory for exploring the large-scale structure of the ISM in a galaxy. In addition, it is known to host a large number of H II regions, bubbles, and superbubbles of various sizes (Henize 1956; Davies et al. 1976). The well-known and best studied extended emission region in the LMC is the giant H II region 30 Doradus and the region south of it, which harbor star formation sites, superbubbles, and SNRs. *ROSAT* data of the superbubbles in the LMC have been studied in detail by, e.g., Chu et al. (1995) and Dunne et al. (2001). *XMM-Newton* observations of superbubbles in the LMC have allowed more detailed studies of the origin, evolution, and the present physical state of the superbubbles, as well as their mass and energy inputs into their environment owing to the improved spatial and spectral resolution (e.g. Sasaki et al. 2011). There is a large number of superbubbles in the LMC out of which only a few have already been observed with XMM-Newton or Chandra. Especially interesting for our studies are more evolved superbubbles like N167 or N204 (Fig. 5.5.4), which have a well defined, almost circular shape in projection and have an extent of ~80 - 150 pc, comparable to, e.g., the Monogem Ring.

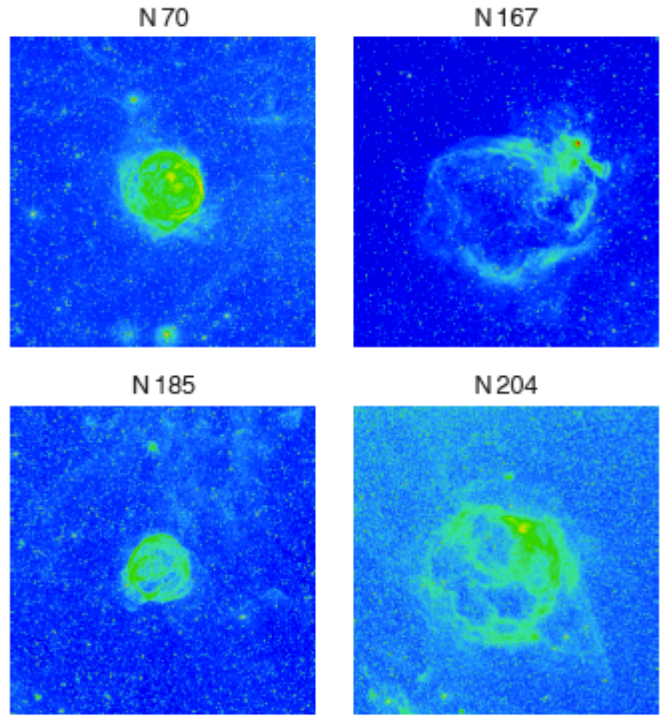

*Figure 5.5.4: MCELS H images of selected bubbles and superbubbles in the LMC. The images have a size of 30' x 30'. The sources in the center of the images are the superbubbles.*

The LMC is located only a few degrees distant from the South Ecliptic Pole. The planned scan geometry of *eROSITA* will achieve a very high exposure at the ecliptic poles while the rest of the sky will have an average exposure of 3 ks. Therefore, the LMC in particular will be observed with exposures from ~10 up to ~140 ks. This will allow us to study the complete sample of SNRs, bubbles, and superbubbles as well as the hot ISM globally in a galaxy. We will obtain images of all the interesting structures observable in soft X-rays, which will give information about, e.g., typical morphologies and sizes of interstellar bubbles, as well as spectra, from which we will obtain the distributions of temperature, density, pressure, abundances, or volume filling factor. In the last decade, new data became available that have revealed the complete distribution of warm and cold interstellar medium in the Magellanic Clouds (e.g., the Magellanic Cloud Emission Line Survey [MCELS], Smith et al. 2005; the Spitzer legacy project "Surveying the Agents of Galaxy Evolution", Meixner et al. 2006). Combining the eRASS data with these tracers of the cooler matter we will study the effects of turbulence on the evolution of the interstellar medium and search for evidence of delayed recombination in plasma that shows significant signs for radiative cooling.

## 5.5.4 Shock physics and non-equilibrium effects in interstellar plasmas

In interstellar plasmas ionization is caused by two-body collisions between electrons and ions, which under the assumption of collisional ionization equilibrium (CIE) is roughly balanced by the rate of radiative recombinations. SNRs and bubbles, in which these hot thin plasmas are produced due to shock waves, are highly dynamical objects, which makes is difficult to reach CIE in the hot ISM. Before CIE can be established in an optically thin, shock-heated plasma, electrons and ions have to establish each a Maxwellian distribution, and subsequently also ions and electrons must come into equilibrium through Coulomb collisions. In systems like SNRs and bubbles, it takes $10^3 - 10^4$ yrs to establish a Maxwellian distribution between electrons and ions, with the ionization structure still being out of equilibrium. If, however, the cooling time of the plasma becomes shorter than the recombination time, the ionization structure will be driven out of equilibrium. In SNR shells, for instance, the outer shock often appears under-ionized, because ionization lags behind. Further downstream, when the plasma becomes thermalized, fast adiabatic expansion can efficiently cool the electrons, and delayed recombination mimics an over-ionized gas (Breitschwerdt & Schmutzler 1994). Recent calculations have shown that such an over-ionized plasma can emit X-rays even at temperatures below $10^5$ K (de Avillez & Breitschwerdt 2010). Moreover, in SNRs and bubbles, which consist of hot gas surrounded by a dense shell of cold material, turbulent motions are especially strong near the shell. Due to this turbulent mixing of the hotter and colder material there are also regions with intermediate temperatures. Different initial conditions in an inhomogeneous ISM (temperature, density, or pressure) and the different history of the plasma can result in totally different plasma conditions and cooling functions even in small regions of the ISM. Therefore, for a better understanding of an evolving plasma in the ISM and a correct interpretation of its emission, the assumption of non-equilibrium ionization is inevitable. A systematic study of the hot ISM with *eROSITA* will provide us with a complete picture of the physics and the evolution of the hot interstellar plasma.

## 5.5.5 The Fermi bubbles

One of the most spectacular results of the *Fermi*-LAT Gamma Ray Observatory is the discovery of two extensive regions of gamma-ray emission above and below the Galactic plane, each 40 - 50 degrees in diameter, and roughly circular in shape (Su et al. 2010). These have been named the "Fermi Bubbles". They are symmetric around the direction towards the Galactic centre, strongly suggesting an origin connected with activity there. The spectrum is hard and the emission rather uniform within the bubbles. The origin is unknown, but plausible explanations include AGN-type activity in the Galactic centre a few million years ago (e.g., Yang et al. 2012), or starburst activity. X-ray emission associated with parts of the Fermi Bubbles is evident in the *ROSAT* all-sky 1.5 keV maps, and this may originate from shocked hot gas on the bubbles' boundaries.

*eROSITA*, with its full-sky coverage and broad energy range, is uniquely suited to probe the origin of these remarkable features. It will also be essential in distinguishing them from other large-scale phenomena like the Local Hot Bubble, which cover similar regions of the sky.

## 5.6 Solar system studies

The solar system is a very rewarding place for X-ray research with *eROSITA*, as it is the location for studying an X-ray emission process which was discovered only 16 years ago: charge exchange between highly charged ions and neutrals. During the recent years, the importance of this process has gained increased general attention due to its potential impact on various fields of astrophysics, because:
- Any X-ray observation made from within the solar system may be affected by charge exchange induced X-rays originating in the heliosphere;
- Charge exchange induced X-rays are expected to occur in all kinds of hot-cold gas interfaces, where they may mimic thermal plasma emission of higher temperature and/or elemental abundance anomalies;
- Comets appear to be better suited than any terrestrial laboratory for deducing specific fundamental parameters of atomic physics by high resolution X-ray spectroscopy, thus improving remote plasma diagnostics in general.

*eROSITA* will be a perfect instrument for studying charge exchange induced X-rays in the solar system. By utilizing its high sensitivity to soft X-rays, its wide field of view, its low soft X-ray background, and its superb spectral capabilities at soft X-ray energies,

- We will be able to investigate the **heavy ion content of the solar wind**, by using comets as natural spaces probes, which convert the energy stored in the solar wind heavy ions into X-rays;
- We will get this unique information over more than half of the solar activity cycle, and **in full 3D**, because the paths of comets are not restricted to the ecliptic plane, where all the currently operational solar wind satellites are located;
- We will investigate the **interaction of the hot solar wind with the cold cometary gas** and will deduce the properties of the cometary bow shock, which would otherwise only be accessible by in-situ measurements;
- We will obtain **clean maps of the diffuse soft X-ray sky**, including emission of the heliosphere, free from geocoronal contamination;
- We will **separate the X-ray emission of the heliosphere** from other sources of diffuse X-ray emission (mainly the Local Hot Bubble) by analyzing its time variability over eight all-sky surveys and by investigating the subtle spectral changes between charge exchange and thermal emission.

### 5.6.1 Historical context

The history of X-ray astronomy is closely related to solar system studies: the first attempts ever to detect X-rays from a celestial object concentrated onto the Sun (Friedman et al. 1951), and the (unsuccessful) attempt to detect X-rays from the Moon, in 1962, is generally considered as the birth of X-ray astronomy (Giacconi et al. 1962). In the recent 16 years, our knowledge about the X-ray properties of the solar system was considerably enhanced. While before 1996 the only solar system objects known to be X-ray sources were the Sun and Jupiter, we know today that our solar system is full of X-ray sources, including all planets from Venus to Saturn, comets, and even the heliosphere itself.

The enormous progress in recognizing the X-ray properties of our immediate astronomical environment has started in 1996 with the discovery of comets as a new class of X-ray sources. This *ROSAT* discovery (Lisse et al. 1996, Dennerl et al. 1997) has revealed the importance of a fundamental process for generating X-rays: charge exchange between highly charged ions and neutrals (Cravens 1997). While this process itself was well known, its capabilities for generating 'standalone' X-ray emission seem to have been completely overlooked before (see Dennerl 2010 for a recent review). The discovery of cometary X-ray emission has also opened a conceptual breakthrough for the understanding of unexplained properties of the soft X-ray background, which could then be attributed to charge exchange reactions in the geocorona and the heliosphere. This has already had a considerable impact on our understanding of the Local Hot Bubble (see Sect. 5.5.1) and may lead to a revision of our view of other astrophysical objects. In the recent years, evidence for charge exchange induced X-ray emission was reported to have also been found outside the solar system, in the *galactic ISM* (North Polar Spur: Lallement 2009, Cygnus Loop: Katsuda et al 2011; Carina Nebula: Townsley et al. 2011a, Giant H II regions: Townsley et al. 2011b), *stars* ($\zeta$ Ori: Pollock 2007), *nearby galaxies* (M31: Liu et al. 2010, M82: Tsuru et al. 2007, Ranalli et al. 2008; Liu et al. 2011, Konami et al. 2011, Liu et al. 2012), and may even be present in *clusters of galaxies* (Lallement 2004; Fabian et al. 2011). The general interest in the topic of charge exchange is reflected in a special issue of the Astronomical Notes of April 2012, which contains 22 additional refereed articles.

## 5.6.2 ROSAT and eROSITA

It was no coincidence that charge exchange induced X-ray emission was discovered with *ROSAT*. In retrospect, this satellite appears to have been specifically tailored to revealing this yet undiscovered fundamental process for the generation of celestial X-rays. The key mission design of *ROSAT* was to perform the first ever all-sky X-ray survey with an imaging telescope. For this purpose, the X-ray telescope was optimized for a large field of view and a large sensitivity at energies below 2 keV. During its all-sky survey, the satellite was continuously scanning the sky along great circles, providing an essentially unlimited field of view. Thus, it was perfectly suited to detect the diffuse soft X-ray emission of comets and the geocorona. In fact, already 28 days after its launch, when *ROSAT* was still in its performance verification phase, the Moon was observed (Schmitt et al. 1991). The image exhibited X-rays which appeared to originate from the dark side of the Moon, which are now understood as the result of charge exchange between highly charged solar wind ions (SWCX) and neutrals in the geocorona. Soon afterwards, just one day after the regular *ROSAT* all-sky survey had started, the first comet happened to get into the field of view, emitting X-rays due to SWCX with cometary neutrals (Dennerl et al. 1997), and during the survey occasional brightenings of the X-ray sky were observed (Snowden et al. 1994), which are now understood as the consequence of temporarily variable SWCX in the geocorona. The fact that SWCX at the geocorona and in comets generates diffuse X-ray emission, mainly below 1 keV, made *ROSAT* the ideal instrument to discover it.

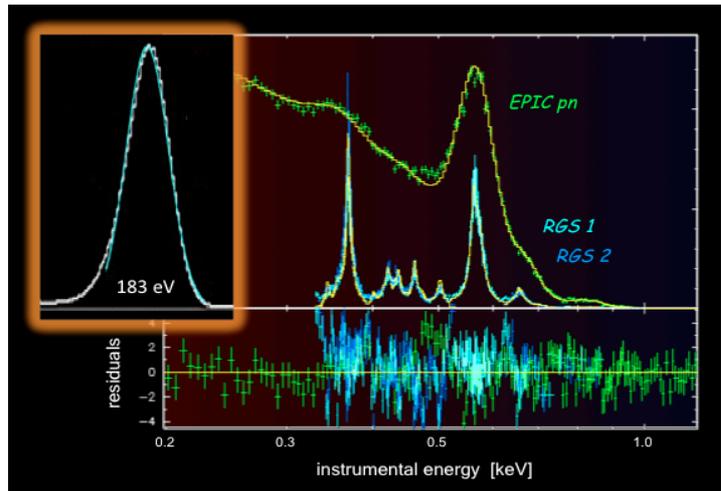

*Figure 5.6.1: XMM-Newton spectra of Comet C/2000 WM1, obtained with EPIC pn, RGS 1, and RGS 2. The inset shows, in the same energy scale, the energy resolution which was obtained at 183 eV with an early prototype of the eROSITA CCD. If this resolution will be available after launch, eROSITA will pioneer X-ray spectroscopy in the range 150 – 350 eV, which is poorly explored to date.*

*eROSITA* has an enormous potential for advancing our understanding of SWCX, as its general properties, effective area at low energies (in particular in the 0.3-0.4 keV range, where the ROSAT PSPC was almost blind due to absorption by carbon in the polypropylene entrance window), its spectral and spatial resolution, are even better than those of *ROSAT*. This, however, requires that it will be possible to use the excellent sensitivity and spectral resolution at low energies after launch. Whether this will be possible will crucially depend on the available telemetry and the properties of the optical blocking filter. In the following it is assumed that the sensitive energy range of *eROSITA* will extend down to 200 eV or even less, and that the energy resolution will be similar to that before launch (Fig. 5.6.1). In this case, the study of charge exchange is likely to become the key science topic of solar system studies with *eROSITA*. This process is important in many respects, as will be shown in the next section.

## 5.6.3 Charge exchange

Compared to other processes which lead to the generation of X-rays, charge exchange is fundamentally different, because the X-rays are not produced by hot electrons, but by ions picking up electrons from neutral gases. This implies that this process enables us to detect cool gas in X-rays (as demonstrated by comets), while all the other processes require very hot and extreme conditions. The fact that the cross sections for charge exchange ($\sim 10^{15}$ cm$^2$) are orders of magnitude higher than for the processes involving hot electrons makes charge exchange induced X-rays a most efficient tracer of tenuous amounts of cool gas. After having been overlooked for a long time, the astrophysical importance of charge exchange is now beginning to receive general attention. This is in contrast to the situation in atomic physics, where charge exchange studies have been performed from its very beginning. Despite this long history of research, revealing the details of this fundamental process still continues to be a significant challenge, both to theoretical and laboratory studies.

Due to these challenges, several approximate methods have been developed in order to treat this problem theoretically, ranging from the classical over-the-barrier model, the classical trajectory Monte Carlo technique (using classical three body dynamics with some quantum mechanical ingredients) and the Landau-Zener approximation, to the quantum-mechanical close-coupling method, the most reliable, but also very demanding theoretical approach. Also laboratory studies are facing challenges and limitations. Therefore, the required information needs to be extracted from a variety of experiments, e.g., the electron beam ion trap, the translational energy spectroscopy, or the cold target recoil ion momentum spectroscopy. Despite all these efforts, some results exhibit considerable uncertainties and discrepancies, in particular between the theoretical calculations and the laboratory measurements.

In this context, a third option becomes particularly important: the study of charge exchange by investigating the X-ray emission of comets. As this emission is the direct result of charge exchange reactions, comets represent an ideal laboratory for studying this process:

- Highly charged ions interact with cold neutrals at low density;
- Complications due to magnetic effects are unimportant;
- There is no other major emission component (like electron bremsstrahlung, scattered solar X-rays, thermal emission);
- The 'experimental setup' is particularly clean;
- The solar wind interaction with comets represents a textbook example of a system which is far away from thermal equilibrium;
- The full range of currently available X-ray observing techniques can be used, providing spatial, temporal, and spectral resolution.

Thus, X-ray observations of comets can be considered as benchmarking experiments for testing our understanding of the physics of charge exchange. They may provide fundamental data for atomic physics, which, when fed into plasma emission codes, are likely to improve plasma diagnostics in general. In addition to this general importance, the X-ray emission of comets is also important for specific applications:

- Calibration: the fact that comets emit a pure line spectrum with no continuum, that they are moving objects with extended emission, irradiating a large detector area, makes them ideal sources for the calibration of the energy response at low energies;
- Solar Wind: the X-ray emission of comets allows us to probe the heavy ion content of the solar wind in three dimensions at various states of the solar cycle, and to study the solar wind interaction with the cometary coma;
- Comets: as the importance of dust for charge exchange is much smaller than that of gas, X-ray observations of comets give us direct remote access to the gas content in the cometary comae, not affected by dust (in contrast to the visual region); furthermore, the X-ray morphology makes a remote observation of the cometary bow shock possible.

While charge exchange induced X-ray emission can be best studied at comets, there are also other objects where this kind of emission can or might be observed:

- Planets: charge exchange occurs also in planetary exospheres, where it is, due to its very high cross section, a very sensitive tracer of tenuous amounts of gas, thus enabling remote, global studies of the outgassing of planetary atmospheres, and linking X-ray astronomy to astrobiology;
- Heliosphere: also the heliosphere is glowing in X-rays due to charge exchange between the solar wind and the stream of interstellar matter flowing through the solar system and due to charge exchange at the heliopause, implying that any observation made from within the solar system may be affected by charge exchange.

In addition, there is an increasing number of various locations outside the solar system where evidence for charge exchange induced X-rays may have been detected (Sect. 5.6.1). Thus, the study of charge exchange in the solar system has an impact on many astrophysical research areas far beyond the solar system. The object which is most severely affected is the Local Hot Bubble, where part of the emission which was originally attributed to it is now understood as solar system charge exchange, requiring that many previous conclusions and models now have to be revisited.

## 5.6.3 Solar system targets for *eROSITA*

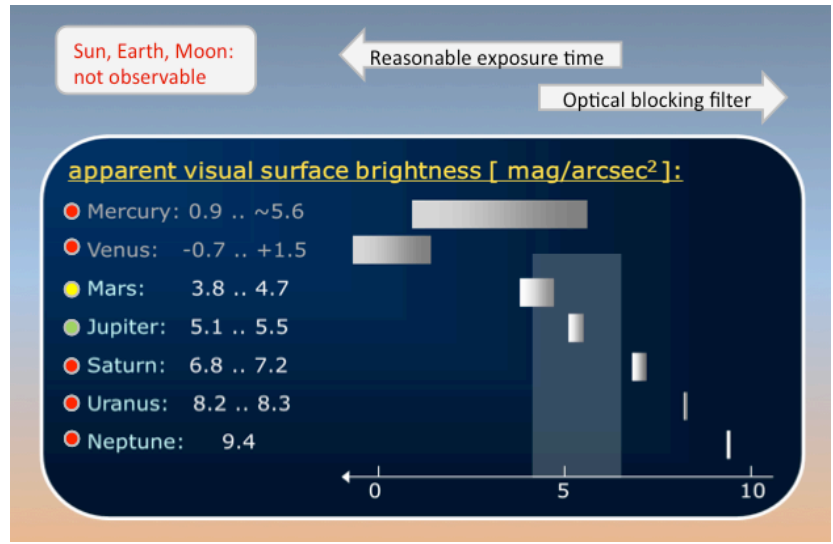

*Figure 5.6.2: Summary of the visibility of the main solar system bodies with eROSITA. The Earth, Moon, Mercury, Venus (and the Sun itself) will not be observable because they violate the solar angle constraint. Uranus and Neptune, and probably also Saturn, will be too X-ray faint. Whether Mars and Jupiter will be observable will depend on the properties of the optical blocking filter. While this is more likely for Jupiter due to its lower visual surface brightness, there will be very favorable observing conditions of Mars in 2016 and 2018.*

While charge exchange, in particular at comets, is likely to become the main research topic of *eROSITA* concerning solar system studies, there are also other processes which can be investigated there, like fluorescence and bremsstrahlung. Due to their lower cross sections, however, they require sufficient densities, which are only found in planets or in planetary systems. Bremsstrahlung, in addition, requires the presence of energetic electrons and thus of planets with a sufficiently high magnetic field.

Not all major bodies in the solar system, however, will be observable with *eROSITA* (see Fig. 5.6.2): the Earth, Moon, Mercury, Venus (and the Sun itself) all violate the solar angle constraint, while Uranus and Neptune will be too X-ray faint. Saturn may be detectable, but only as a very faint source in long pointed observations. The most promising planets, concerning their X-ray flux, will be Mars and Jupiter. With a visual surface brightness of $3.8 - 4.7$ mag/arcsec$^2$ and $5.1 - 5.5$ mag/arcsec$^2$, respectively, they will require an optical blocking filter of sufficient density. Due to its lower visual surface brightness, its (on average) higher X-ray flux, and the more constant observing conditions, Jupiter will be the easier target. Mars, however, will be very favorably placed for an observation with *eROSITA* in 2016 and 2018.

While Mars and Jupiter will probably require pointed observations to get enough flux, it is likely that comets will be well observable during the all-sky survey, as it was demonstrated in the *ROSAT* all-sky survey. The observing efficiency, however, could be enhanced by adjusting the sequence of survey scans to maximize the number of comet passes (Fig. 5.6.3), without affecting the all-sky exposure. Extrapolating from the *ROSAT* all-sky survey, where four comets were detected on seven occasions within half a year (Dennerl et al. 1997), it is well possible that more than 20 comets will be serendipitously observed during the whole *eROSITA* survey phase, providing a good coverage of the solar wind properties over the declining part of the solar activity cycle. By adjusting the sequence of the survey scans, it is likely that this scientific harvest could be considerably improved.

In addition to comets, Mars, and Jupiter, there is an important *eROSITA* target which does not require any pointed observation, because it will be continuously observed: the heliosphere. The main challenge here is to separate the heliospheric X-ray emission from that of the Local Hot Bubble and other diffuse sources beyond the solar system. This will be a difficult task, which, however, will be facilitated by the following fortunate and unique properties of *eROSITA*:

- *eROSITA* will perform eight all-sky surveys; this will be most valuable for separating time variable structures from persistent ones;

- *eROSITA* will provide a considerably enhanced energy resolution compared to *ROSAT*, which should be adequate for distinguishing between thermal and charge exchange induced X-ray emission;
- *eROSITA* will become the first satellite to observe the X-ray sky from L2, sufficiently far away from Earth to avoid the additional contribution due to geocoronal X-ray emission.

Thus, by using all these properties, it might become possible to map the X-ray emission of the heliosphere (including the heliopause). This would be of immediate value for many studies of the diffuse X-ray emission from beyond the solar system, in particular from the Local Hot Bubble.

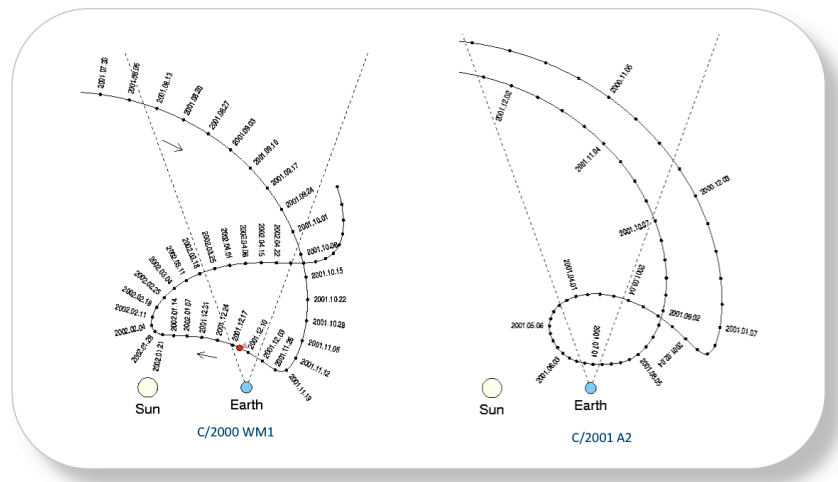

*Figure 5.6.3: Path of two comets with respect to the Sun and Earth. The dashed lines enclose the range of solar angles which will be accessible to eROSITA. Depending on the geometrical circumstances, a comet may pass this range several times, or stay there for a long time, or miss it completely.*

With its unique combination of capabilities, the expectations are high that *eROSITA* will make a substantial contribution to solar system studies, in particular to the physics of charge exchange.

## 5.7 Studying the variable sky with *eROSITA*

Following initial work, e.g., with the Vela 5B monitors, the past two decades have seen significant progress in the study of the variable and transient X-ray sky due to the availability of dedicated instrumentation. The WATCH and SIGMA instruments aboard the Russian *GRANAT* satellite discovered and monitored the most prominent X-ray transients in the inner Galaxy until 1998. They were followed by the *RXTE* All Sky Monitor (*RXTE-ASM*), that yielded light curves and X-ray colors of the ~100 brightest sources in the 2-10 keV band from 1999 until early 2012 (see Figure 5.7.1 for a selection of X-ray binary light curves).

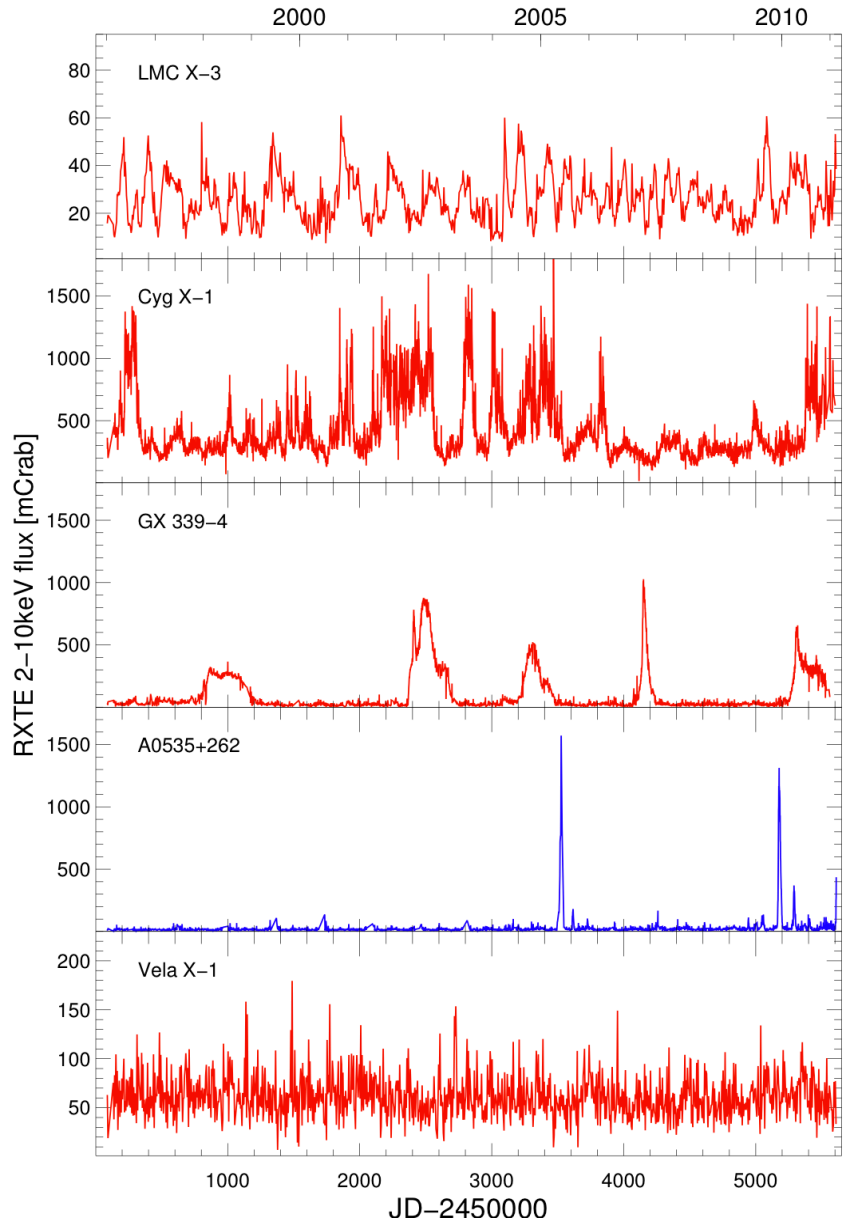

*Figure 5.7.1 Example long-term lightcurves of different types of accreting X-ray binaries as measured with the RXTE-ASM. From top to bottom: LMC X-3: a luminous HMXB characterized by typical variability timescales of months. Cyg X-1: a HMXB who randomly switches between the low-luminosity hard state and the high-luminosity soft state. GX 339-4: a black hole in a LMXB which shows bright outbursts and months to year long phases of quiescence. A0535+35: a neutron star with a Be companion showing very bright outbursts. Vela X-1: erratic variability caused by variations of the mass accretion rate in this wind accreting HMXB*

The most recent addition was the *MAXI* instrument at the International Space Station which provides information about the soft X-ray sky since 2009. Starting in 2005 and 2008, respectively, *Swift*-BAT and *Fermi*-GBM have been adding hard X-ray information to this database. Together, these monitors have revealed strong variability originating from a rich zoo of phenomena: stellar flares, neutron star and black hole outbursts, disrupted stars in the vicinities of supermassive black holes, shock-breakouts of core collapse supernovae, and gamma-ray bursts to name a few (Figure 5.7.2). They have also probed the long-term variability of active galactic nuclei, cataclysmic variables and other, less violent source populations.

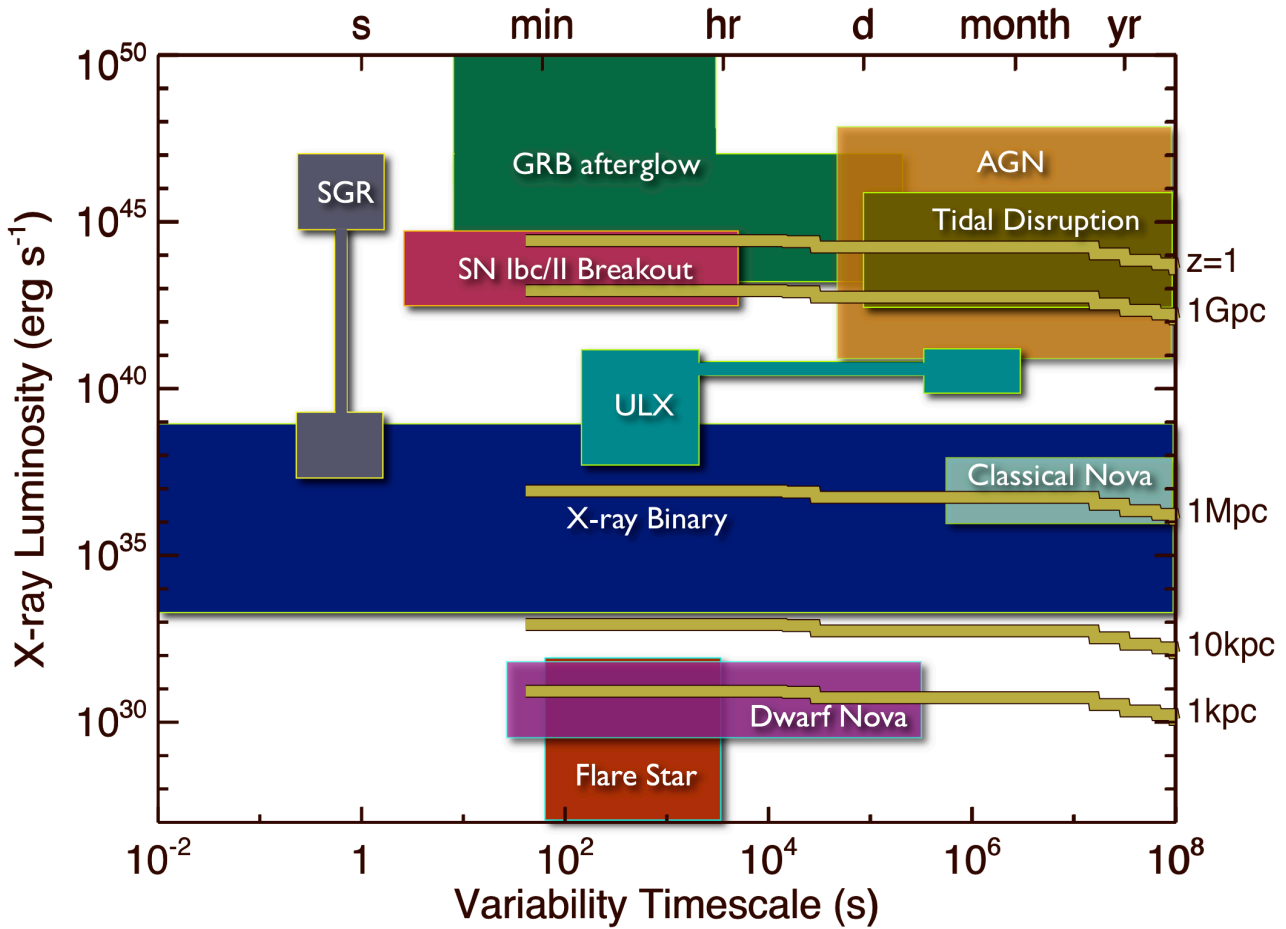

*Figure 5.7.2: Peak X-ray luminosity of the most prominent source populations as function of their characteristic variability time scales (adopted from Soderberg et al. 2009). Yellow tracks indicate the an approximate eROSITA 0.5-2 keV sensitivities for various source distances. The sensitivities at the shortest time scales (within a single scan) are omitted as they are currently subject to strong uncertainties due to the varying instrument PSF as function of off-axis angle.*

## 5.7.1 *eROSITA* discovery space

During the *eROSITA* survey phase most of the sky will be revisited approximately every 6 months, with the poles being monitored with higher cadence. So, each location in the sky gets visited (at least) eight times during the *eROSITA* survey phase. The area around the ecliptic poles, on the other hand, is visited more often, due to the overlapping scans, as discussed in section 3. Figure 5.7.3 shows a "cadence" map of the celestial sphere observed in eRASS:8, i.e. the number of such daily visits as a function of position in the sky. This assumes simple sun-pointing scans, and is therefore just an approximation of more realistic scenarios (still under investigation) within which the final number of daily visits around the ecliptic poles will be reduced to avoid source confusion (see discussion in Section 3.2).

In fact, for a given sky location, each of these visits will consist of a few ≈30s long scans, separated by 6 hours (i.e., one "eRoDay"). Such a pattern results in a wide range of variability timescales to which eROSITA will be sensitive, from (tens of) seconds to months and years (see Fig. 5.7.2). Overall, this specific scanning strategy implies that *eROSITA*'s capabilities for studying variability will be at the same time more complex and more constrained than experienced with very wide field instruments such as *RXTE* or *MAXI*, but it will likely be the only X-ray mission capable of regularly monitoring the full X-ray sky in the period 2015-2018 and beyond. The possible exception here could be the Brazilian *MIRAX* instrument.

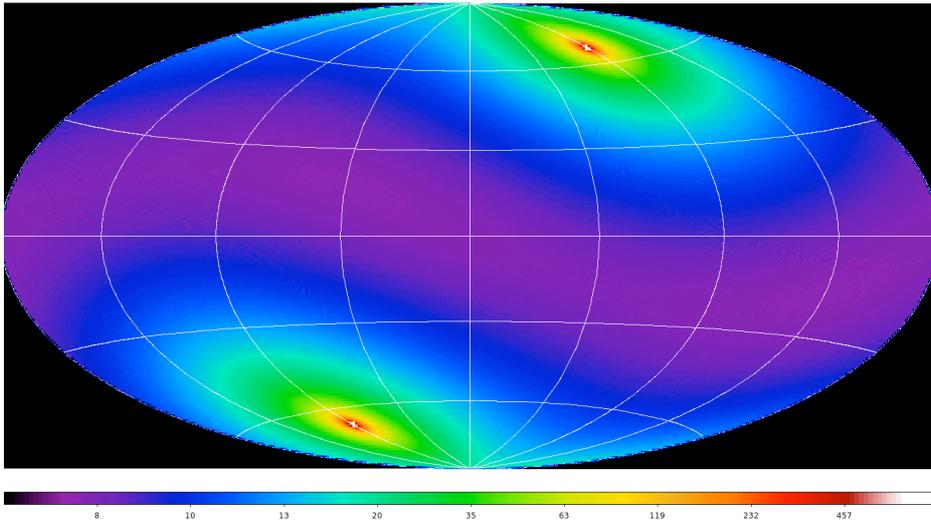

*Figure 5.7.3: eRASS:8 "cadence" map. The celestial sphere in equatorial coordinates is color-coded by the number of daily visits of eROSITA during the 4-years all-sky survey. Each daily visit, amounting to a total exposure of ~250 s is in fact the sum of ~6 (overlapping) scans of about 30 seconds each. Note that about 1,000 deg$^2$ around the poles will be visited more than 30 times.*

In this chapter, we shortly describe which role *eROSITA* can play for the exploration of the X-ray time domain science. Here, we extrapolate from the results of previous and existing all sky monitors and from the earlier *ROSAT* survey.

*eROSITA's* sensitivity for individual scans will be in the mCrab range (1mCrab= $2.4 \times 10^{-11}$ erg/s/cm$^2$) during the survey phase. Thus, on 30s time scales it will surpass that of *RXTE*-ASM and *MAXI* and it will be comparable to the *RXTE*-PCA slews. Figure 5.7.4 shows the number of ASM sources brighter than a certain flux averaged over the 14 years of *RXTE*. At a single-scan sensitivity of approximately $10^{-13}$ erg cm$^{-2}$ s$^{-1}$ in the 0.5-10 keV band, *eROSITA* will be able to follow 40-50 sources on eRoDay scales and a significantly larger population on a 6 months time scale. We stress here that reliable simulation tools that are based on the currently available plans for the all-sky survey only became available during the time of writing, and that the results of the simulations are not yet contained here. They will be incorporated in a future version of this science book.

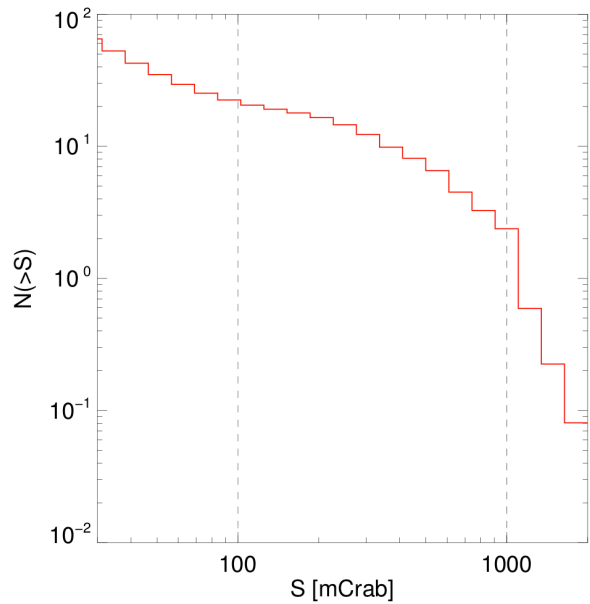

*Figure 5.7.4: The number of RXTE-ASM sources on average brighter than a given flux S as a function of S (in mCrab; 1 mCrab=2.4 x 10$^{-11}$ erg/s/cm$^2$)*

## 5.7.2 Stellar Flares

Stellar flares occur on timescales of seconds to weeks (Kuerster et al., 1996). They are caused by magnetic reconnection events in the stellar corona that heat the coronal plasma, leading to soft X-ray emission. The typical brightness of stellar flares is such that *eROSITA* will only be able to pick up the brightest flares. For these flares, however, *eROSITA* will allow triggering follow-up observations in other wavebands and with pointed instruments. Furthermore, through measuring stellar X-ray fluxes through the survey *eROSITA* will allow to determine the statistics of stellar flaring. For further details see Section 5.4.3.

## 5.7.3 Neutron Stars and Black Holes

The most luminous Galactic X-ray emitting population is formed by X-ray binaries, stellar mass black holes and neutron stars accreting matter from a companion star (see also Section 5.3). The erratic nature of mass accretion leads to X-ray variability with a wide range of amplitudes and timescales. In the persistent HMXB this variability is mainly due to variations of the photo-ionized stellar wind, which is often focused onto the compact object. In Be systems X-ray outbursts are observed when the excretion disk around the donor star has grown large enough that accretion onto the compact object - usually a neutron star - can occur. Be X-ray binaries are therefore transient sources and are typically characterized by several weeks long outbursts separated by month to decade long quiescence periods.

In LMXB the system accretes from a late type donor star via Roche Lobe overflow. These systems can be persistent or transient. In case of transient systems an accretion disk is built up by Roche Lobe overflow and an instability in the accretion disk then triggers the accretion onto the compact object. In the case of black hole transients, the resulting X-ray outburst will first be characterized by a phase of increasing luminosity combined with a hard Comptonization spectrum, which then softens until the thermal spectrum from the accretion disk dominates. After this soft state phase the system becomes fainter, but remains soft, until at some critical lower luminosity the X-ray spectrum hardens and remains hard until the source disappears. The total duration of such an outburst can be up to several months. The quiescent phases between outbursts can last months to decades.

*eROSITA's* capabilities to find new transients during the eRASS can be estimated from the *RXTE* Galactic Bulge Scans, where *RXTE* scanned the Galactic Center region twice weekly from 1999 onwards. The scans yielded effectively 20 seconds of on source data with an effective area comparable to *eROSITA*. In these scans 50 new sources were discovered and strong variability was observed on all timescales. 45% of all sources were active for less than 25% of the time, 40% of all sources were active for more than 50% of the scans. Based on this experience we expect that *eROSITA* will detect several new transients during its scans over the Galactic Plane and the Galactic Center.

## 5.7.4 Tidal disruption events and Gamma-Ray Bursts

Stars passing within a distance of about $5 \cdot M_7^{-2/3}$ Schwarzschild radii of a supermassive black hole of mass $M_{BH}=10^7 M_7 M_\odot$ will be torn apart by the strong tidal gravitational field (e.g., Hills 1975; Rees 1988). For $M_7<20$ the disruption of a main-sequence star happens outside the black hole event horizon and should give rise to a detectable flare of emission. Until recently, only a few candidate tidal disruption events were known, primarily from the *ROSAT* All Sky Survey archive (e.g., Komossa & Bade 1999; Komossa & Greiner 1999) and from rest frame-UV monitoring of otherwise dormant galaxies (e.g., Renzini et al. 1995; Gezari et al. 2006, 2008). Now, new time domain surveys (e.g., SDSS Stripe 82; Palomar Transient Factory, PTF, Rau et al. 2009) and *Pan-STARRS* have added candidates discovered at optical wavelengths (Cenko et al. 2010; van Velzen et al. 2011; Gezari et al. 2012) and X-ray observations with *Swift* uncovered the first members of a possible population of relativistic tidal disruption flares (Bloom et al. 2011; Burrows et al. 2011; Cenko et al. 2012).

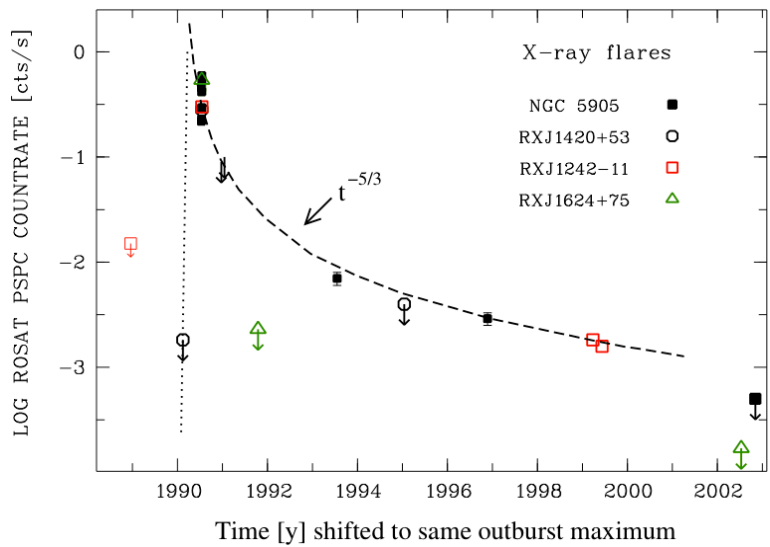

*Figure 5.7.5: Lightcurves of tidal disruption events as seen in the RASS (Komossa et al. 2004).*

The flare emission is predicted to be dominated by an optically thick accretion disk reaching a temperature of about $T_{eff} \approx 3\times10^5 M_7^{-1/4}$ K (Ulmer 1999), thus peaking in the X-rays to far-UV. Characteristic for a tidal disruption event, according to the models, is that the flux declines as $t^{-5/3}$. This makes these events visible in all sky or slew surveys for years (see Figure 5.7.5, and Komossa et al. 2004; Esquej, Read, Saxton 2010) and gives confidence that *eROSITA* should be able to pick up further events based on their typical light curves. These discoveries can not only provide a dramatic increase in the number of ironclad events but can also provide a window into the demographics of SMBHs where detailed follow-up observations could allow for an entirely new way of measuring black hole mass, thus offering an independent test of the $M_{BH}$-$\sigma^*$ relation of galaxies.

The most luminous extra-galactic transients that *eROSITA* could detect are Gamma-Ray Bursts (GRBs) and their afterglows. However, their rarity (about 1 per day, over the whole sky), coupled with the short duration (seconds for the prompt emission to hours and days for the X-ray afterglows) reduces the detection probability during the *eROSITA* all sky survey dramatically. Recent simulations by Khabibullin et al. (2012) predict the detection of about 4-8 events per year, identified mainly through their characteristic power-law decline. Some of those events might lack a corresponding prompt gamma-ray signal, whether the collimated jet was not directed at the Earth, no gamma-ray detector was observing the right position at the right time, or no gamma-rays have been produced in the first place (i.e. 'failed gamma-ray burst'). In this way, *eROSITA* can compile a sample of events that is unbiased with respect to the prompt gamma-ray emission opening a new view into these energetic explosions.

The most critical aspect will be the rapid identification of the new X-ray source and the communication to the ground-based follow-up community. Experience from *ROSAT* shows that a sparsely sample light curve is often not sufficient for an indisputable identification. A search for afterglows of un-triggered GRBs in the *ROSAT* all-sky survey by Greiner et al. (2000) revealed two dozen candidates, most, if not all, found to be misidentified flares from late type stars.

## 5.7.5 AGN Variability

All AGN show variability on long time scales. Detailed studies of this variability through Power Spectral Density (PSD) have shown an almost ubiquitous PSD shape characterized by a steep power law shape ($\alpha \sim 2$), above a special frequency $\nu_{br}$, and a flatter ($\alpha \sim 1$) power law slope below $\nu_{br}$ (McHardy et al. 2006; Gonzalez-Martin & Vaughan 2012). In particular, it has been observed that the characteristic break frequency $\nu_{br}$ scales with BH mass and (possibly) with accretion rate (McHardy et al. 2006; Gonzalez-Martin & Vaughan 2012) and that the PSD amplitude appears almost constant at $\nu_{br}$. For these reasons high mass AGN are dominated by variability on weeks to years timescales, compared to Galactic black holes typically highly variable on fractions of a second.

The eRASS will have, on average, 8 flux measurements over 4 years (many more, indeed, in the small areas around the ecliptic poles, see Fig. 5.7.3) during which the all sky survey will be performed, thus allowing to characterize the variability of the brighter AGN on such time-scales. How many variable AGN will eROSITA detect? An attempt to get a rough estimate of this number can be made under the assumption of a ubiquitous PSD with shape: $P(\nu) = A(\nu/\nu_{br})^{-2}$ for $\nu > \nu_{br}$ and $P(\nu) = A(\nu/\nu_{br})^{-1}$ for $\nu < \nu_{br}$, assuming that the break frequency $\nu_{br}$ is inversely correlated to the black hole mass $M_{BH}$ ($\nu_{br}=B/M_{BH}$, where B is a constant) and a constant PSD amplitude at $\nu_{br}$ ($PSD_{ampl}=A\cdot \nu_{br}$). Then, applying the Parseval theorem, we can estimate:

$$\sigma^2_{rms} = \int_{\frac{1}{T}}^{\frac{1}{t_{min}}} P(\nu)d\nu = PSD_{ampl} \cdot [ln(T\nu_{br}) - t_{min}\nu_{br} + 1]$$,

where $t_{min}$ is the exposure time of each single *eROSITA* daily exposure (250 s) and T is the time difference between the first and last observations (4 yr). Assuming, as observed by O'Neill et al. (2005), $PSD_{ampl}=0.02$ and B=40 (see also Papadakis et al. 2004; Ponti et al. 2012) we expect, for $M_{BH}=10^9$ $M_\odot$, a $\sigma^2_{rms}=0.05$, that corresponds to 23% variability (a higher variability is expected for smaller mass BH, e.g. 40% for $M_{BH} = 10^6$ $M_\odot$). Thus, if the systematic uncertainties associated with the fluxes measured between the different scans (PSF modelling, etc.) are known to better than a few per cent, the systematics should add little contribution to the measured variability.

Imposing that $\sigma^2_{rms} > Err(\sigma^2_{rms})$, we can use eq. (11) of Vaughan et al. (2003) to estimate the limiting count rate, to detect variability. For $M_{BH} = 10^8$ $M_\odot$ this corresponds to a count rate of about 0.02 photons/s. Such a low count rate corresponds to an average of just 5 photons in a daily visit (250 s). We can then conservatively fix as a limiting count rate 0.08 photons/s, in order to ensure the applicability of Gaussian statistics (>20 counts), as in Vaughan et al. (2003) approach. Such a count rate corresponds to a flux of $f_{0.5-2}= 1.1 \times 10^{-13}$ and $f_{2-10}= 2.2 \times 10^{-12}$ erg cm$^{-2}$ s$^{-1}$ in the soft and hard bands, respectively. Thus, we expect to detect, in the 2-10 keV band, about 0.05 variable AGN per deg$^2$ and ~1,500 in the full sky. In the soft band, we expect about 1.5 variable AGN per deg$^2$ and 45,000 in the full sky.

We caution the reader on the several limitations of this first rough estimate. For example: i) the months to years AGN variability has been measured only for few tens of sources, thus extrapolating the PSD observed at higher frequencies to these timescales might not be accurate; ii) AGN at higher redshift might be different than the local ones; iii) more accurate consideration of the observed BH mass distribution and variation of observed time-scales with redshift will be performed; iv) the large gaps between the different *eROSITA* scans might introduce large, spurious, scatters in the variability determinations (the impact of these effects is still under investigation); and v) the level of systematic uncertainties on the flux determinations (e.g. PSF model), possibly contaminating the variability measurements (e.g. PSF and background variations), are still uncertain and they might have a significant impact on these estimates.

### 5.7.6 *eROSITA* detection strategy for transients

Based on the discussion presented above, *eROSITA*'s capabilities to study variability can be roughly divided into two categories: 1) prompt and possibly short-term events which justify immediate observations with pointed X-ray instruments and follow-up in other wavebands, and 2) long-term variability, which will only be studied when data from several years of survey observations are available. While the latter class of sources is easily studied and identified in the data products of the survey, the detection of transient phenomena is time critical.

The current strategy for the detection of transient events during the survey phase is that directly after the downlink, the science data are analyzed as part of the Near Real-Time Data Analysis software (NRTA). During the run of the NRTA on each telemetry downlink a temporary archive is populated in a way that the data are immediately analyzable with a special version of the *eROSITA* SASS. This version of the SASS will determine the fluxes of the brightest sources and will find new and bright sources, as well as strong changes in the source luminosity with respect to a database. These changes will initially be based on catalogue fluxes of the brightest 50-100 X-ray sources, during later stages of the eRASS survey these measurements will be based on a comparison of earlier *eROSITA* measurements. Interesting phenomena in the part of the sky accessible to the German team for scientific analysis will then announced to the community either as Astronomer's Telegrams and/or through special email lists.

# 6. *eROSITA* in the context of multi-wavelength large area surveys

A major advantage for *eROSITA* as the first all sky X-ray survey after *ROSAT* is that in the intervening 25 years there has been dramatic progress in sky surveys in the optical, NIR, radio and mm-wave. Moreover, there have been dramatic advances in the use of galaxy clusters for cosmological studies and in the understanding of AGN and the special role that X-ray selected AGN studies can play. These developments should allow a quick progress from the eRASS X-ray sky toward new discoveries in structure formation and cosmology.

As the eRASS will proceed, our goal is to produce catalogues of X-ray selected *eROSITA* sources that will include state of the art identification of X-ray sources by making use of as much information as possible from the wealth of multi-wavelength data discussed below. The tremendous impact of the *ROSAT* catalogues (Voges et al. 1999) is a powerful reminder of the importance of such task.

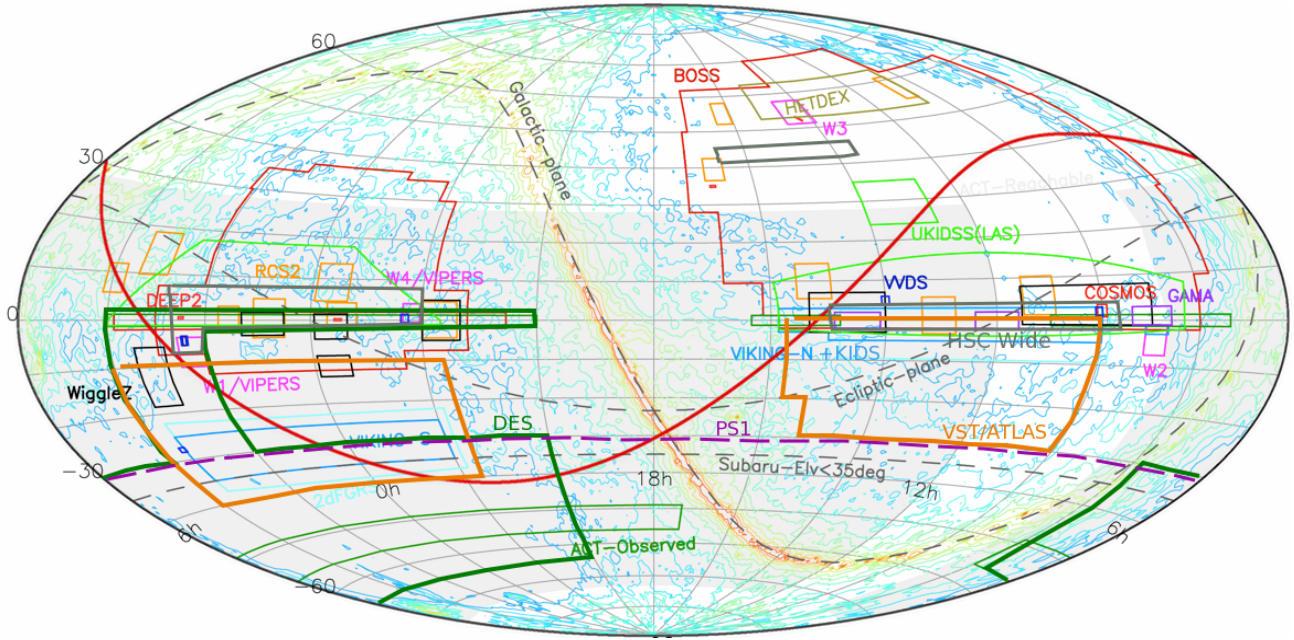

*Figure 6.1.1: Multi-band wide area optical imaging surveys (figure courtesy A. Nishizawa, IPMU) are displayed in equatorial coordinates. The thick red line mark the separation between the German and the Russian eROSITA sky, with the former being the southernmost one. Existing and planned optical/NIR surveys are outlined with colored boxes. Pan-STARRS (PS1) survey will cover all area above the dashed magenta line ($\delta > -30°$). Together, DES and PS1 provide the multi-band photometric data needed for cluster and AGN confirmation in the extragalactic German eROSITA sky, as well as the cluster photometric redshift estimation and weak lensing mass constraints.*

Below we provide a brief overview of the key datasets that will be coupled with the *eROSITA* X-ray sky. The following descriptions are divided into the Optical/IR (OIR) Imaging Followup, OIR Spectroscopic Followup, mm-wave Surveys, Radio Surveys and X-ray Followup. Depending on the long term status of *XMM-Newton* and *Chandra*, the X-ray followup will largely come from the *eROSITA* pointed mission phase.

## 6.1 Multi-band optical/IR imaging followup

A first step in the analysis of the extended and point sources identified in the *eROSITA* X-ray survey will be cross comparison with multi-band OIR imaging data. These photometric data allow for tests of the presence of associated galaxies in a cluster and could be used in combination with X-ray data in the classification of *eROSITA* sources (clusters or AGN or galactic compact objects). Accessing the multi-band OIR data is also critical for estimating photometric redshifts for the clusters and AGN. Moreover, these data provide galaxy targets for additional spectroscopy that is needed, and will also provide important shear information for background galaxies, enabling more calibration of the galaxy cluster masses through weak lensing shear analyses.

There are a variety of surveys that cover the sky, including digitized versions of photographic plate surveys that date from the 1950s. An important issue is the wavelength coverage and depth of these surveys, and over the past few months we have busied ourselves with an attempt to try to characterize the depth of these multi-band imaging survey in a uniform way. We have attempted to assess the 10σ photometric depths within an approximately 2" diameter circular aperture, which is particularly useful for characterizing the depth of the survey for faint galaxy photometry and photometric redshifts. Such a depth measure is less sensitive to the assumed and delivered image quality (seeing) than the more commonly adopted point source sensitivity. Moreover, at the 10σ depth it is still possible to estimate galaxy photometric redshifts, whereas below the 10σ limit the color uncertainties lead the redshift estimates to become very crude. Finally, a rough rule of thumb is that at the 10σ limit a galaxy catalog is approximately 90% complete for typical ranges of seeing (i.e. <1.5") and within extragalactic fields where the stellar density isn't extremely high.

Table 6.1.1 shows the key optical/NIR surveys, their operational timescales, their achieved or projected photometric depths and a designation of whether the telescope lies in the northern or southern celestial sphere, where the depths are meant to reflect 10σ photometric depths. In some cases the point source depths have been replaced with depths for extended sources within 2" diameter apertures, but that has not been possible in all. While this isn't warranted for space based imaging (*GAIA*, *Euclid*), it is helpful to avoid gross overstatements of expected depth due to optimistic assumptions about the image quality. The bands shown are the common broad band filters that are typically used in these surveys. Note that *GAIA* and *Euclid* both employ broad band filters, and so only a single optical band depth is limited. In the case of *GAIA* this limit is not a 10σ depth. The end of mission photometric error for stars with G=20 mag is estimated to be 0.003 mag!

In addition, the *WISE* IR survey has covered the sky to depths of 19/18.8/16.4/14.45 mag in the 3.4/4.6/12/22 μm bands (5σ point source; see http://wise2.ipac.caltech.edu/docs/release/allsky/). The AKARI satellite has imaged the whole sky in four FIR bands (50μm-180μm) and two MIR bands (9μm and 18μm). These catalogs have already been released and are available to the collaboration.

| Survey | Lat | Date | Ω | u | g | r | i | z | Y | J | H | K |
|---|---|---|---|---|---|---|---|---|---|---|---|---|
| SDSS | +30 | -'10 | 10000 | 21.6 | 22.6 | 22.4 | 21.6 | 20.1 | - | - | - | - |
| PS1 | +20 | '10-'12 | 30000 | - | 22.6 | 22.4 | 22.1 | 21.1 | - | - | - | - |
| SkyMapper | -30 | 11- | 30000 | - | 22.5 | 22.0 | 20.9 | 20.6 | - | - | - | - |
| KIDS+VIKING | -20 | 11- | 1500 | 24.8 | 25.4 | 25.2 | 24.2 | 22.4 | 21.6 | 21.4 | 20.8 | 20.5 |
| DES+VHS | -30 | '12-'16 | 5000 | - | 24.6 | 24.1 | 24.3 | 23.8 | 21.5 | 20.2 | 20.1 | 19.5 |
| ATLAS+VHS | -20 | 11- | 4500 | 22.0 | 22.2 | 22.2 | 21.3 | 23.8 | 21.5 | 20.5 | 19.9 | 19.3 |
| HSC | +20 | '12-'16 | 1500 | - | 25.5 | 25.2 | 25.5 | 24.3 | 23.3 | - | - | - |
| PS2 | +20 | 14- | 10000 | - | 24.5 | 24.5 | 24.5 | 24.5 | - | - | - | - |
| GAIA | - | '13- | 41253 | | | 20 | | | | | | |
| Euclid | - | '19-'24 | 15000 | | | | 24.5 | | 24.0 | 24.0 | 24.0 | - |
| LSST | -30 | '20-'30 | 18000 | 24.0 | 26.0 | 26.0 | 26.0 | 26.0 | 26.0 | - | - | - |

*Table 6.1.1 Overview of key optical/NIR wide area imaging surveys. Surveys include Sloan Digital Sky Survey (SDSS), Pan-STARRS1 (PS1), Kilo-degree Survey (KIDS), VISTA Kilo-degree Infrared Galaxy Survey (VIKING), Dark Energy Survey (DES), Vista Hemisphere Survey (VHS), VST Atlas Survey (ATLAS), Hyper-Suprime-Cam Survey (HSC), Pan-STARRS 2 (PS2), Euclid Space Mission (Euclid) and the Large Synoptical Survey Telescope (LSST). 'Lat' encodes the latitude of the observatory, 'Date' encodes the range of years over which the survey takes place, and Ω encodes the solid angle of the survey.*

For the *eROSITA* cluster cosmology experiment, we will focus on the extragalactic portion of the PS1 and DES surveys that lie within the German half of the *eROSITA* sky. Together, these datasets will cover approximately 10,000 deg² of extragalactic sky at galactic latitudes |b|>30° and 13,160 deg² at |b|>20°. The SDSS is completely covered by the PS1 survey, which is deeper in i/z and similar depth in g/r, and so we expect the SDSS survey will be particularly valuable in the calibration of the PS1 survey. In addition, there are two surveys much deeper than PS1 that overlap the PS1 and SDSS surveys. These include KIDS and HSC. The overlap with the German side of the *eROSITA* sky is expected to be about 750 deg² for each of these.

These multi-band photometric data will be used to reliably identify cluster and AGN counterparts and to estimate photometric redshifts for both classes of sources. According to our previous experience (both in XMM-COSMOS, e.g. Brusa et al. 2010, and in the 2XMM catalog, e.g. Pineau et al. 2011) we can infer that at fluxes of the order of the all-sky survey limit in the soft band, 90% of AGN counterparts have i<23 and K<22, and more than 80% are detected at r~22 (see Figure 6.1.2). While none of the existing surveys will cover the entire sky to the desired depth - but eventually *Euclid* and *LSST* will approach this goal - the *eROSITA* sources detected in the 5,000 deg² DES area will constitute a fairly large and statistically significant sub-field for which it will be possible to provide an almost completely identified sample of sources. These numbers do not consider, however, the further uncertainty in the counterpart identification due to the poorer angular resolution of *eROSITA* in the all-sky survey (HEW ~30", see section 2.1) as compared to *XMM-Newton*. In order to assess such an incompleteness, we have performed a test on the XMM-COSMOS sample, by applying probabilistic counterpart identification algorithms (Brusa et al. 2007) to the *XMM-Newton* AGN catalog, using optical catalogs at the i=24 magnitude depth, but where the positional accuracy of the X-ray point-sources above the *eROSITA* flux limit is worsen by a (pessimistic) factor of 6. Still, we found that 87% of the sources could be reliably associated to a single counterpart, and 5% more had an ambiguous counterpart (2 or more optical sources). Only slightly more than 8% of the AGN remained unidentified.

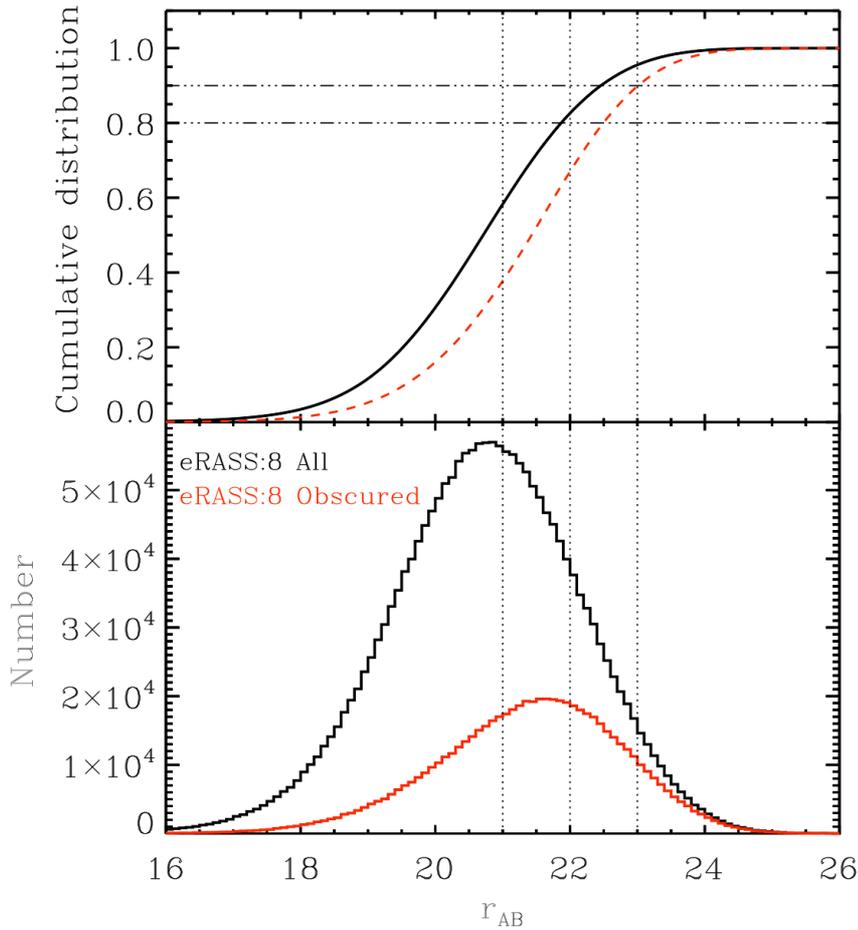

*Figure 6.1.2: Bottom panel: expected distribution of rband optical magnitudes for the counterparts of eROSITA AGN detected at the end of the 4-year all-sky survey, based on the observed distribution for the XMM-COSMOS soft-band detected AGN counterparts. In black we show the entire 0.5-2 keV selected sample, in red only the obscured AGN. Top panel: Cumulative distribution. Vertical lines at $r_{AB}$= 21, 22, 23 and horizontal ones at 0.8 and 0.9 help guiding the eye.*

Photometric redshifts of X-ray selected clusters using SDSS depth griz photometry show excellent performance $\delta z/(1+z)$=0.018 out to $z$~0.6 (Song et al. 2010). Photometric redshifts of SZE selected clusters (i.e. massive clusters with $M_{200}$>4 × $10^{14}$ $M_\odot$) extending to redshift $z$~1.1 show performance of $\delta z/(1+z)$~0.025 (see Figure 6.1.3). At the high redshift end the i/z band photometry is about 0.5mag/1.5mag shallower than DES, and so we should have no problems estimating robust photometric redshifts of the bulk of the *eROSITA* clusters over the DES region and a large fraction (>80%) of the clusters over the PS1 region.

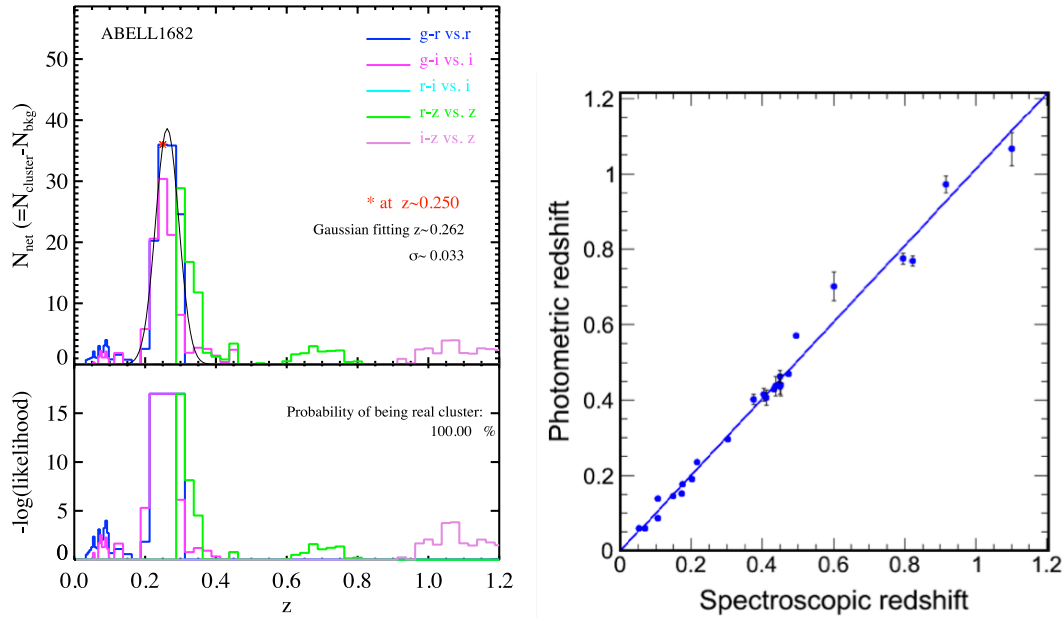

*Figure 6.1.3: Left: For massive clusters in eROSITA one can obtain a red sequence redshift by examining the overdensity of galaxies with appropriate red sequence colors as a function of redshift (Song et al. 2011). Right: We have applied this technique widely within the SPT project, measuring photo-z with griz photometry extending to z=1. Characteristic photometric redshift uncertainties are measured to be δz/(1+z)~0.025 (figure courtesy Alfredo Zenteno, LMU).*

AGN redshifts computed through SED fitting, on the other hand, present more of a challenge, due to the impossibility to break a number of degeneracies when only a few photometric bands are available. We should be able to measure single galaxy redshifts in the DES-VHS area, using multi-epoch photometry (to take into account variability), with a tuned library of templates and with the use of priors, we estimate to obtain an averaged accuracy of 0.08 with a fraction of 20% outliers (Salvato et al. 2011). A smaller number of bands and no correction for variability will increase the fraction of outliers up to 40% of the cases (Saglia et al. 2012). The biggest limitation of SED fitting is that only the optical/NIR bands can be used for the fit, as templates beyond 5 micron are not well characterized. For this reason, to further improve on this figure, we are currently testing machine-learning techniques that can take into account also information available from other bands such as radio, MIR and X-rays itself (Salvato et al., in prep).

## 6.2 Optical/NIR spectroscopic followup

OIR Spectroscopic followup is important for several reasons. On the galaxy cluster side, there is the need to obtain spectroscopic redshifts for testing and tuning of the photometric redshifts. In principle, this is a very limited need that could be accomplished with ~$10^2$ cluster spectroscopic redshifts. However, moving from a typical cluster photometric redshift uncertainty of δz/(1+z)~0.025 to a typical spectroscopic redshift uncertainty that is at least an order of magnitude more accurate enables additional science within the study of the evolution of the clustering of the galaxy clusters. One can, for example, use redshift space distortions as a test of the growth of structure and the Alcock-Paczynski test to obtain additional constraints on the expansion history of the Universe $h(z)$. In a recent analysis (Pillepich et al. 2012) it was shown that for *eROSITA* there are improvements of up to 40% in the constraints on most cosmological parameters, e.g., $\sigma_8$, when spectroscopic redshifts are available. Moreover, spectroscopy of cluster galaxies can provide mass constraints through the kinematics of these galaxies, and provides an Intracluster Medium (ICM) independent method of calibrating the relationships between the cluster virial mass and the X-ray luminosity $L_X$, the X-ray temperature $T_X$, the ICM mass $M_{ICM}$ or a combination thereof $Y_X$ (Evrard et al 2008, White et al 2010, Saro et al., in prep). These cluster velocity dispersion mass constraints can be used along with weak lensing constraints for the overall calibration of the amplitude, slope and redshift evolution of the X-ray mass—observable relations. Finally, other cosmological tests that relies on the measure of the clusters power spectrum would strongly benefit from the higher accuracy of spectroscopic redshift determination.

Currently there are several spectroscopic survey missions being planned, and almost all of them would be relevant for a portion of the *eROSITA* sources, depending the decisions of how to deploy fibers.

| Spectrograph | Telescope | Hemisphere | FoV [deg$^2$] | # of Fibers | Depth | Timescale |
|---|---|---|---|---|---|---|
| SDSS I-III | SDSS 2.5m | N | 7.07 | 600-1000 | r=17.8,19.5 | Complete |
| AAOmega | AAT 3.9m | S | 3.14 | 392 | r=20.5 | Complete |
| LAMOST | LAMOST 4m | N | 19.6 | 4000 | r=20.5 | 2012+ |
| AS3- eBOSS/ SPIDERS | SDSS 2.5m | N | 7.07 | 1000 | i=21.5 | 2014+ |
| 4MOST | VISTA 4.1m | S | 7.1/3.5 | 1500-3000 | r=21.5 | 2018+ |
| WEAVE | WHT 4m | N | 3.14 | 1000 | r=21.5 | 2017+ |
| BigBOSS | Mayall 4m | N | 7.07 | 5000 | r=21.5 | 2017+ |
| DESpec | Blanco 4m | S | 3.8 | 4000 | r=21.5 | proposed |
| PFS/SuMIRE | Subaru 8.2m | N | 1.78 | 2400 | i=22.7 | 2017+ |
| Euclid | 1.2m/Space | - | 0.55 | Slitless Spec | Emission lines | 2019+ |

*Table 6.2.1 Overview of key optical/NIR wide area multi-object spectrographs.*

For cluster spectroscopy the required depths depend sensitively on redshift. For nearby clusters the galaxies are luminous with high central surface brightness and widely distributed on the sky, and the spectroscopy is straightforward with a wide field fiber spectrograph. By the time one reaches redshifts $z>0.8$, the clusters are very compact on the sky and spectroscopy of all but the single one or two brightest galaxies requires the high sensitivity optical spectrographs (i.e. *VLT+FORS2*). Calculations we have done to plan for a wide field fiber spectrograph on a 4-meter class telescope suggest that spectroscopy to depths of r=21.5 would map well onto cluster spectroscopic redshift measurements reaching out to $z\sim0.7$. The ability of such a survey to deliver cluster dispersions for mass calibration depends sensitively on the close-packing constraints on the fiber positioner.

In the case of AGN, the many challenges to an accurate photometric redshift determination outlined above mean that massive spectroscopic follow-up campaigns are often mandatory. However, the optical magnitude distribution of *eROSITA* AGN counterparts cannot be fully accessed with 2-meter class telescopes such as the Sloan one, and the next generation of wide FoV, multi-object spectrograph on 4-meter class telescopes has to be exploited.

Current plans of the German *eROSITA* Consortium in this direction revolves around a proposed new instrument for the ESO *VISTA* telescope, the optical multi-object spectrograph 4MOST. The strong synergies between 4MOST and *eROSITA* are readily summarized: (i) there will be no other X-ray mission in the foreseeable future to assemble the high statistics needed to address the key science goals of *eROSITA*, and (ii) the full characterization of the AGN population relies on a complete as possible spectroscopic identification of the X-ray sources. Ongoing (e.g. SDSS/BOSS) and planned (e.g., BigBOSS) large-area spectroscopic surveys provide, due to their optical color selection, redshifts and BH mass measurements mainly for the luminous un-obscured QSOs. Within the After-SDSS-III (AS3) program, an "early" spectroscopic follow-up survey of *eROSITA* sources (SPIDERS) is currently under definition. This, however, will cover a relatively small area of the northern sky (~1,500 deg$^2$) and is designed to follow-up only those sources detected in the first three or four *eROSITA* all-sky surveys, i.e. down to an X-ray depth at least a factor of two shallower than the final eRASS:8, that will be the target of 4MOST. In this respect, SPIDERS can be considered as a key preliminary step that will help us refine the strategy for the optimal 4MOST follow-up of *eROSITA* sources.

Not only clustering and BAO studies can be carried out with the large eROSITA spectroscopic AGN samples (see Section 5.2.2). Classification on the basis of optical emission lines diagnostics and line widths is a key tool to separate unobscured and obscured AGN. Experience from ongoing X-ray survey projects at the expected depth of the *eROSITA* survey (e.g. COSMOS, 2XMM-SDSS) suggests that a spectroscopic survey to depths down to r=22 would provide an excellent match to the bulk of the *eROSITA* AGN sample. In particular, for unobscured AGN, BH masses over a very wide redshift range ($0<z<5$) can be derived using the so-called virial relations, based on the width of the different broad emission lines (H$\beta$, MgII, and CIV) redshifted into the optical spectra/observed wavelength window. For obscured AGN up to $z\sim1$, instead, spectral features from the host galaxy such as emission line intensities and Balmer break can be used to constrain the star formation rates, the reddening by dust, the stellar population ages and metallicities of the AGN host galaxies. Spectroscopy follow-up will be also crucial to confirm high-redshift ($z>3$) QSO candidates, preselected on the basis of photometric information.

## 6.3 Millimeter-wave and radio followup

Surveys at frequencies around 150GHz are now becoming available (Schaffer et al. 2011), and these provide interesting scientific opportunities when combined with the *eROSITA* sky survey. Direct selection of clusters from mm-wave surveys using the Sunyaev-Zel'dovich effect will provide samples that can be cross-compared to the *eROSITA* sample (i.e. Suhada et al. 2010). A combination of X-ray and SZE measurements on individual clusters leads to direct distance measurements, which would provide an interesting cosmological test of the Hubble parameter and the expansion history (Reese et al. 2002; Bonamente et al. 2009). Finally, the large scale cluster mass calibration effort using weak lensing, velocity dispersions and deep X-ray observations within the *SPT* survey would provide a dataset that could be used to calibrate the *eROSITA* mass-observable relations at $z$~1 (i.e. Andersson et al. 2011, Bazin et al., in prep.).

There are currently three surveys that have probed deeply enough to select galaxy clusters. The *Planck* all-sky survey provides an excellent match to *eROSITA* because of the all-sky coverage. An analysis of the *Planck* all-sky sample has led to the creation of a sample of ~200 clusters and almost all of those are previously known clusters selected from the *ROSAT* survey (Planck Collaboration 2011). This low redshift weighting in the *Planck* cluster sample comes from the poor angular resolution of 8 arcmin at 150GHz, which makes it difficult to select the arcminute-scale clusters at higher redshift because of the signal dilution within the background CMB.

Another survey is the Atacama Cosmology Telescope (*ACT*) which uses a 6-meter telescope to image the sky at a resolution of around 1.5 arcmin. The smaller beam size is better matched to cluster scales, and the survey site has enabled surveys to about 50 μK-armin at 150GHz over approximately 800 deg$^2$ of sky. This survey has turned up a few dozen very interesting clusters, including systems at redshifts approaching $z$ ~1 (Marriage et al. 2011).

The South Pole Telescope (*SPT*) is a 10-meter telescope operating at the South Pole where the conditions allow it to stay on-sky surveying for periods extending to 9 months (Carlstrom et al. 2011). The *SPT* survey has 1 arcmin resolution at 150GHz and has just completed a 2,500 deg$^2$ survey to 18μK-armin depths, with associated depths at 90GHz and 220GHz of 40μK-arcmin and 60μK-arcmin, respectively (Vanderlinde et al. 2010). *SPT* is picking up all clusters at $M_{200}>4\times10^{14}$ M$_\odot$, independent of redshift, and their highest redshift cluster to date is at $z$ =1.35. The sample of clusters over the full survey is ~600, and this survey will increase the sample of known massive clusters at $z$>0.8 by more than an order of magnitude over previous samples. Extensive X-ray followup using *XMM-Newton* and *Chandra* pushing out to $z$ =1.35 is revealing the fluxes, ICM masses, temperatures (and redshifts!) of these systems, and interestingly the *SPT* detection limit at $z$ ~1.1 corresponds to an X-ray flux of around 5×10$^{-14}$ ergs/s/cm$^2$, corresponding to a source with a few tens of photons at the typical depth of the *eROSITA* all sky survey. This suggests that the *eROSITA* cluster sample will have a tail of massive systems extending to beyond $z$ ~1. A challenge will be identifying these and determining their redshifts.

A fraction of the X-ray sources to be discovered with *eROSITA* will also emit at radio frequencies. A cross comparison between X-ray and radio emission will enable (i) improved identification of sources, (ii) subdivision of source classes (e.g., radio load vs. radio quiet AGN), and (iii) possibly even constraints on source redshifts (e.g., through HI and polarization measurements).

Several new large-area sensitive radio surveys in both hemispheres are currently being performed or planned. They include, e.g., *LOFAR*, *WSRT*/WODAN, *ASKAP* (especially the surveys EMU, WALLABY, and POSSUM), and *MeerKAT* (especially Mightee). See, e.g., Norris et al. (2011) for a short description of these and other surveys as well as a detailed account of EMU. These surveys will provide similar increases in sensitivity compared to previous radio surveys as *eROSITA* will improve upon *ROSAT*. Therefore, excellent opportunities for exploiting the radio/X-ray synergy in a timely manner exist.

## 6.4 X-ray followup

X-ray followup of galaxy clusters and AGN will be enormously helpful in the study of these sources. For example, with *Chandra* and *XMM-Newton* it would be possible to get detailed spectra of a subset of high redshift galaxy clusters that lie near the *eROSITA* detection limit. With followup one can push from a hundred or so *eROSITA* photons to the few thousand required for an accurate temperature measurement, robust ICM mass $M_{ICM}$ measurement or X-ray Compton-Y proxy $Y_X$. These then provide mass proxies with scatter in the 10% to 15% range as compared to the much higher scatter (45%) X-ray luminosity $L_X$ mass proxy (i.e. Vikhlinin et al. 2009). Regardless of the state of the aging X-ray observatories *XMM-Newton* and *Chandra*, the *eROSITA* mission has its own pointed phase. Roughly speaking, plans have been to focus on a sample of ~$10^3$ systems in such a pointed phase to enable improved mass proxies for these systems and to carry out astrophysical evolution studies of the ICM entropy and metallicity. An interesting possibility would be to coordinate with the upcoming *Astro-H* mission which will carry a high energy resolution calorimeter. Spectra with a few eV resolution would open a new window on ICM astrophysics, enabling studies of turbulence in the ICM of these clusters and detailed single element abundances.

## 6.5 Followup of galactic X-ray sources

The *eROSITA* X-ray sky will be populated by several 100,000 coronal emitters and several 10,000 accreting binaries. The source composition will be a strong function of the galactic latitude. Identification work in the galactic plane is challenged by strong extinction and confusion with background AGN. Efficient preselection of optical counterparts for multi-fiber spectroscopy with e.g. 4MOST will require a combination of optical and IR survey data. Fortunately, the VPHAS+ and VHS surveys to be performed at the ESO VST and VISTA telescopes will have released their data before *eROSITA*. The identification rate will nevertheless be rather low, <30% (see e.g. Motch et al. 2010). At high galactic latitudes a distinction between stellar coronal and extragalactic X-ray sources is much simpler due to the soft X-ray spectra of stellar coronae and the small flux ratio between the X-ray and the optical energy band (see Fig. 6.6.1). *GAIA* will play a very important role to separate galactic from extragalactic sources based on parallax and proper motion.

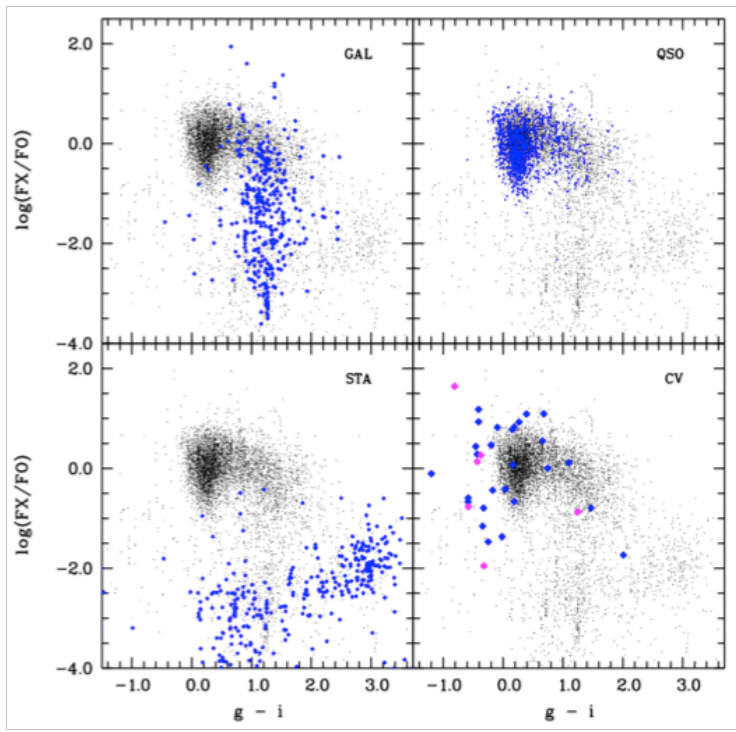

*Figure 6.6.1: X-ray to optical vs. optical color of unresolved XMM-Newton X-ray sources with SDSS counterparts. The association with the main source classes QSO, GALAXY, and STAR follows Pineau et al. (2011). CVs are identified from Szkody et al (2011, and references therein), and own compilations (magenta indicates ultra-compact binaries with two degenerate stars). Small black symbols indicate associated sources with SDSS photometry, whereas colored symbols indicate a secure identification based on SDSS-spectroscopy or parallax or other criteria.*

The identification of compact stellar accreting sources with white dwarf accretors (summarized under the CV-label) is not possible from X-ray and optical imaging data alone. The X-ray morphology, the X-ray color, optical colors and the $F_X/F_O$ ratio makes CVs almost indistinguishable from AGN. Spectroscopy over large sky areas is of utmost importance to build sufficiently large samples of the various binaries classes. As a matter of fact, the similarity of AGN and CVs will reveal CVs in large number with the planned wide-field spectroscopic AGN surveys without further preselection. Assigning a CV subclass will not be unique in many cases, if just one identification spectrum is available. It is thus foreseen to put a focus on imaging surveys with a built-in timing component. Apart from the mentioned Pan-STARRS, SkyMapper, and LSST surveys, the Catalina Real-Time Transient Survey (crts.caltech.edu) and the PTF are of particular importance since both cover a large piece of the sky in many epochs and, in particular, construct long-term light curves. These are important to search for periodic variability, and to study previous outburst evolutions of most of the compact stellar objects, as well as of AGN.

# List of acronyms

4MOST: 4-meter Multi-Object Spectroscopic Telescope
$\Lambda$CDM: $\Lambda$-Cold Dark Matter
AAAC: Astronomy and Astrophysics Advisory Committee
ABRIXAS: A BRoad-band Imaging X-ray All-sky Survey
ACT: Atacama Cosmology Telescope
ADC: Analog-Digital Converter
AGN: Active Galactic Nuclei
ASCA: Advanced Satellite for Cosmology and Astrophysics
ASIC: Application-Specific Integrated Circuit
ASKAP: Australian Square Kilometer Array Pathfinder
ASM: All-Sky Monitor
AXP: Anomalous X-ray Pulsar
BAO: Baryonic Acoustic Oscillations
BAT: Burst Alert Telescope
BH: Black Hole
BOSS: Baryonic Oscillation Spectroscopic Survey
CAMEX: CMOS Analog MultiplEXing
CCD: Charge-Coupled Device
CFHT: Canada France Hawaii Telescope
CIE: Collisional Ionization Equilibrium
CL: Confidence Limit
CMB: Cosmic Microwave Background
CRTS: Catalina Real-time Transient Survey
CT: Compton Thick
CTI: Charge Transfer Inefficiency
CXRB: Cosmic X-ray Background
CV: Cataclysmic Variable
DD: Double Degenerate
DE: Dark Energy
DES: Dark Energy Survey
DETF: Dark Energy Task Force
DLR: Deutschen Zentrums für Luft- und Raumfahrt
DOE: Department of Energy
EMU: Evolutionary Map of the Universe
EPIC: European Photon Imaging Camera
eRASS: eROSITA All-Sky Surveys
eROSITA: extended ROentgen Survey with an Imaging Telescope Array
EW: Equivalent Width
FIR: Far-Infrared
FoM: Figure of Merit
FoV: Field of View
FPGA: Field-Programmable Gate Array
FWHM: Full Width Half at Maximum
GBM: Gamma-ray Burst Monitor
GFRP: Glass Fibre Reinforced Polymer
GRB: Gamma-Ray Burst
GRXE: Galactic Ridge X-ray Emission
HEAO-1: High Energy Astronomy Observatory-1
HEPAP: High Energy Physics Advisory Panel
HEW: Half-Energy Width
HMXB: High-Mass X-ray Binary
HRD: Hertzsprung-Russell Diagram
HSC: Hyper-Suprime Cam
ICM: Intra-Cluster Medium
IMF: Initial Mass Function
INS: Isolated Neutron Star
INTEGRAL: International Gamma-Ray Astrophysics Laboratory

IP: Intermediate Polar
ISM: Interstellar Medium
ISS: International Space Station
KIDS: KIlo-Degree Survey
LAB: Leiden/Argentine/Bonn
LAT: Large Area Telescope
LHB: Local Hot Bubble
LISA: Laser Interferometer Space Antenna
LISM: Local Interstellar Medium
LMC: Large Magellanic Cloud
LMXB: Low-Mass X-ray Binary
LOFAR: LOw-Frequency ARray
LRG: Luminous Red Galaxy
LSST: Large Synoptical Survey Telescope
MCELS: Magellanic Cloud Emission Line Survey
MCV: Magnetic Cataclysmic Variable
MIR: Mid-Infrared
MOS: Metal Oxide Semi-conductor
MW: Milky Way
NASA: National Aeronautics and Space Administration
NEI: Non-Equilibrium Ionization
NEP: North Ecliptic Pole
NGO: New Gravitational wave Observatory
NIR: Near Infrared
NRTA: Near Real-Time Data Analysis software
NSF: National Science Foundation
NuSTAR: Nuclear Spectroscopic Telescope ARray
OIR: Optical-Infrared
Pan-STARRS: Panoramic Survey Telescope and Rapid Response System
PCA: Proportional Counter Array
PNG: Primordial Non-Gaussianity
PSD: Power Spectral Density
PSF: Point Spread Function
PSPC: Position Sensitive Proportional Counters
PTF: Palomar Transient Factory
QSO: Quasi- Stellar Object
RASS: ROSAT All-Sky Survey
RBS: ROSAT Bright Survey
REXCESS: Representative XMM-Newton Cluster Structure Survey
RN: Recurrent Nova
RRAT: Rotating Radio Transient
RXTE: Rossi X-ray Timing Explorer
SASS: Science Analysis Software System
SDSS: Sloan Digital Sky Survey
SED: Spectral Energy Distribution
SFRs: Star-Forming Regions
SGR: Soft Gamma-ray Repeater
SMBH: Supermassive Black Hole
SMC: Small Magellanic Cloud
SMEX: SMall EXplorer
NS: Neutron Star
S/C: Spacecraft
SNIa: Supernova Ia
SNR: Supernova Remnant
SPIDERS: SPectroscopic IDentification of ERosita Sources
SPT: South Pole Telescope
SRG: Spektrum Roentgen Gamma
SSS: Super-Soft X-ray Source
SWCX: Solar Wind Charge Exchange
SXRB: Soft X-ray Background

SZE: Sunyaev-Zeldovich Effect
ULX: Ultra-Luminous X-ray source
UCB: Ultra-Compact Binary
VHS: Vista Hemisphere Survey
VIKING: VISTA Kilo-degree Infrared Galaxy Survey
VLT: Very Large Telescope
VPHAS+: VST Photometric H-alpha Survey of the Southern Galactic Plane
VST: VLT Survey Telescope
WD: White Dwarf
WISE: Wide-Field Infrared Survey Explorer
WMAP: Wilkonson Microwave Anisotropy Probe
WSRT: Westerbork Synthesis Radio Telescope
XDINS: X-ray Dim Isolated Neutron Stars
XIS: X-ray Imaging Spectrometer
XLF: X-ray Luminosity Function
XRB: X-ray Binary
XSPEC: X-Ray SPECtral fitting package